\documentclass[twocolumn, twocolappendix, floatfix]{aastex631}
\usepackage{amsmath}

\newcommand{\rowname}[1]

\shorttitle{Supernova Energies and Explosion Probability}
\shortauthors{Steinwandel \& Goldberg}

\graphicspath{{./}{figures/}}

\begin{document}

\title{Some stars fade quietly: Varied Supernova explosion outcomes and their effects on the multi-phase interstellar medium}

\author[0000-0001-8867-5026]{Ulrich P. Steinwandel}
\affiliation{Center for Computational Astrophysics, Flatiron Institute, 162 5th Avenue, New York, NY 10010, USA}
\email{usteinwandel@flatironinstitute.org}

\author[0000-0003-1012-3031]{Jared A. Goldberg}
\affiliation{Center for Computational Astrophysics, Flatiron Institute, 162 5th Avenue, New York, NY 10010, USA}
\email{jgoldberg@flatironinstitute.org}

\begin{abstract}

We present results from galaxy evolution simulations with a mutiphase Interstellar medium (ISM), a mass resolution of $4$ M$_{\odot}$ and a spatial resolution of 0.5 pc. These simulations include a stellar feedback model that includes the resolved feedback from individual massive stars and accounts for heating from the far UV-field, non-equilibrium cooling and chemistry and photoionization. In the default setting, individual supernova (SN) remnants are realized as thermal injections of $10^{51}$ erg; this is our reference simulation WLM-fid. Among the remaining seven simulations, there are two runs where we vary this number by fixing the energy at $10^{50}$ erg and $10^{52}$ erg (WLM-1e50 and WLM-1e52, respectively). We carry out three variations with variable SN-energy based on the data of Sukhbold et al. (2016) (WLM-variable, WLM-variable-lin, and WLM-variable-stoch). We run two simulations where only 10 or 60 percent of stars explode as SNe with $10^{51}$ erg, while the remaining stars do not explode (WLM-60prob and WLM-10prob). We find that the variation in the SN-energy, based on the tables of Sukhbold et al. (2016), has only minor effects: the star formation rate changes by roughly a factor of two compared to the fiducial run, and the strength of the galactic outflows in mass and energy only decreases by roughly 30 percent, with typical values of $\eta_m \sim 0.1$ and $\eta_e \sim 0.05$ (measured at a height of 3 kpc after the hot wind is fully decoupled from the galactic ISM). In contrast, the increase and decrease in the canonical SN-energy has a clear impact on the phase structure, with loading factors that are at least 10 times lower/higher and a clear change in the phase structure. We conclude that these slight modulations are driven not by the minor change in SN-energy but rather by the stochasticity of whether or not an event occurs when variable SN-energies are applied.

\end{abstract}


\section{Introduction} \label{sec:intro}

Galactic winds are ubiquitous in galactic systems and play a major role in regulating their baryon cycle, setting their star formation rate and determining their chemical evolution \citep[see, e.g.,][for recent reviews]{Somerville2015, Naab2017}. Some forms of modeling for galactic winds in large-scale cosmological simulations \citep[e.g.,][]{Vogelsberger2014, Hirschmann2014, Schaye2015, Pillepich2018, Nelson2019} and cosmological zoom-in simulations \citep[e.g.,][]{Guedes2011, Agertz2013, Hopkins2014, Hopkins2018, Hopkins_fire3, Wang2015} have been proven to successfully reproduce important observed galaxy scaling relations such as the stellar-halo mass relation or the mass-metallicity relation \citep[see, e.g.,][for more details]{Moster2018,Behroozi2018}. While such models mark an incredible success in the numerical modeling of the formation of galaxies, all such models rely on some form of sub-grid model for the treatment of SN-feedback that, at typical mass resolutions of $10^3$ to $10^6$ M$_{\odot}$ (where thermal energy injection falls victim to the ``overcooling'' problem), remains \textit{unresolved} in the majority of the simulation domain. Often, these simulations use some mechanism to handle that issue by inputting the terminal momentum found in high-resolution simulations of individual SN-explosions at the end of the Sedov-Taylor phase \citep[e.g.,][]{Cioffi1988, Thornton1998, Petruk2006, Kim2015, Haid2016, Steinwandel2020}; this is typically offset by some factor (from 3 to 5, depending on the exact parameterization of the feedback implementation) to account for the momentum gain in the pressure-driven snowplow phase of the remnant. It is interesting to note that \citet{Kim2015} and \citet{Steinwandel2020} showed in that context that the difference between simulations in homogeneous vs. turbulent media seems to make little difference in terms of momentum gain until shell formation. 

Although  such ``momentum feedback models'' in galaxy formation simulations have had tremendous success in regulating star formation via the injection of turbulence into the ISM, they do somewhat lack the ability to establish a resolved hot phase of the ISM, which is responsible for the launching of outflows from the ISM. However, the latter point is still under debate, and it has been shown, for instance for the Fire-2 simulations, in \citet{Pandya2021} that there is a clear hot phase in the ISM and in galactic outflow. 

Recently, it has become possible to simulate individual galaxies at such high resolution that individual SN-explosions from single massive stars can be resolved. This transition, from marginally resolved remnants to fully resolved remnants, needs a mass resolution of the order of $\sim 1$ M$_{\odot}$; this has been shown explicitly in \citet{Hu2017} and \citet{Steinwandel2020} for particle codes that use either smoothed particle hydrodynamics (SPH) or the newer meshless finite mass (MFM) method \citep{Gaburov2011, Hopkins2015}.

These simulations have been especially useful in dwarf galaxies \citep[e.g.,][]{Hu2016, Hu2017, Hu2019, Emerick2019, Steinwandel2020, Hislop2022, Gutcke2021, Smith2021, Steinwandel2022, Lahen2023, Hu2023} and have more recently also been applied to more complex environments such as merger simulations \citep[e.g.,][]{Lahen2020a, Lahen2020b}, more massive galaxies \citep{SteinwandelLMC},  cosmological environments \citep{Gutcke2022}, and higher-metallicity stratified box-type simulations \citep[e.g.,][]{Hu2021, Hu2022}. These models are an important step forward and allow for detailed studies of the origin of multiphase galactic winds in global galactic disc simulations, as well as detailed studies of the ISM itself. However, since these models incorporate the detailed results of individual stellar evolution, the results of these simulations do depend to some extent on the results obtained from the stellar evolution community. In fact, it has been become quite popular in recent years for ``galaxy simulators''  to treat the results obtained from stellar evolution as ``ground truth'' and just implement them into their star-by-star feedback models. This has led to a number of interesting papers studying the effects of runaway stars \citep[e.g.,][]{Andersson2022, Steinwandel2022}, variable SN-energy \citep[e.g.,][]{Gutcke2021}, SN-explosion mass cutoff \citep[e.g.,][]{Keller2022}, FUV heating and photoionizing radiation from individual sources \citep[e.g.,][]{Hu2017, Smith2021} and stellar winds \citep[e.g.,][]{Lahen2023}.

In this paper, we dive deeper into one aspect of these previous works: namely, we explore in detail the variability of the SN-feedback energy and discuss the caveats one should keep in mind when implementing these as a sub-grid model in numerical simulations of galaxy formation and evolution.
Predicting which stars will explode as SNe and identifying the SN explosion energy scale are areas of active ongoing research in stellar astrophysics. On the observational end, there is the so-called ``Missing Red Supergiant Problem," which is the statement that in observed hydrogen-rich supernovae with observed progenitors, the luminosities of those progenitor stars are systematically lower, compared to the distribution of red supergiant luminosities observed in nearby galaxies \citep[see, e.g.,][]{Smartt2009, Smartt2015,VanDyk2017,Kochanek2020}, with a suggested cutoff of $\sim 10^{5.2}L_\odot$. This might imply an upper mass limit on  hydrogen-rich stars that explode (though there is also some debate over the statistical significance of the ``problem;" see, e.g., discussions in \citealt{Davies2018,Davies2020a,Davies2020b}).  
On the theoretical end, modeling efforts that analyze stellar evolution models and their structure compared to various parameterized models of ``explodability" \citep[e.g.,][]{Sukhbold2016, Sukhbold2018, Ertl2016, Ertl2020} as well as self-consistent 3D simulations \citep[e.g.,][]{Glas2019a, Glas2019b, Burrows2020} find so-called ``islands of explodability" in which the explosion outcome of a star (i.e., whether or not it explodes, and with how much energy) varies with the initial stellar mass. 
Moreover, predicting the explosion outcome of a star is sensitive to the stellar structure at the time of core collapse, which is in turn sensitive to the input physics and modeling choices made in stellar evolution models \citep[see, e.g.,][]{Farmer2016}, including nuclear reaction rates \citep[such as C burning; e.g.,][]{Sukhbold2020}, core-boundary mixing \citep[e.g.,][]{ADavis2019}, stellar winds \citep[e.g.,][]{Renzo2017}, and stellar evolution in a binary system \citep[e.g.,][]{Laplace2021, Zapartas2021}. Though different metrics make different predictions for the final fates of different stars \citep[e.g.,][]{Patton2020, Patton2022}, there is a growing consensus that some stars within the range of $\sim$ 10-100 $M_\odot$ will explode, while others will not. In this work, we aim to discuss more openly the advantages and caveats involved when results from 1D stellar evolution are adopted in galaxy formation simulations, since these results have become more and more relevant as sub-grid models in simulations of galaxy evolution in the dwarf galaxy regime \citep[e.g.,][]{Hu2016, Hu2017, Hu2019, Emerick2019, Smith2021, Gutcke2021, Hislop2022, Steinwandel2022, SteinwandelLMC, Lahen2023}.

The structure of this paper is as follows. In Sec.~\ref{sec:methods}, we will briefly discuss the numerical methods used and describe the initial conditions adopted. In Sec.~\ref{sec:results}, we will present the main findings of our study concerning the time evolution of outflow rates and loading factors and their spatial properties. In Sec.~\ref{sec:discussion}, we will discuss our results and compare to previous work where appropriate. In Sec.~\ref{sec:conclusions}, we will summarize our findings and discuss pathways for future improvements.

\section{Numerical Methods and Simulations} \label{sec:methods}

\begin{table*}
    \caption{Overview of the physics variations adopted in our different simulations.} 
    \label{tab:one}
    \begin{tabular}{lcccccc}
      \hline
	  Name		& Supernovae & PI-radiation & PE-heating & SN-energy\\
	  \hline
      WLM-fid & \checkmark  & \checkmark & \checkmark & $1\mathrm{e}51$ erg\\
      WLM-1e50 & \checkmark & \checkmark & \checkmark & $1\mathrm{e}50$ erg\\
      WLM-1e52 & \checkmark &  \checkmark & \checkmark & $1\mathrm{e}52$ erg\\
      WLM-variable & \checkmark & \checkmark & \checkmark & \citet{Sukhbold2016} (nearest neighbors)\\
      WLM-variable-lin & \checkmark & \checkmark & \checkmark & \citet{Sukhbold2016} (linear interpolation)\\
      WLM-variable-stoch & \checkmark & \checkmark & \checkmark & \citet{Sukhbold2016} (probabilistic)\\
      WLM-10prob & \checkmark & \checkmark & \checkmark & $1\mathrm{e}51$ erg (only for 10\%)\\
      WLM-60prob & \checkmark & \checkmark & \checkmark & $1\mathrm{e}51$ erg (only for 60\%)\\
      \hline
      \end{tabular}
\end{table*}

\subsection{Gravity and hydrodynamics}
All the simulations used in this paper are carried out with the Tree-multi-method code P-Gadget3 based on the codes P-Gagdet2 \citep{springel05} and Gizmo \citep{Hopkins2015} with important updates by \citet{Hu2014} and \citet{Steinwandel2020}. We solve gravity using a tree code \citep[][]{Barnes1986, springel05}. The code has the ability to use the particle-mesh (PM) method to speed up that calculation, but for the simulations presented here, we do not use this capability. Additionally, our code can use either the pressure-energy formulation of SPH, which is stabilized by a high-resolution shock-capturing method (time-dependent artificial viscosity) and a high-resolution maximum-entropy method (time-dependent artificial conduction), or the Riemann-based MFM method \citep{Gaburov2011, Hopkins2015} to solve for the fluid flows. For the latter method, we use an HLLC Riemann solver to reconstruct fluid fluxes between Lagrangian tracer particles. The surface areas for the flux computation are estimated based on smoothing lengths at the midpoint between the single tracer particles. We compute fluxes pairwise to avoid strong diffusion in the simulation domain. We adopt the MFM method for our simulations as it allows a resolution that is higher than that of SPH at fixed mass resolution by a factor of 2.5.

\subsection{Cooling and chemistry network}

We adopt a cooling and chemistry network similar to the tallbox framework used in the ``SILCC'' simulations \citep[e.g.,][]{Walch2015, Girichidis2016, Girichidis2018, Gatto2017, Peters2017}, which is based on the non-equilibrium cooling and heating prescription of the work of \citet{Nelson1997}, \citet{Glover2007a}, \citet{Glover2007b} and \citet{Glover2012}. Relevant rate equations have been updated following \citet{Gong2016}, making the framework relatively similar to the recent TIGRESS-NCR framework \citep{Kim2022} when considering the non-equilibrium chemistry aspect of the two simulation frameworks. Our network is tracing six individual species: molecular hydrogen (H$_{2}$), ionized hydrogen (H$^{+}$), and neutral hydrogen (H), as well as carbon monoxide (CO), ionized carbon (C$^{+}$) and oxygen. Additionally, we follow the free electrons and assume that all silicon in the simulation is stored as Si$^{+}$. The individual chemical reactions we follow are summarized in Table 1 of \citet{Micic2012}. We follow the generation of molecular hydrogen over dust grains by directly solving the non-equilibrium rate equations to obtain $x_{H_{2}}$ and $x_{H^{+}}$ and then obtain $x_H$ based on the conservation law:
\begin{equation}
    x_\mathrm{H} = 1 - 2 x_{\mathrm{H}_2} - x_{\mathrm{H}^{+}}.
\end{equation}
In the network we track x$_e$, which we compute via:
\begin{equation}
    x_\mathrm{e} = x_{\mathrm{H}^{+}} + x_{\mathrm{C}^{+}} + x_{\mathrm{Si}^{+}}
\end{equation}
Non-equilibrium cooling rates are obtained from the local density, temperature and chemical abundance of individual tracer particles. The leading cooling processes are the fine-structure line cooling of atoms and ions (C$^{+}$, Si$^{+}$ and O), the vibration and rotation line cooling due to molecules (H$_{2}$ and CO), the Lyman-$\alpha$ cooling of hydrogen, the collisional dissociation of H$_{2}$, the collisional ionization of hydrogen and the recombination of H$^{+}$ (in the gas phase and on dust grains). The main heating  processes are photoelectric heating from dust grains, the contribution from polycyclic aromatic hydrocarbonates (PAHs), cosmic ray ionization at a constant  rate of $10^{-19}$ s$^{-1}$, photodissociation of H$_{2}$, UV-pumping of H$_{2}$ and formation of H$_{2}$. For high temperatures (T  $> 3 \times 10^{4}$ K), we follow the  equilibrium cooling formalism first described by \citet{Wiersma2009} as implemented by \citet{Aumer2013}, including the elements H, He, C, N, O, Ne, Mg, Si, S, Ca, Fe, and Zn. 

\subsection{Star formation and stellar feedback in our fiducial model} \label{sec:default}
The star formation and stellar feedback models that we use here have been described in a number of previous works \citep[e.g.,][]{Hu2016, Hu2017, Hislop2022, Steinwandel2022, SteinwandelLMC}. Thus we  give only a short overview of the inner workings of each component. 
Individual stars are formed by sampling the initial mass function (IMF), following \citet[][]{Hu2016} and \citet[][]{Hu2017}, who adopt a  Schmidt-type star formation scaling with a fixed star formation efficiency of 2 percent per free-fall time. Further, we compute the Jeans mass for all gas cells and instantly convert any gas cell with less than 0.5 M$_\mathrm{J}$ in the MFM kernel radius  to a star particle in the next time step. The Jeans mass is given via: 
\begin{equation}
    M_{\mathrm{J},i} = \frac{\pi^{5/2}{c_{\mathrm{s},i}^3}}{6G^{3/2}\rho_{i}^{1/2}}.
\end{equation}  
Star particles above the resolution limit of 4 M$_{\odot}$ are treated as single stars. These single stars follow a variety of feedback channels, including photoelectric heating and photoionizing radiation as well as the metal enrichment from AGB stellar winds. The core of the stellar feedback scheme is SN-feedback as implemented in \citet{Hu2017} for the SPH version of the code  and in \citet{Steinwandel2020} for the MFM version of the code;  at our simulation's target resolution of  4 M$_{\odot}$, this allows both a self-consistent build-up of the hot phase and the momentum of individual SN-events in a resolved Sedov-Taylor phase \citep[see][]{Hu2021, Hu2022} with the deposition of the canonical SN-energy of $10^{51}$ erg into the ambient ISM. Moreover, we adopt lifetimes of massive stars based on \citet{Georgy2013}, allowing us to self-consistently trace the locations in the ISM where SNe explode. Thus, all the outflow properties in these simulations can be interpreted as being self-consistently launched from the ISM.
We follow the FUV radiation in a spatially dependent and time-dependent fashion, which we update on the gravity tree. Photoionizing radiation is implemented based on a Stroemgren approximation following the method of \citet{Hu2017}, which is itself based on the method used in \citet{Hopkins2012}. We call this fiducial model, with the canonical $10^{51}$-erg explosions, WLM-fid.

\subsection{Variable SN-Explosion energies}\label{sec:models}

In total, we perform eight simulations, which we discuss below and summarize in Table~\ref{tab:one}: three with fixed explosion energies (including the fiducial $10^{51}$-erg model discussed in Sec.~\ref{sec:default}), three with variable explosion energies as a function of stellar mass, and two with fixed explosion energies but varied explosion outcomes (i.e., not all massive stars explode).

The major difference between our default model of Sec.~\ref{sec:default} and the variable SN-energy implementation is that in the latter, we adopt the W20 stellar tables of \citet{Sukhbold2016} to select the explosion energy of each individual event. These tables identify different explosion energies for different stellar masses, ranging from around $3 \times 10^{50}$ erg to around $2 \times 10^{51}$ erg. (It is worth noting that if we apply a flat prior to the tables, only about 60\% of the stars undergo an explosion to begin with; the other $\sim$ 40\% of stars fail to explode or form direct-collapse black holes.) This tabulation is similar to what is implemented in \citet{Gutcke2021}; however, there are some notable differences. First, \citet{Gutcke2021} sample the IMF from 4 to 120 M$_{\odot}$, whereas we stop sampling at around 50 M$_{\odot}$ (although sampling out to 120 M$_\odot$ would not be a technical issue for our code). The reason for that is twofold. First, we do not believe that in a dwarf galaxy such as WLM, sampling the full IMF of supermassive stars of the order of 100 M$_{\odot}$ is meaningful. This is because, as has been shown in \citet{Hislop2022}, in dwarf galaxies such as our simulated WLM system, most star clusters that form are of rather low mass (between 100 and 1000 M$_{\odot}$). Star clusters of that size cannot sample the full IMF, as they originate from relatively small molecular clouds (see \citealt{Grudic2023} for a more detailed discussion). Second, the tables of \citet{Sukhbold2016} are very well sampled between 8 and 30 M$_{\odot}$ (in 0.25-M$_{\odot}$ bins from 8 to 13 M$_{\odot}$ and in 0.1-M$_{\odot}$ bins from 13 to 30 M$_{\odot}$), still well sampled between 30 and 50 M$_{\odot}$ (in 1-M$_{\odot}$ bins from 30 to 35 M$_{\odot}$ and in 5-M$_{\odot}$ bins out to 50 M$_{\odot}$), but very poorly sampled beyond 50 M$_{\odot}$. 

We do note, however, that we tested various methods of mapping the variable tabulated SN-energies to the massive stars. Essentially, we came up with three methods to implement interpolations across the tables:
\begin{enumerate}
    \item Nearest-grid-point mapping to the host star (hereafter called WLM-variable)
    \item Linear (first-order) interpolation from the tables to the host star (hereafter called WLM-variable-lin)
    \item Probability mapping to the host star (hereafter called WLM-variable-stoch); i.e., if the host star's mass is sitting exactly in the middle of two mass bins, it will have a 50 percent chance of ending up in each bin. If it is closer to one bin edge than to the other, the probability will be higher proportionally. This is enforced by drawing one random number and checking if it is closer to the left or the right bin and by what percentage, which is then applied as the probability for the left or the right bin, respectively. 
\end{enumerate}
As we see in Sec.~\ref{sec:results}, in the case of variable explosion energies, it does not really matter which of these interpolation schemes is adopted: our results for star formation and outflow loading factors are insensitive to this choice. 

As a further experiment, in addition to our fiducial model ($10^{51}$ erg, WLM-fid), we conducted two more runs with fixed supernova explosion energies for every massive star: $10^{50}$ erg (WLM-1e50) and $10^{52}$ erg (WLM-1e52). 
Moreover, we conducted two runs in which all SN explosions occur with fixed explosion energies of $10^{51}$ erg, but only 10\% (WLM-10prob) or 60\% (WLM-60prob) of massive stars explode. For the stars that do not explode in these runs and in the WLM-variable runs, we reject the energy feedback event but we still allow the mass transfer. We do that because we want to isolate the effect of SN-energy at fixed metal yield. The latter is then only offset by the galaxies' star formation rate.

\subsection{Initial conditions and simulations}

The initial conditions for the isolated galaxy simulations in this paper are similar to the initial conditions used in \citet{Hu2016}, \citet{Hu2017} (the simulations labeled  ``cmp'') and \citet{Smith2021} (the simulations in their main paper). We use the method of \citet{springel05b}. The dark matter halo is set up as a Hernquist profile with concentration parameter c = 10, virial radius R$_\mathrm{vir}$ = 44 kpc and virial mass M$_{vir}$ = $2 \times 10^{10}$ M$_{\odot}$.  The initial gas mass of the system is $4 \times 10^{7}$ M$_{\odot}$ and the stellar background potential has a mass of $2 \times 10^{7}$ M$_{\odot}$. The initial disk has 4 million dark matter particles, 10 million gas particles and 5 million ``old'' star particles, yielding a dark matter particle mass resolution of $6.8 \times 10^3$ M$_{\odot}$ and a baryonic particle mass resolution of $4$ M$_{\odot}$. The gravitational softening lengths are 62 pc for dark matter and 0.5 pc for gas and stars.
We use these initial conditions for a total of eight simulations, which represent variations of the injected SN-energy discussed below in Sec.~\ref{sec:models} and summarized in Table \ref{tab:one}. We note that for all simulations, we adopt an identical parameterization of the star formation model, with a Jeans mass threshold and an efficiency parameter of 2 percent.

\section{Results} \label{sec:results}

\begin{figure*}
    \centering
    \includegraphics[width=1.1\textwidth]{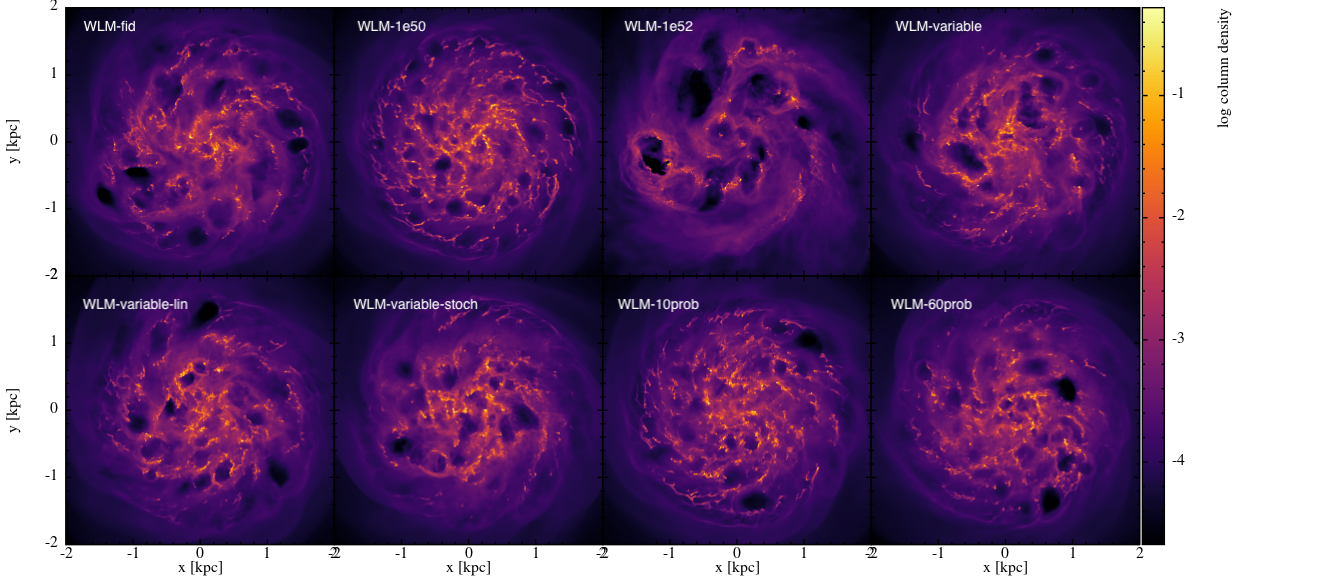}
    \caption{Gas surface-density projections for all our models. Color corresponds to column density in Gadgets' code units of $10^{10}$ M$_{\odot}$ kpc$^{-3}$. 
    \textit{Top row:} WLM-fid (left), WLM-1e50 (center left), WLM-1e52 (center right), WLM-variable (right). \textit{Bottom row:} WLM-variable-lin (left), WLM-variable-stoch (center left),  WLM-10prob (center right) and WLM-60prob (right). We note that the models with the largest difference in the structure of the ISM are WLM-1e50 and WLM-1e52: WLM-1e50 shows smaller bubbles and WLM-1e52 shows bigger bubbles.}
    \label{fig:splash}
\end{figure*}

\begin{figure}
    \centering
    \includegraphics[width=0.49\textwidth]{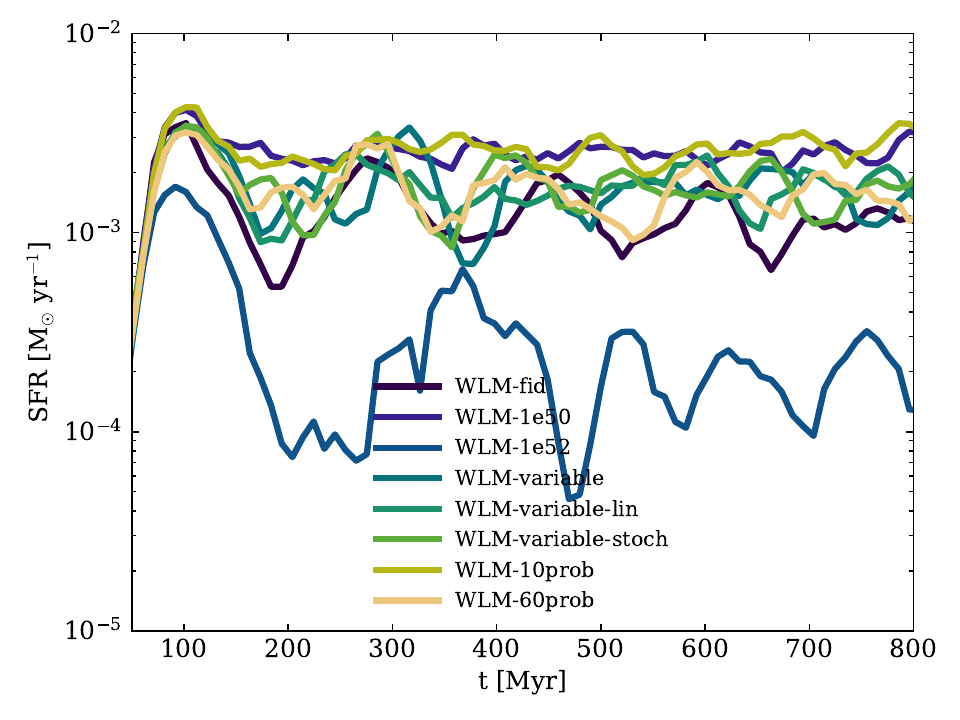}
    \caption{Star formation histories for all models as a function of time, as listed in Table \ref{tab:one}. The mean star formation rate of each model is listed in Table \ref{tab:loadings}. We show the star formation history out to 800 Myr of time evolution. We note that we find only a marginal impact on the star formation history in the runs in which we include variable SN-energies, as in \citet{Sukhbold2016}. }
    \label{fig:sfr}
\end{figure}

\begin{figure*}
    \centering
    \includegraphics[width=0.32\textwidth]{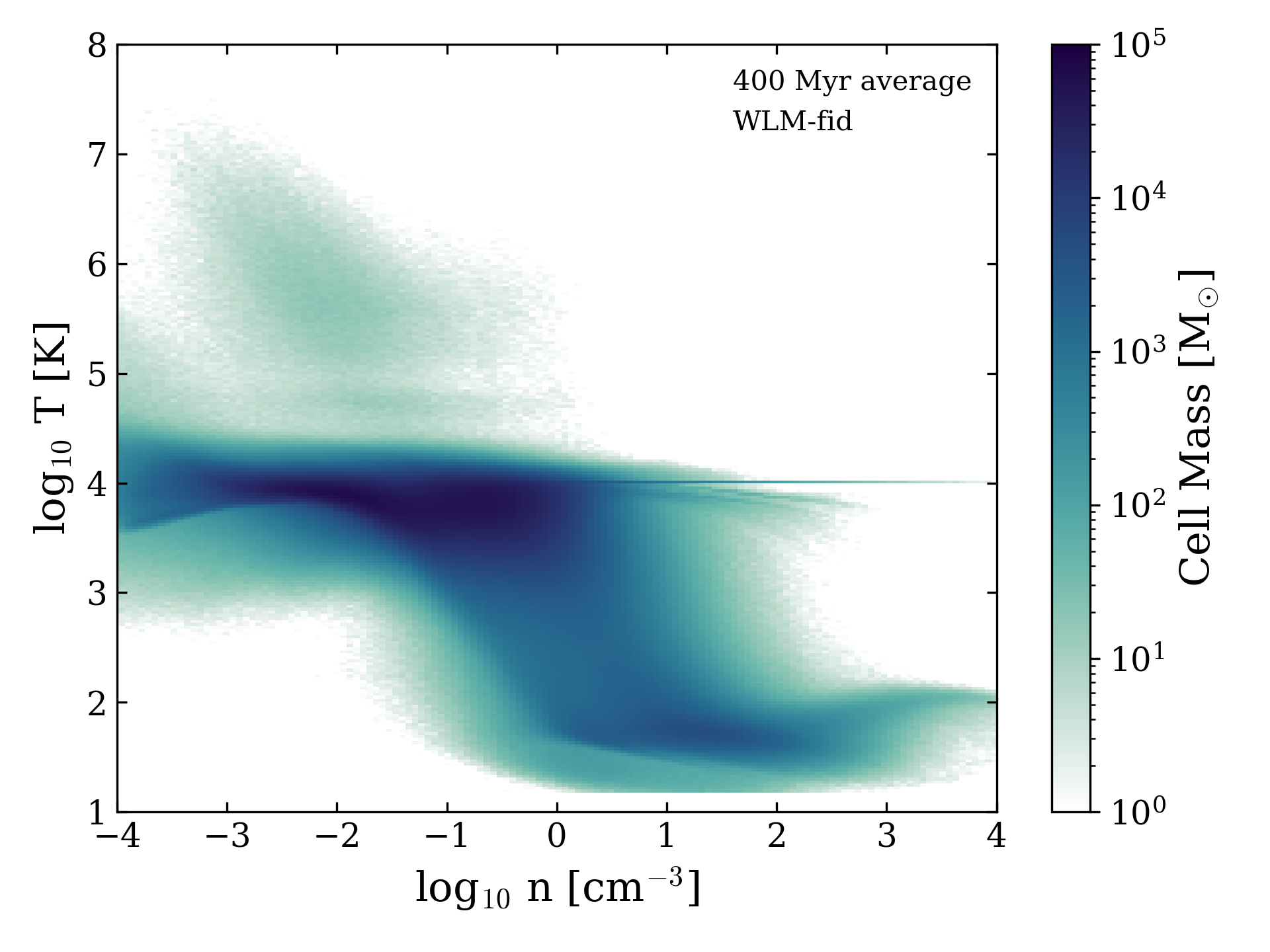}
    \includegraphics[width=0.32\textwidth]{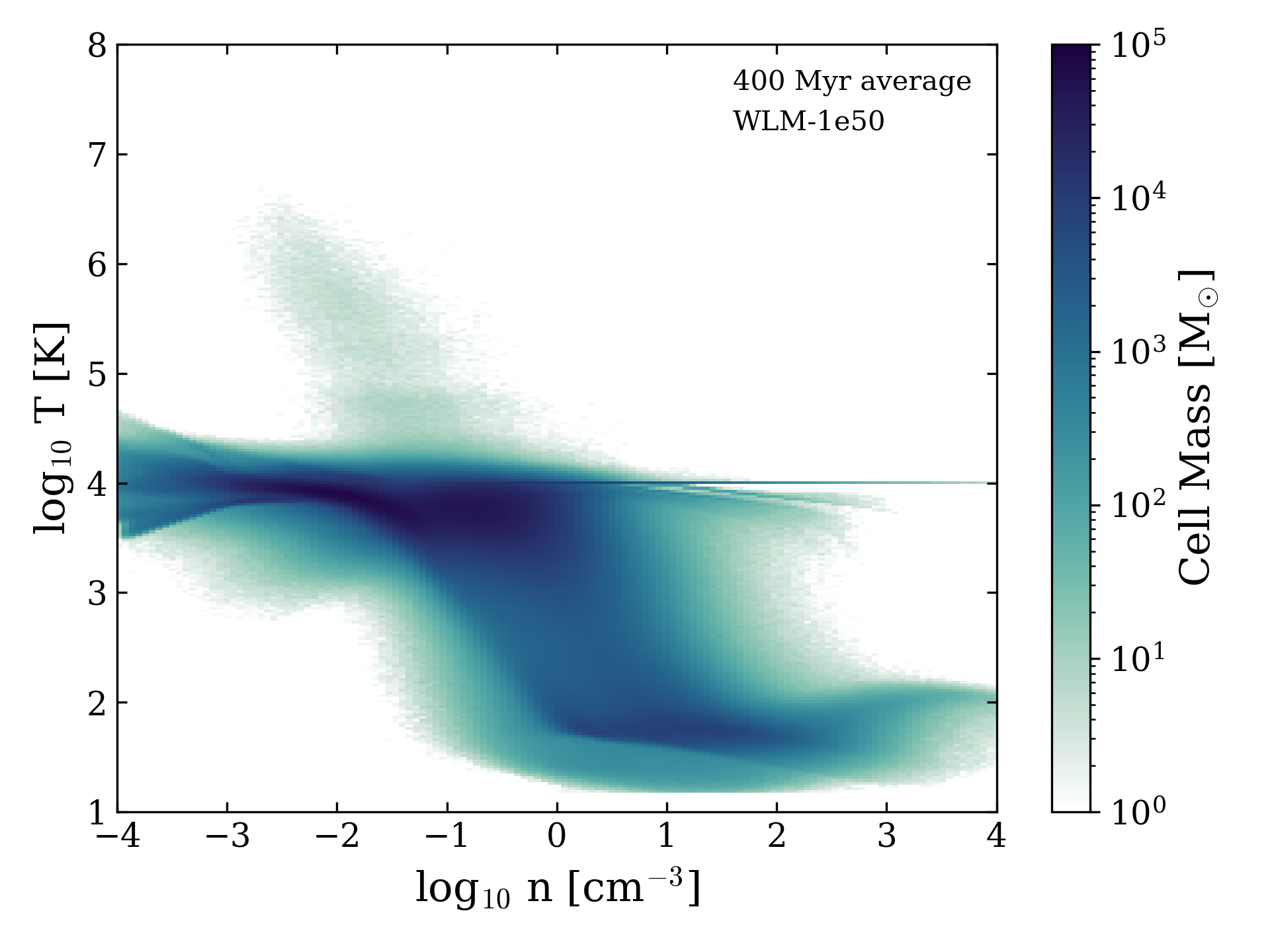}
    \includegraphics[width=0.32\textwidth]{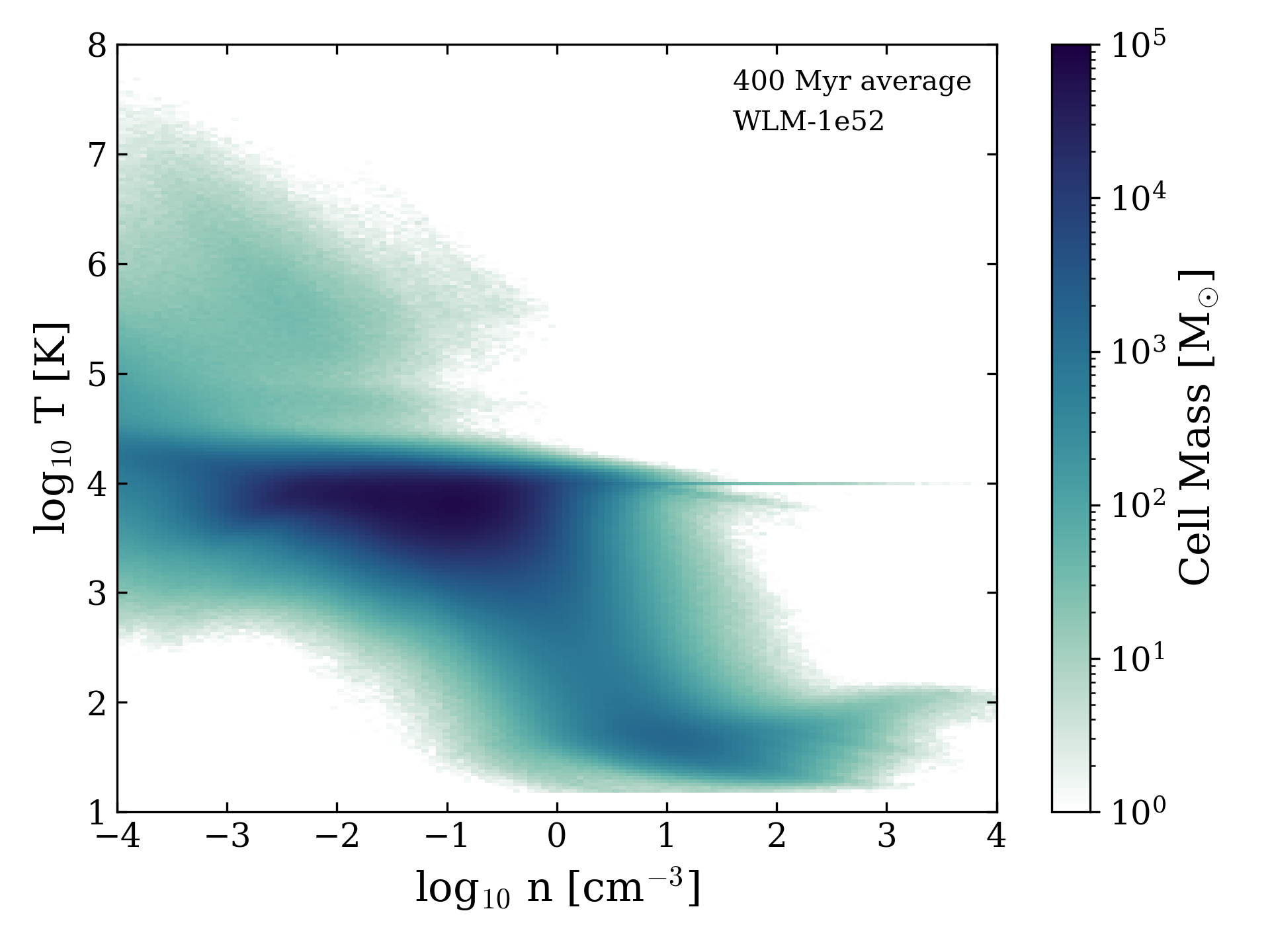}
    \includegraphics[width=0.32\textwidth]{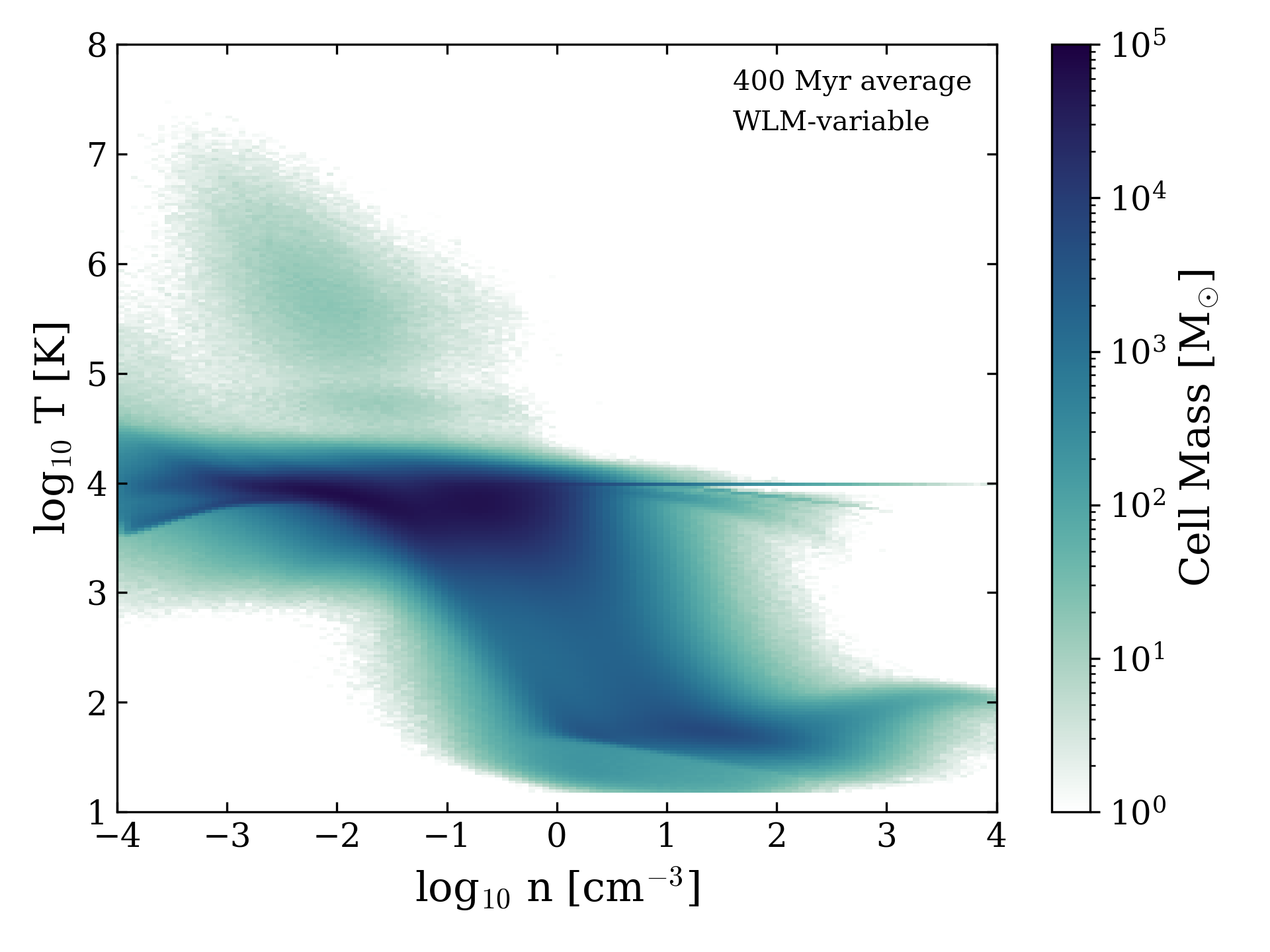}
    \includegraphics[width=0.32\textwidth]{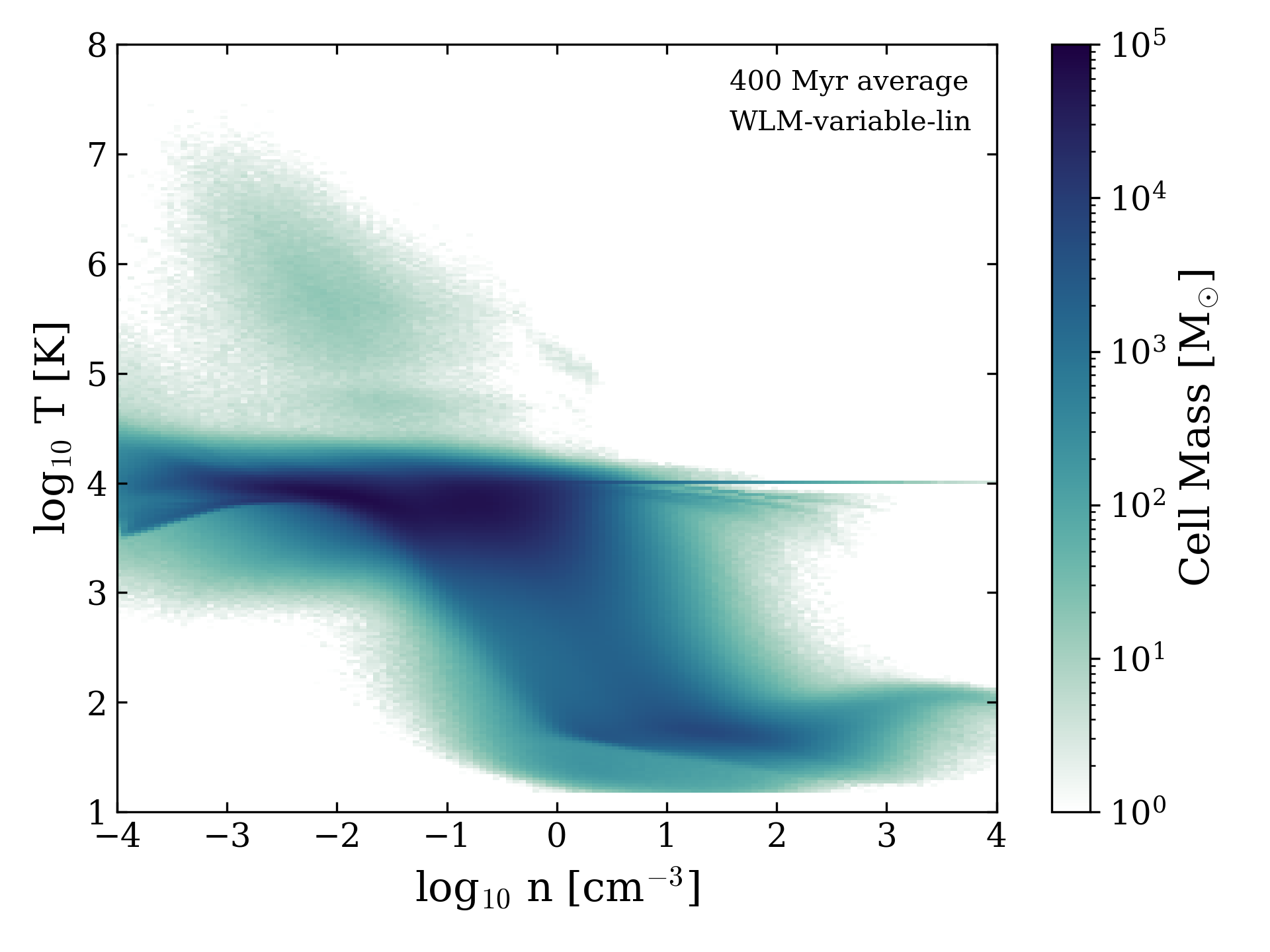}
    \includegraphics[width=0.32\textwidth]{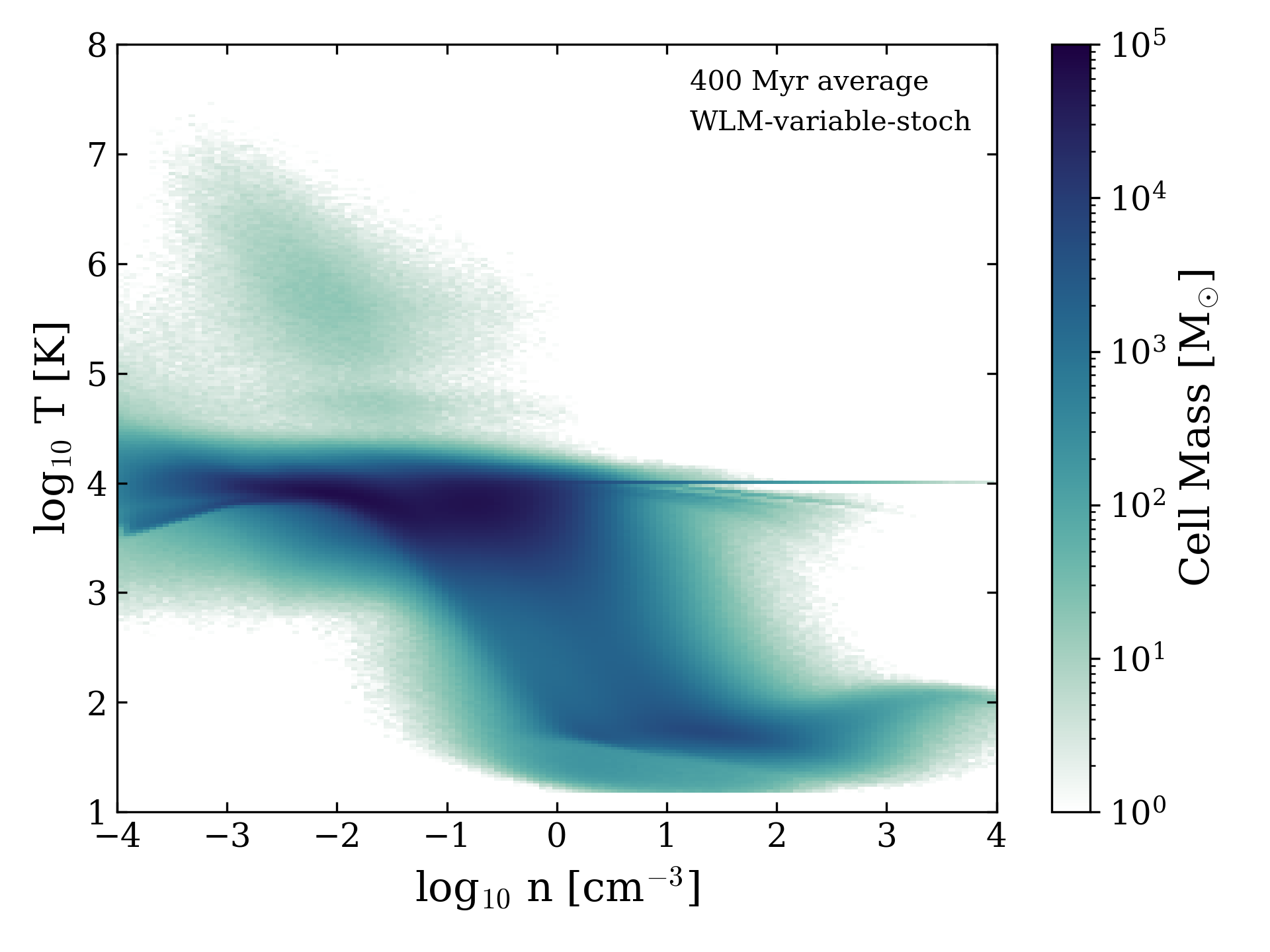}
    \includegraphics[width=0.32\textwidth]{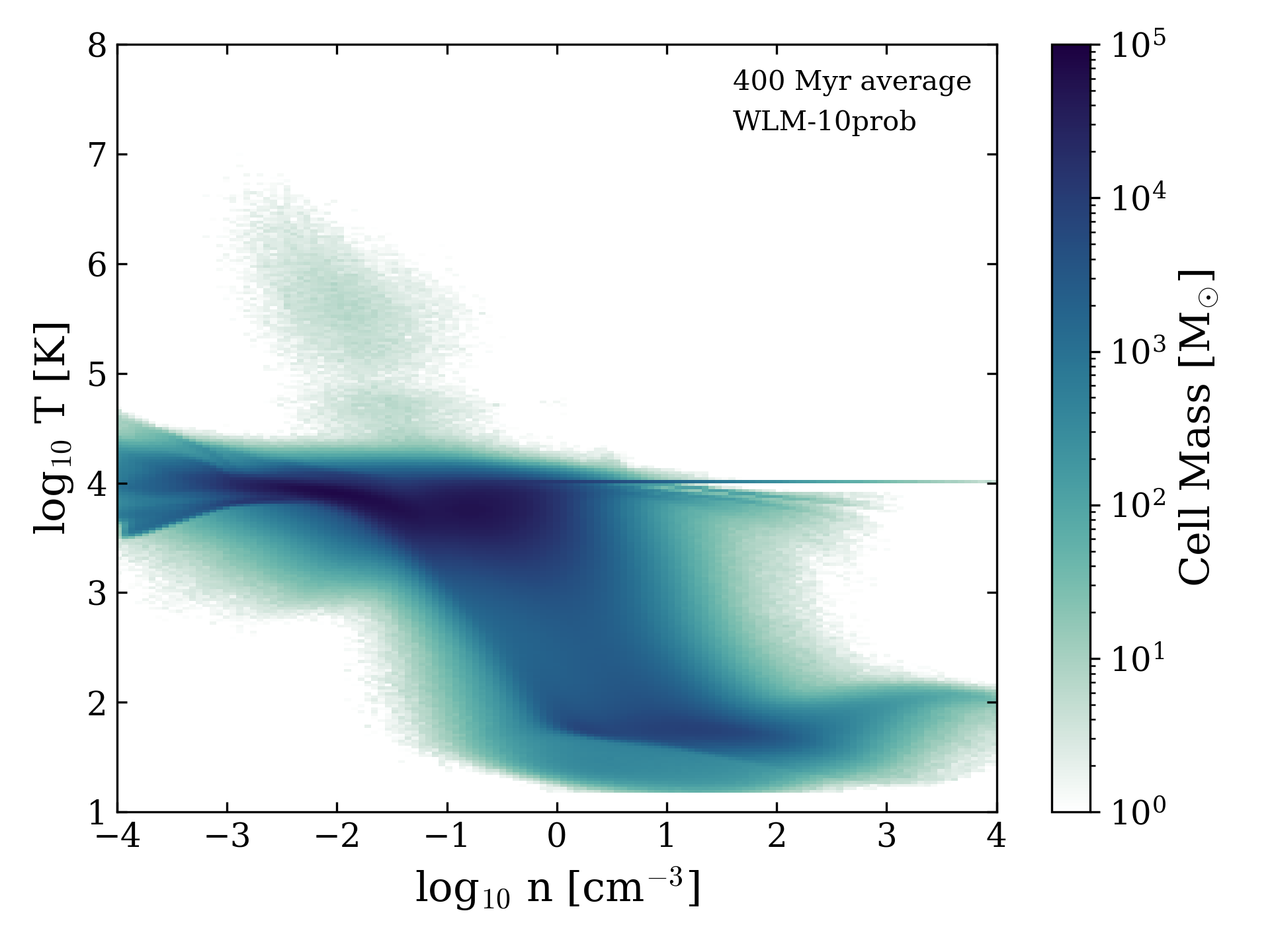}
    \includegraphics[width=0.32\textwidth]{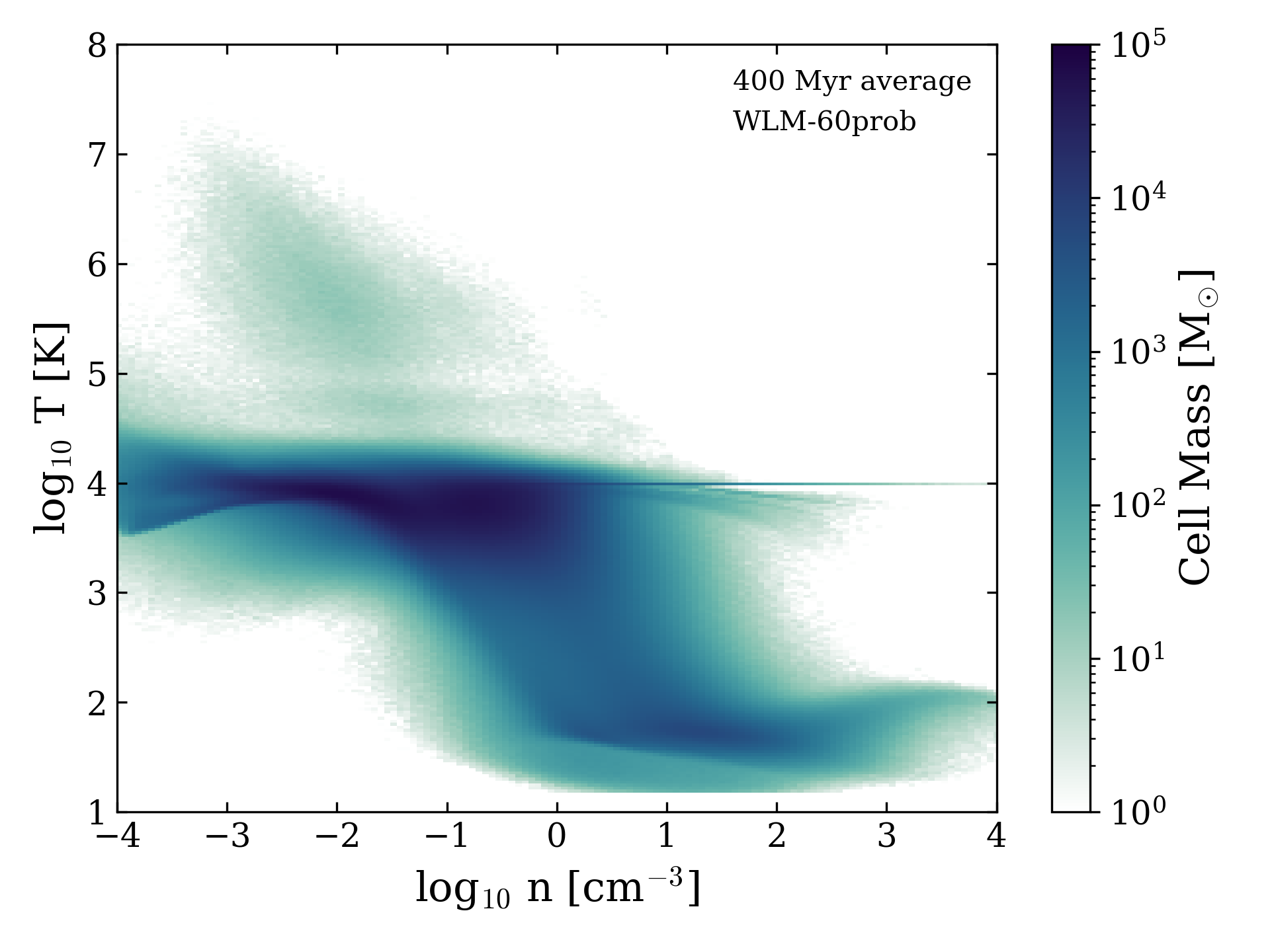}
    \caption{Time-averaged (between 200 and 800 Myrs in 10-Myr intervals) density-temperature phase space diagrams for all eight simulations. \textit{Top row:} WLM-fid (left), WLM-1e50 (center), WLM-1e52 (right). \textit{Center row:} WLM-variable (left), WLM-variable-lin (center), WLM-variable-stoch (right). \textit{Bottom row:} WLM-10prob (left) and WLM-60prob (right). The models that include variable SN-energy (WLM-variable, WLM-variable-lin and WLM-variable-stoch) differ very little from the fiducial model WLM-fid, while the models WLM-1e50 and WLM-10prob show a reduced potential for building the hot phase of the ISM. The model WLM-1e52 shows increased potential for build-up of the hot phase of the ISM, while the model WLM-60prob has a  phase structure that is similar to that of the fiducial model WLM-fid.}
    \label{fig:phase}
\end{figure*}

\begin{figure*}
    \centering
    \includegraphics[width=0.49\textwidth]{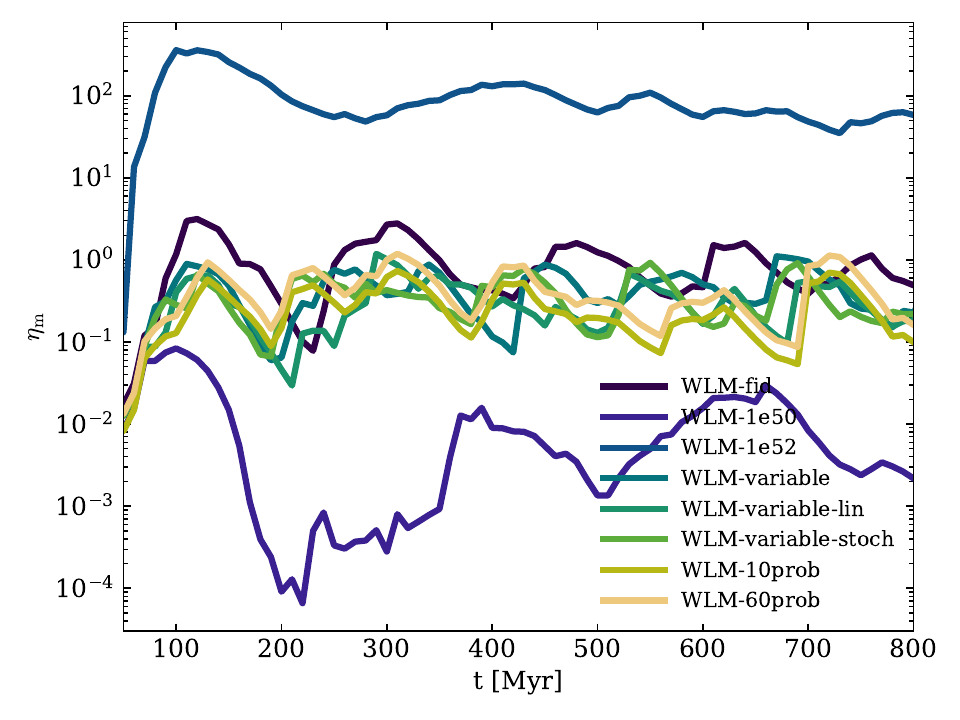}
    \includegraphics[width=0.49\textwidth]{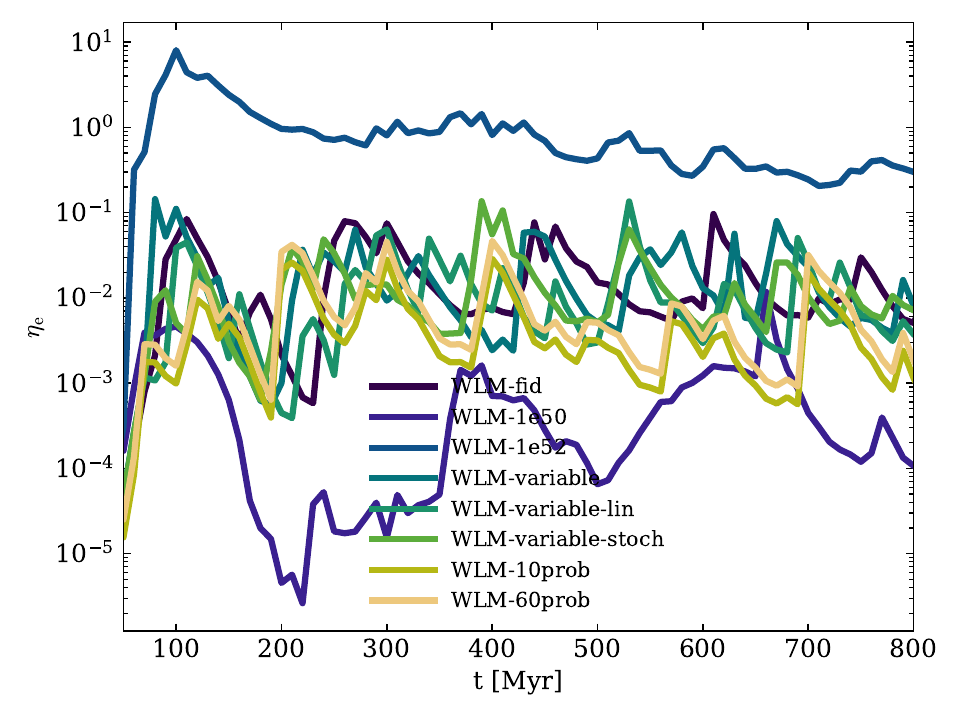}
    \caption{Mass loading (left) and energy loading (right) for all eight simulations. The effect of the the variability of SN-energy is marginal for both quantities: we find a factor-of-two reduction for the mass loading, while the effect is weaker for the energy loading. However, this can be  attributed at least in part to the exact way the energy loading is defined, as the SN-energy normalization factors are the IMF-averaged values of each simulation run. It is interesting that the run WLM-prob60 shows a  reduction in star formation rate that is similar to those of the variable SN-energy runs WLM-variable, WLM-variable-lin and WLM-variable-stoch. This indicates that the SN-energy itself is rather unimportant as long as as it is around $10^{51}$ erg and resolved, while the reduction of outflows appears to be driven by the stochasticity in explosion outcome itself.}
    \label{fig:loadings}
\end{figure*}

\begin{figure*}
    \centering
    \includegraphics[width=0.49\textwidth]{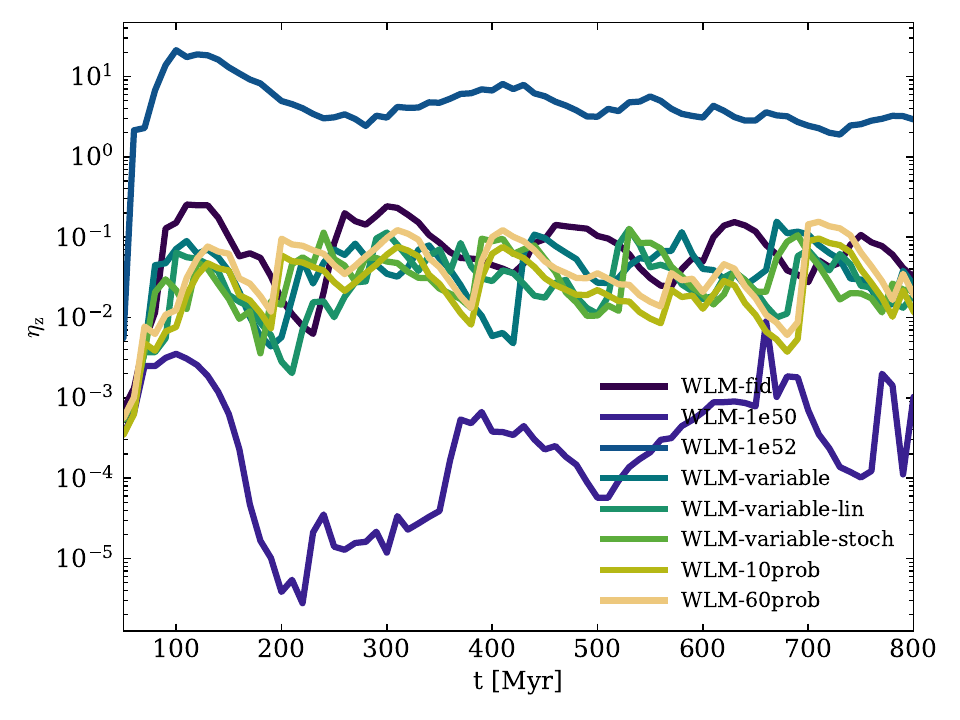}
    \includegraphics[width=0.49\textwidth]{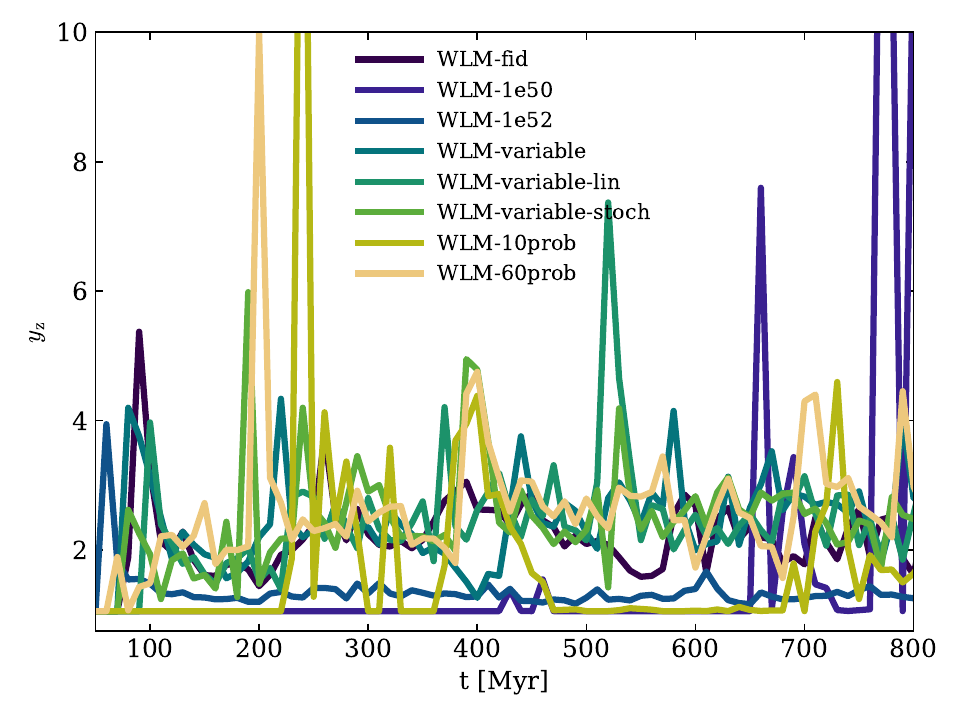}
    \caption{Metal loading (left) and metal enrichment factor (right) for all eight simulations.  Generally speaking, we find that the variability has little effect on these quantities, except in the runs WLM-1e50 and WLM-1e52, where  the SN-feedback energy is changing the mass outflows significantly (see Fig.~\ref{fig:loadings}). To first order, the mass and the metal loading are proportional to one another, which explains why the metal loading simply follows the mass trend. However, the metal enrichment factor shows that the metal outflow of WLM-1e52 is mostly composed of ambient gas and is equally low as in the WLM-1e50 run. The variability models do show some small excess in the averaged metal enrichment factors compared to WLM-fid, but it is a 5 to 10 percent deviation, which is typically within the run-to-run variations of such models.} 
    \label{fig:metal_loadings}
\end{figure*}

In this section, we discuss the definitions of loading factors and present the relevant results for the galaxy structure  and phase structure of the ISM and galactic outflows and for the time-evolution of the loading factors. 

\subsection{Definition of outflow loading factors}

Before discussing our simulation results, we discuss here how we define many quantities of interest.
For outflow rate, mass loading, and enrichment factors, we follow  the definitions presented in \citet{Hu2019}, \citet{Steinwandel2022} and \citet{SteinwandelLMC}. Generally, galactic wind quantities can be split into mass flux (m) and energy flux (e), defined as:
\begin{align}
    {\bf \mathcal{F}}_\mathrm{m} = \rho \textbf{v},
\end{align}
\begin{align}
    {\bf \mathcal{F}}_\mathrm{e} = (\rho e_\mathrm{tot} + P)\textbf{v},
\end{align}
where $\rho$, $v$, and $P$ are the fundamental fluid variables density, velocity, and pressure, respectively, and $e_\mathrm{tot}$ is defined as the total energy per unit mass: $e_\mathrm{tot} = 0.5 \cdot \mathbf{v}^2 + u$, where $u$ is the internal energy per unit mass. In standard code units, $u$ takes the units $\mathrm{km}\,\mathrm{s}^{-2}$. We note this to avoid confusion between these units and proper cgs units in the sections below. 

In this work, we measure the loading factors at a height of 1 kpc in a slab of height 100 pc and at a height of 10 kpc in a slab of height 1 kpc . 

We will adopt the following definitions for outflow rates, with the total flow rates defined as $\dot{M} = \dot{M}_\mathrm{out} - \dot{M}_\mathrm{in}$, $\dot{p} = \dot{p}_\mathrm{out} - \dot{p}_\mathrm{in}$ and $\dot{E} = \dot{E}_\mathrm{out} - \dot{E}_\mathrm{in}$. Outflow is defined by a positive value of the radial velocity given by $\mathbf{v \cdot \hat{n}} = v_{r} > 0$ (this is actually v$_\mathrm{z}$). We note that if we were to take the proper radial velocity over the slab, which would lead to a slightly higher value, that would be incorrect as the radial velocity would have some angle compared to v$_\mathrm{z}$. One could correct that by offsetting the outflow rate computed over the radial velocity with the sine of the angle between the radial and vertical velocities for a given particle. Our testing confirmed that it does not make a difference which of the approaches outlined above is followed, as they give the same results for mass loading and for energy loading (within a few percentage points). Nevertheless, we adopt the definition given above because it is, strictly speaking, correct when v$_\mathrm{z}$ is the outflow velocity over a plane-parallel slab. The reason for this is that in dwarf galaxies, rotational velocity is low, and so the vertical velocity will always dominate the radial velocity contribution at a given height. We expect this to change for larger galaxies, such as the Milky Way, that are more strongly rotationally supported. Hence, we can write down the discrete outflow rates for mass, momentum, and energy:
\begin{align}
    \dot{M}_\mathrm{out} = \sum_{i, v_{i,r} > 0} \frac{m_{i}v_{i,r}}{dr},
    \label{eq:massout}
\end{align}
\begin{align}
    \dot{E}_\mathrm{out} = \sum_{i, v_{i,r} > 0} \frac{m_{i} [v_{i}^2 + \gamma u_{i}]v_{i,r}}{dr},
\end{align}
where we use $P = \rho u (\gamma -1)$ as an equation of state. Furthermore, we will compute metal outflow rates via:
\begin{align}
    \dot{M}_\mathrm{out, Z} = \sum_{i, v_{i,r} > 0} \frac{Z_{i} m_{i}v_{i,r}}{dr},
\end{align}
where $Z_{i}$ is the metallicity of particle $i$.
In this paper we will investigate loading factors that are obtained by normalizing to a set of reference quantities, i.e., mass ($\eta_\mathrm{m}$), metals ($\eta_\mathrm{Z}$), and energy ($\eta_\mathrm{e}$):
\begin{enumerate}
    \item outflow mass loading factor: $\eta_\mathrm{m}^\mathrm{out} = \dot{M}_\mathrm{out} / \overline{\mathrm{SFR}}$,
    \item energy outflow loading factor: $\eta_\mathrm{e}^\mathrm{out} = \dot{E}_\mathrm{out}/(E_\mathrm{SN} R_\mathrm{SN})$,
    \item outflow metal loading factor: $\eta_\mathrm{Z}^\mathrm{out} = \dot{M}_\mathrm{Z, out} / (m_\mathrm{Z} R_\mathrm{SN})$,
\end{enumerate}
where we adopt $E_\mathrm{SN} = 10^{51}$ erg, p$_{SN} = 3 \cdot 10^{4}$ M$_{\odot}$ km s$^{-1}$,  supernova rate R$_{SN} = \overline{\mathrm{SFR}}/(100 M_{\odot})$ and  metal mass (IMF-averaged) $m_\mathrm{Z} = 2.5 M_{\odot}$, which we take for the best comparison with previous dwarf galaxy studies from \citet[][]{Hu2019}.
Additionally, we define metal enrichment factors that not only quantify the amount of metal transported in the outflows but can also quantify the degree to which metals are over- or under-abundant in galaxy outflows relative to the ISM. This can be achieved by normalizing the metal outflow rate to the mass outflow rate, weighted by the background metallicity of the galactic ISM, Z$_\mathrm{gal} = 0.1 Z_{\odot}$: 
\begin{enumerate}
    \item outflow enrichment factor: $y_\mathrm{Z}^\mathrm{out} = \dot{M}_\mathrm{Z, out}/ (\dot{M}_\mathrm{out} Z_\mathrm{gal})$, 
    \item inflow enrichment factor: $y_\mathrm{Z}^\mathrm{in} = \dot{M}_\mathrm{Z, out}/ (\dot{M}_\mathrm{in} Z_\mathrm{gal})$.
\end{enumerate}

\subsection{Dwarf Galaxy Morphology}

Fig.~\ref{fig:splash} shows the face-on surface density projections for all simulated models as summarized in Table~\ref{tab:one}. The morphological structures of the isolated galaxy simulations are quite similar, but there are some distinct differences among  the models that are worth pointing out. First, the models WLM-fid, WLM-variable, WLM-variable-lin and WLM-variable-stoch are quite similar in morphological structure. The models WLM-1e50 and  WLM-1e52 show clear evidence for smaller and larger bubble size, respectively. It appears that the ISM transitions to a smoother gas distribution for stronger explosions and remains more filamentary for lower explosion energies. 

\subsection{Star formation rate and ISM structure}

In Fig.~\ref{fig:sfr}, we show the star formation rate as a function of time for our eight simulated models, WLM-fid (black line), WLM-1e50 (dark blue line), WLM-1e52 (light blue line), WLM-variable (turquoise line), WLM-variable-lin (dark green line), WLM-variable-stoch (light green line), WLM-10prob (olive green line) and WLM-60prob (yellow line). Generally, we find a star formation rate of around 0.001-0.005 M$_{\odot}$ yr$^{-1}$, which is in good agreement with other model variations of such a system \citep[e.g.,][]{Hu2017, Emerick2019, Smith2021, Gutcke2021, Hislop2022, Steinwandel2022, SteinwandelLMC}. We note that the run WLM-1e52 exhibits the lowest star formation rate by almost an order of magnitude, and the star formation history is rather bursty. The low-explosion-energy run WLM-1e50 has the highest star formation rate overall. We find that the runs with variable SN-energy (WLM-variable, WLM-variable-lin, WLM-variable-stoch) exhibit a  star formation rate that is slightly higher (by roughly 25 percent) than that of our fiducial model WLM-fid. The run WLM-60prob is quite comparable to the three runs with variable SN-energy injection, while the run WLM-10prob shows a high star formation rate that is very similar to that of the run WLM-1e50.

In Fig.~\ref{fig:phase}, we show the phase diagrams of density and temperature for all eight models, averaged over the final 400 Myr of the simulation run 
The major difference that we can report is that the prominence of the hot phase of the ISM depends on the adopted numerical value of the SN-energy. While there is little difference in the build-up of the hot phase (high $T$, low $n$ gas present in the upper left region of the phase diagram) among the models WLM-fid, WLM-variable, WLM-variable-lin, WLM-variable-stoch and WLM-60prob, we find that the build-up of the hot phase is strongly diminished in the models WLM-1e50 and WLM-10prob, while the build-up of the hot phase is enhanced in the model WLM-1e52. It is not necessarily surprising that the models with lower feedback energy show less hot gas and the model with the most feedback energy shows the most hot gas. Nevertheless, it is interesting to point out that fact explicitly.

\subsection{Time evolution of the loading factors}

In Fig.~\ref{fig:loadings}, we show the time evolution of the mass loading (left panel) and the energy loading (right panel). We adopt the same color conventions as in Fig.~\ref{fig:sfr}. In contrast to some previous work \citep[e.g.][]{Hislop2022, Steinwandel2022}, we only show these loading factors calculated at a height of 3 kpc. This is because, as shown in previous work \citep{Hu2019, Steinwandel2022, SteinwandelLMC}, the height of true hot wind formation is located somewhere around 2 kpc, so we want to make sure that we measure true outflow loading factors that are not contaminated by the warm fountain flow. For all the models that explode most of their SNe at an energy of $\sim 10^{51}$ erg we find a mass loading of around 0.1-1, which is in good agreement with the loading factors found in related work \citep[e.g.,][]{Hu2017, Hu2019, Smith2017, Gutcke2021, Hislop2022, Steinwandel2022, SteinwandelLMC}. The same is true for the energy loading factors for these models, for which we find values between 0.01 and 0.1. The models WLM-1e52 and WLM-1e50, however, exhibit much higher/lower values than the other simulations. The mass loading factor for the WLM-1e52 is around 100 and its energy loading is around 10, whereas the model WLM-1e50 shows mass loading factors between 0.001 and 0.01 and energy loading factors of around $10^{-5}$. In that context, it is interesting to point out that the model WLM-10prob is energetically quite similar to WLM-1e50 but nonetheless is capable of producing a mass- and energy-loaded wind. WLM-10prob shows the lowest loading factors -- around 0.1 for mass and 0.001 for energy -- but it still drives a galactic wind despite the fact that only 10 percent of the SNe explode with $10^{51}$ erg. This is interesting and a priori not directly clear. Both systems increase their star formation rate to compensate for the decrease in the overall energy input into the ISM. However,  the individual explosions remain ineffective in the case of WLM-1e50, which  means either that they are only able to regulate star formation by driving in a momentum phase or that they are entirely unresolved in our framework. 

\subsection{Phase structure of the ISM}

\begin{figure*}
    \centering
    \includegraphics[width=0.32\textwidth]{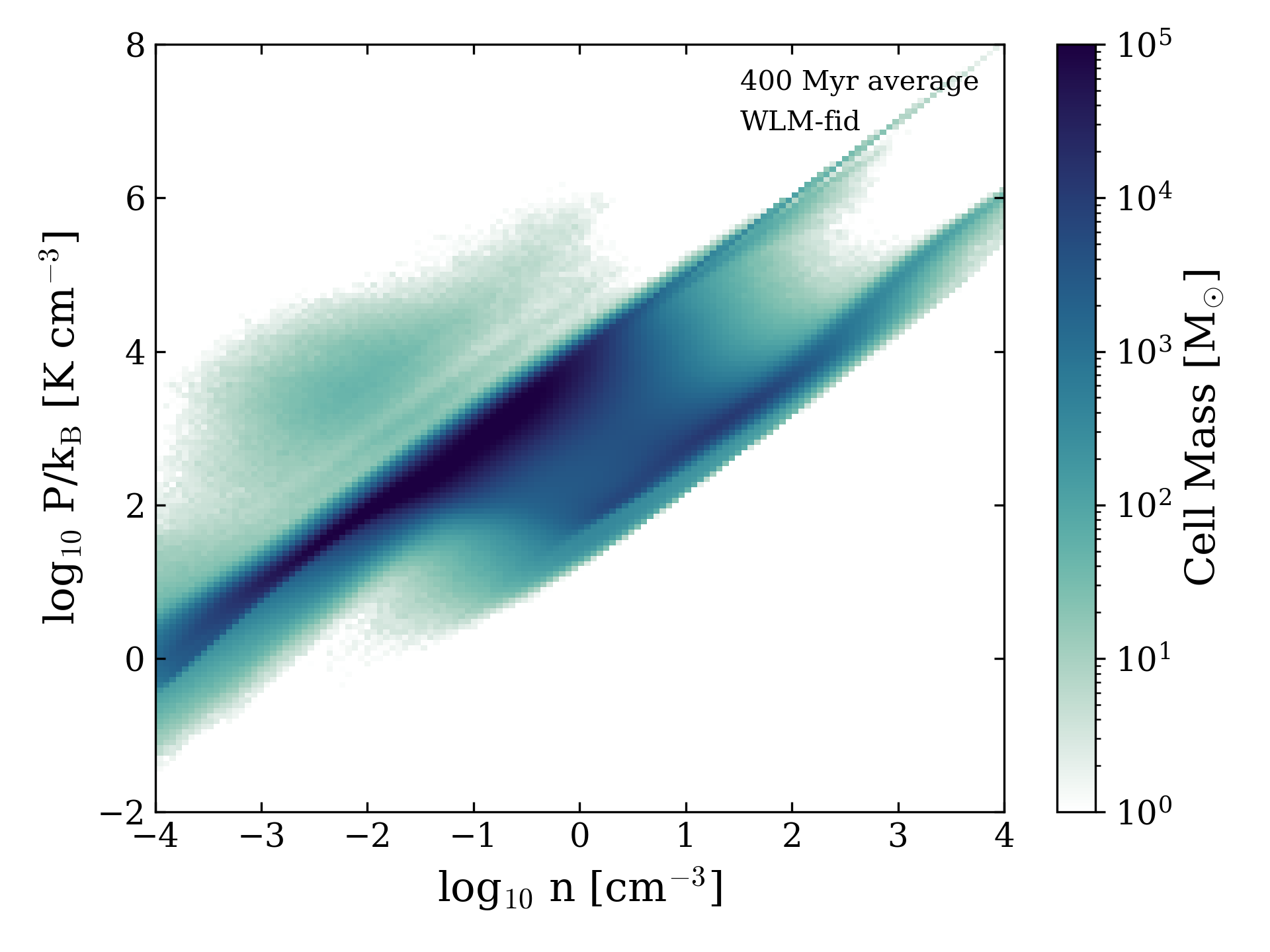}
    \includegraphics[width=0.32\textwidth]{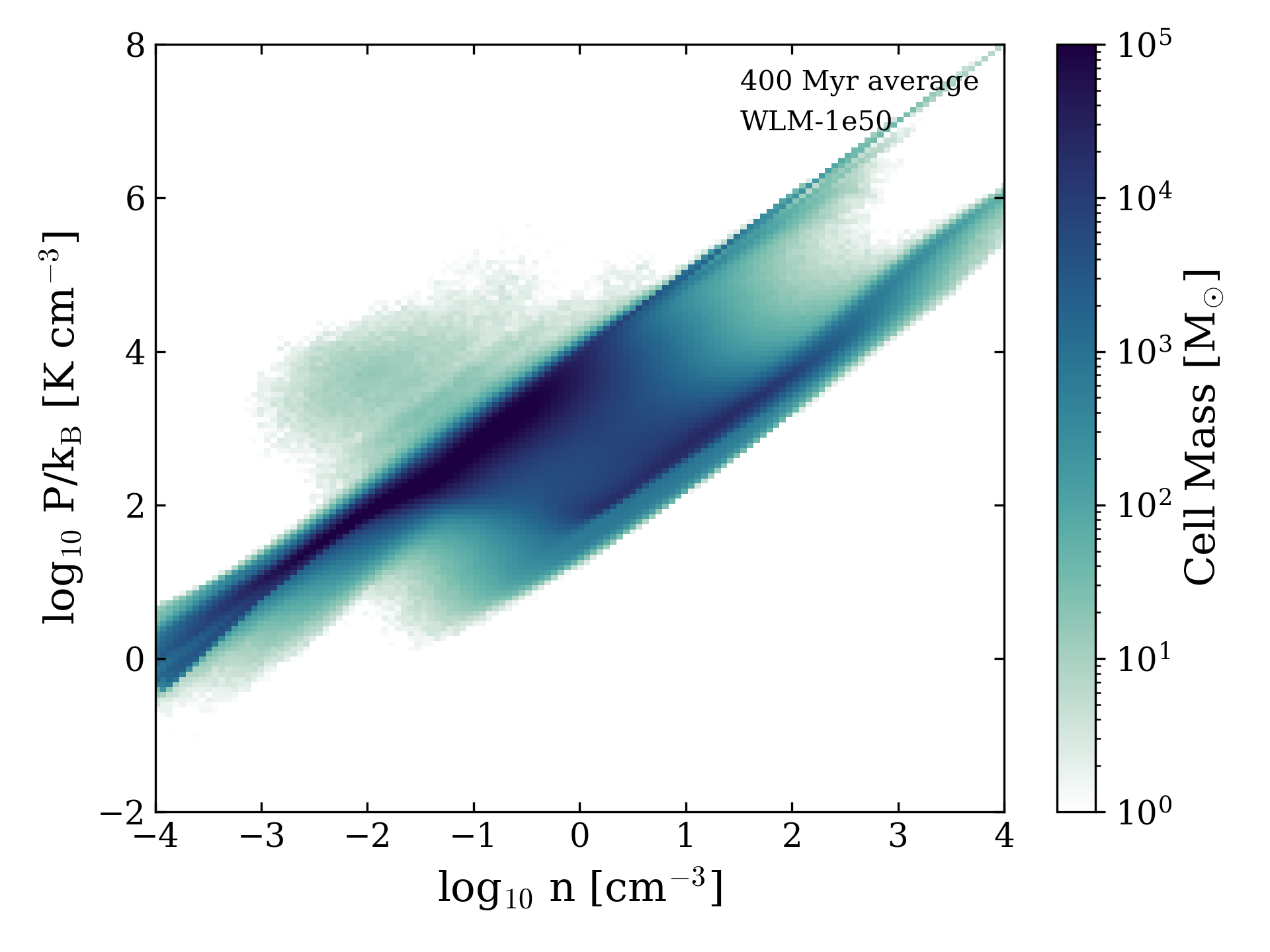}
    \includegraphics[width=0.32\textwidth]{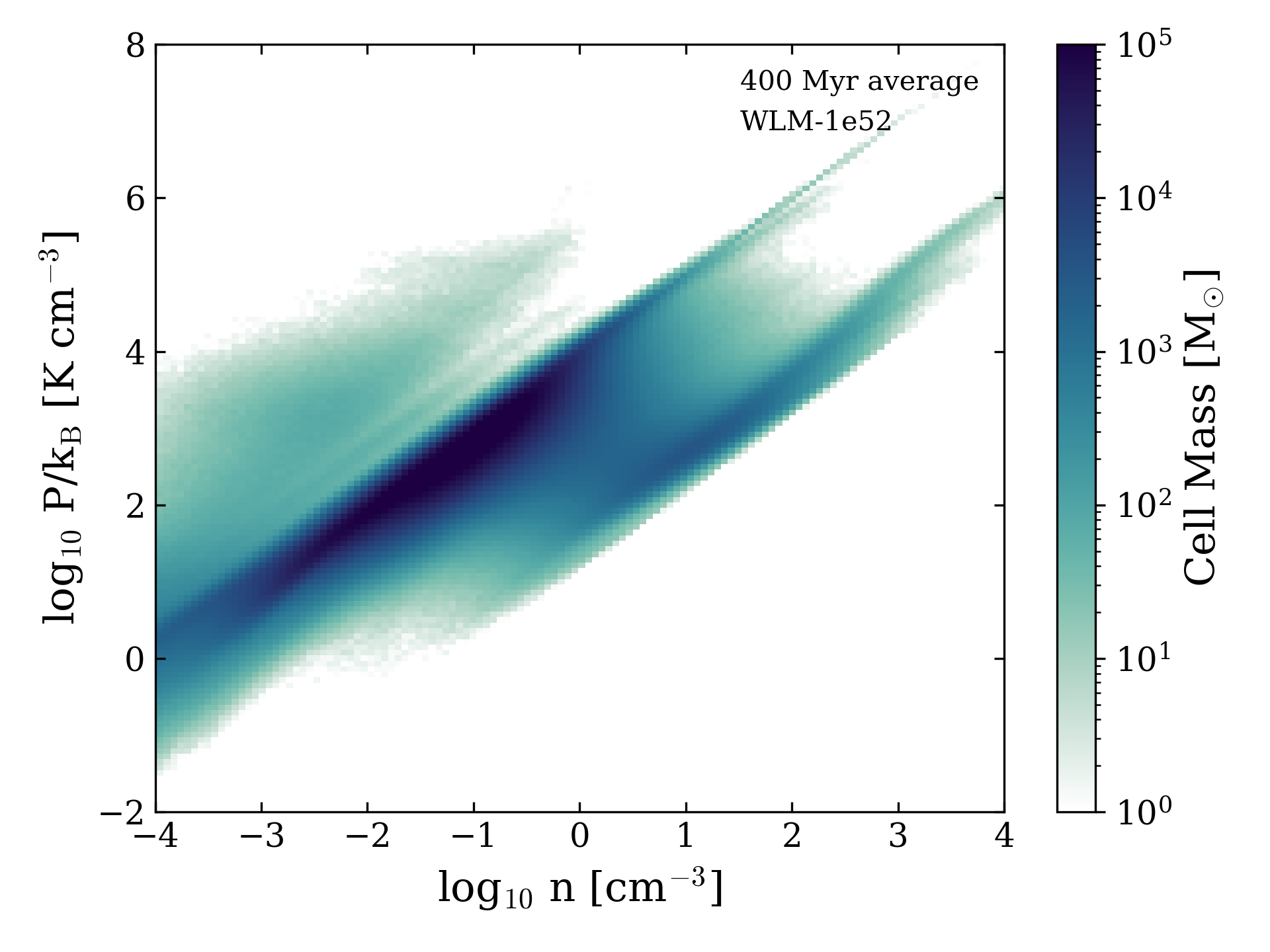}
    \includegraphics[width=0.32\textwidth]{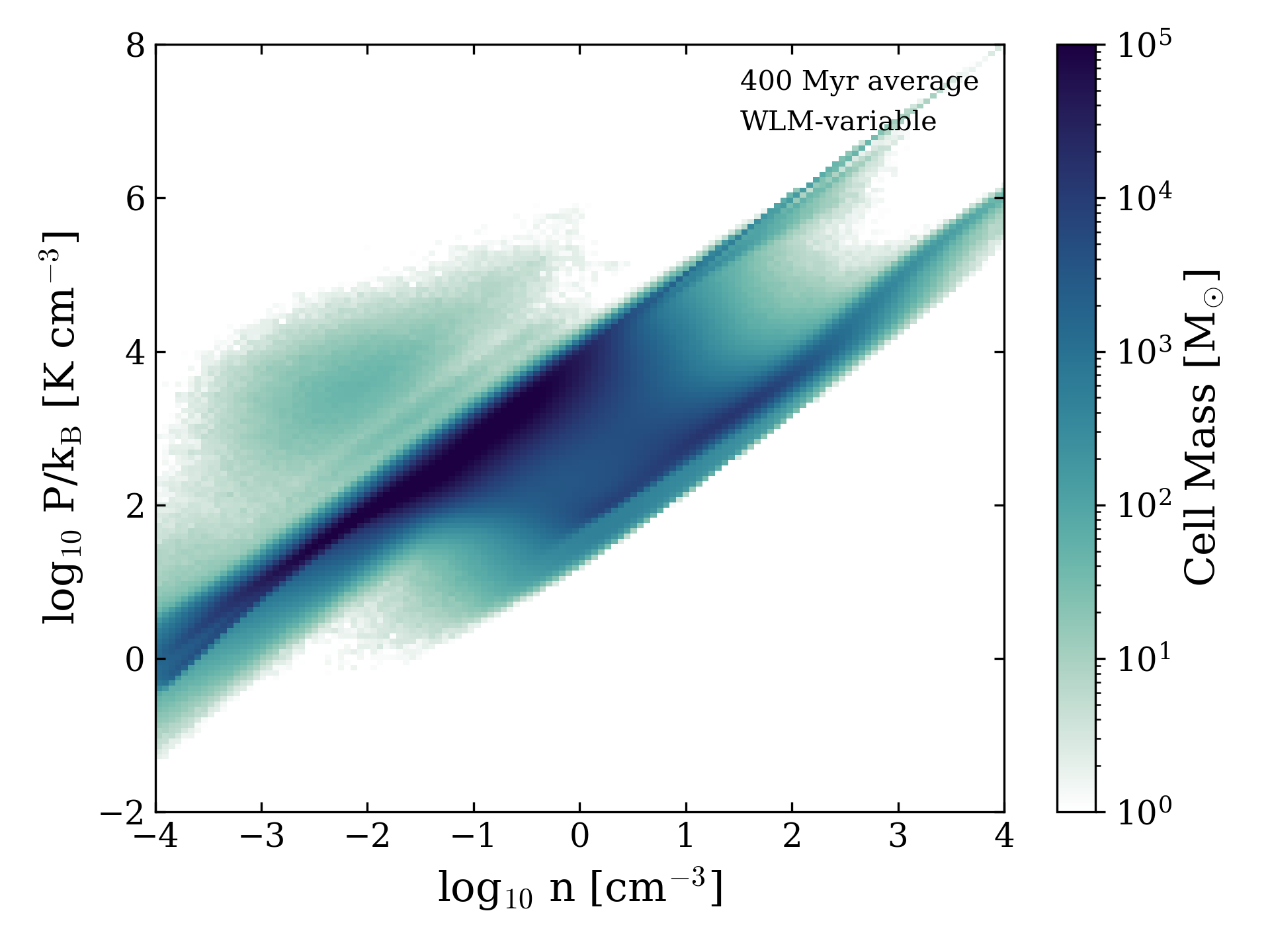}
    \includegraphics[width=0.32\textwidth]{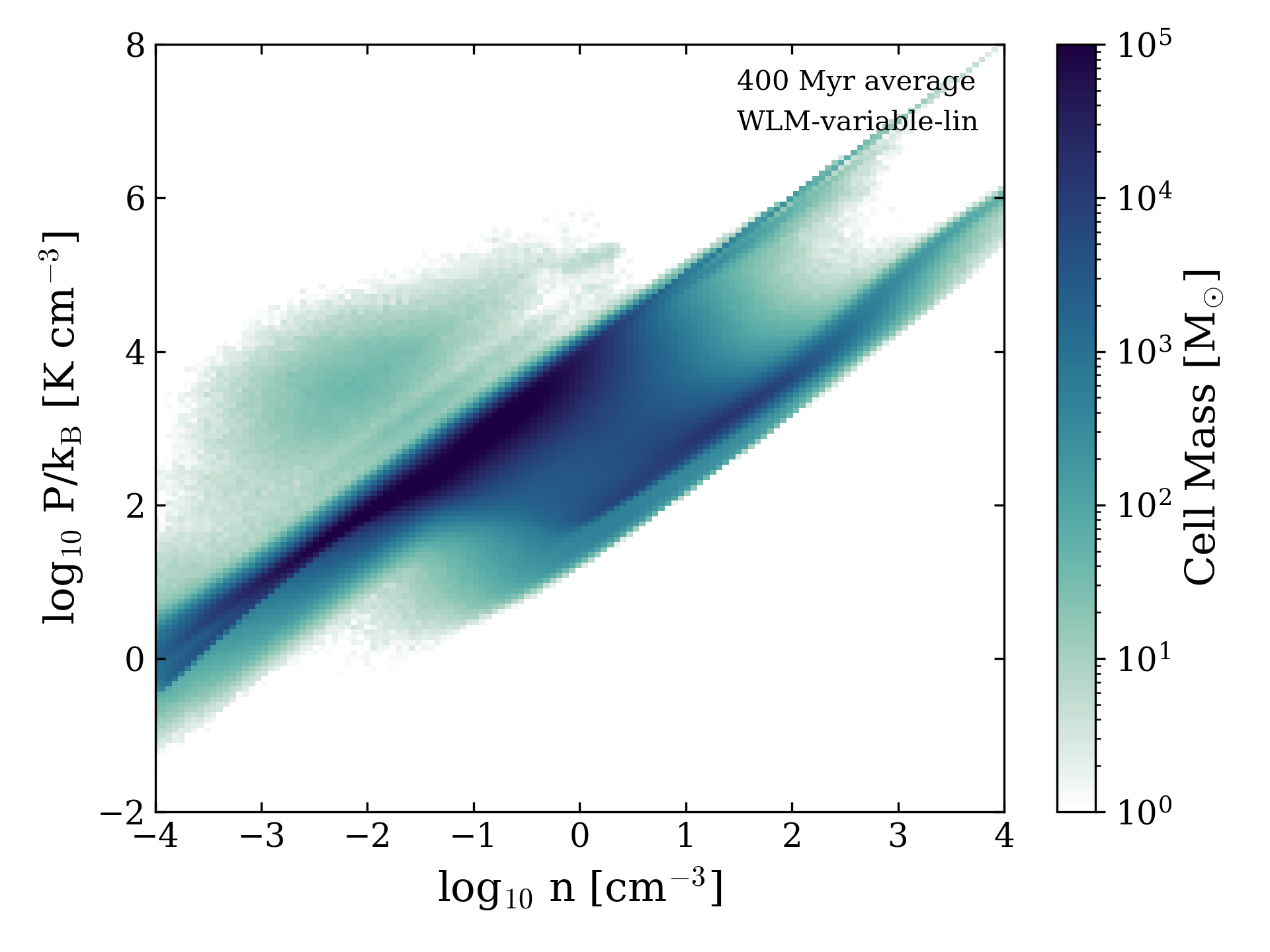}
    \includegraphics[width=0.32\textwidth]{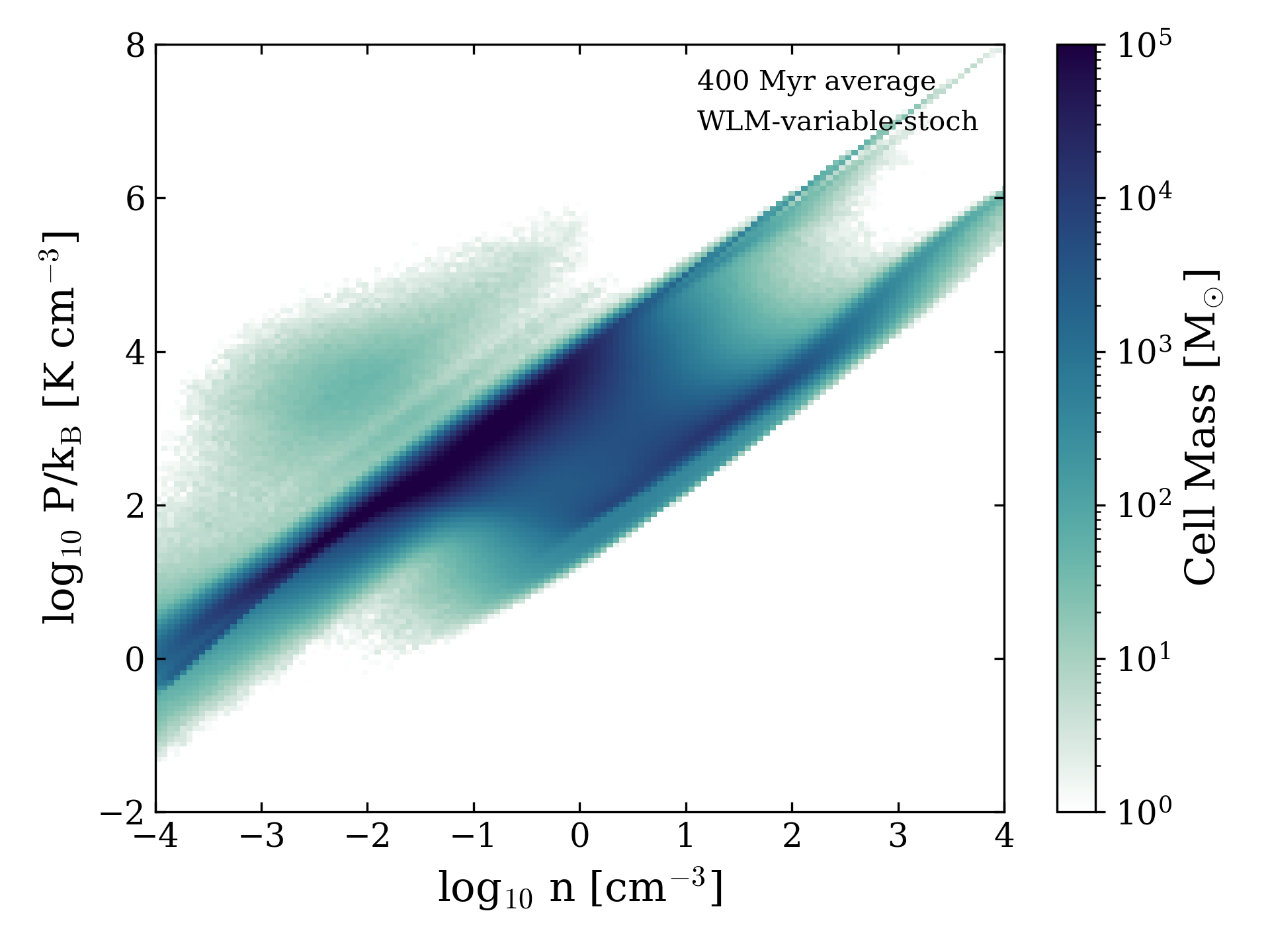}
    \includegraphics[width=0.32\textwidth]{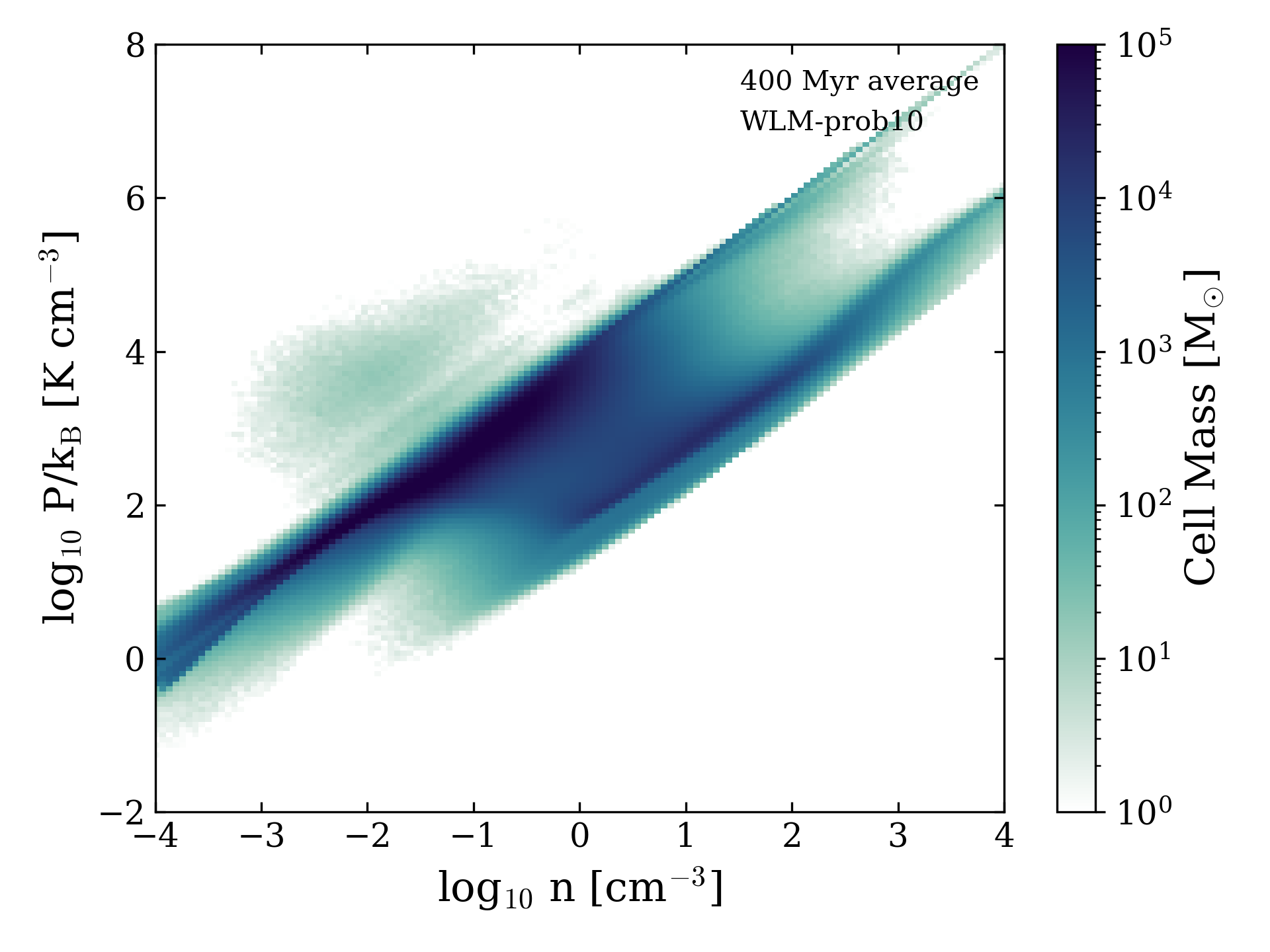}
    \includegraphics[width=0.32\textwidth]{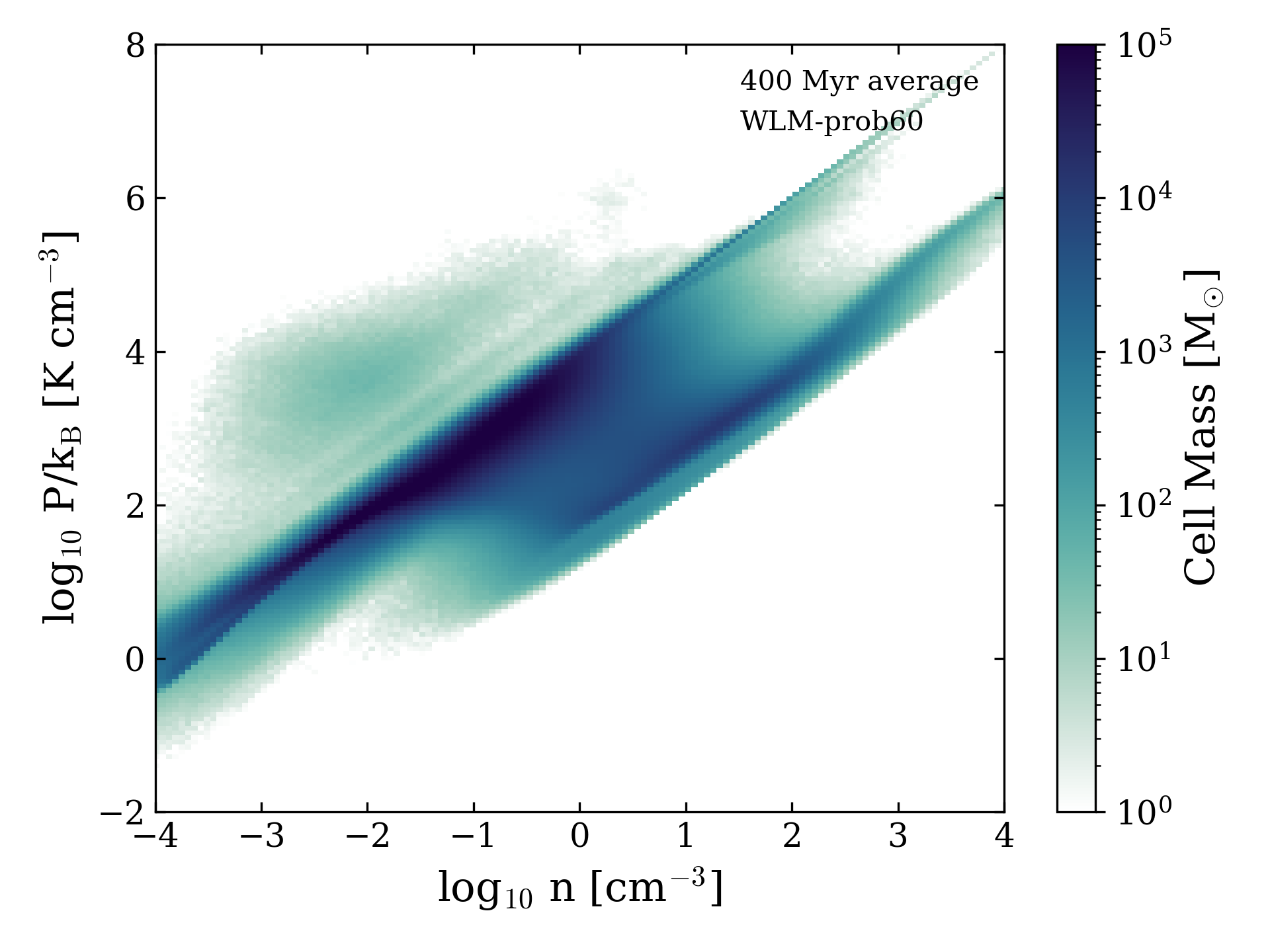}
    \caption{ISM phase space diagrams of density and pressure, color coded by mass, for the models WLM-fid (top row, left), WLM-1e50 (top row, center), WLM-1e52 (top row,  right), WLM-variable (middle row, left), WLM-variable-lin (middle row, center), WLM-variable-stoch (middle row, right), WLM-10prob (bottom row, left) and WLM-60prob (bottom row, right).}
    \label{fig:phase_press}
\end{figure*}

\begin{figure*}
    \centering
    \includegraphics[width=0.32\textwidth]{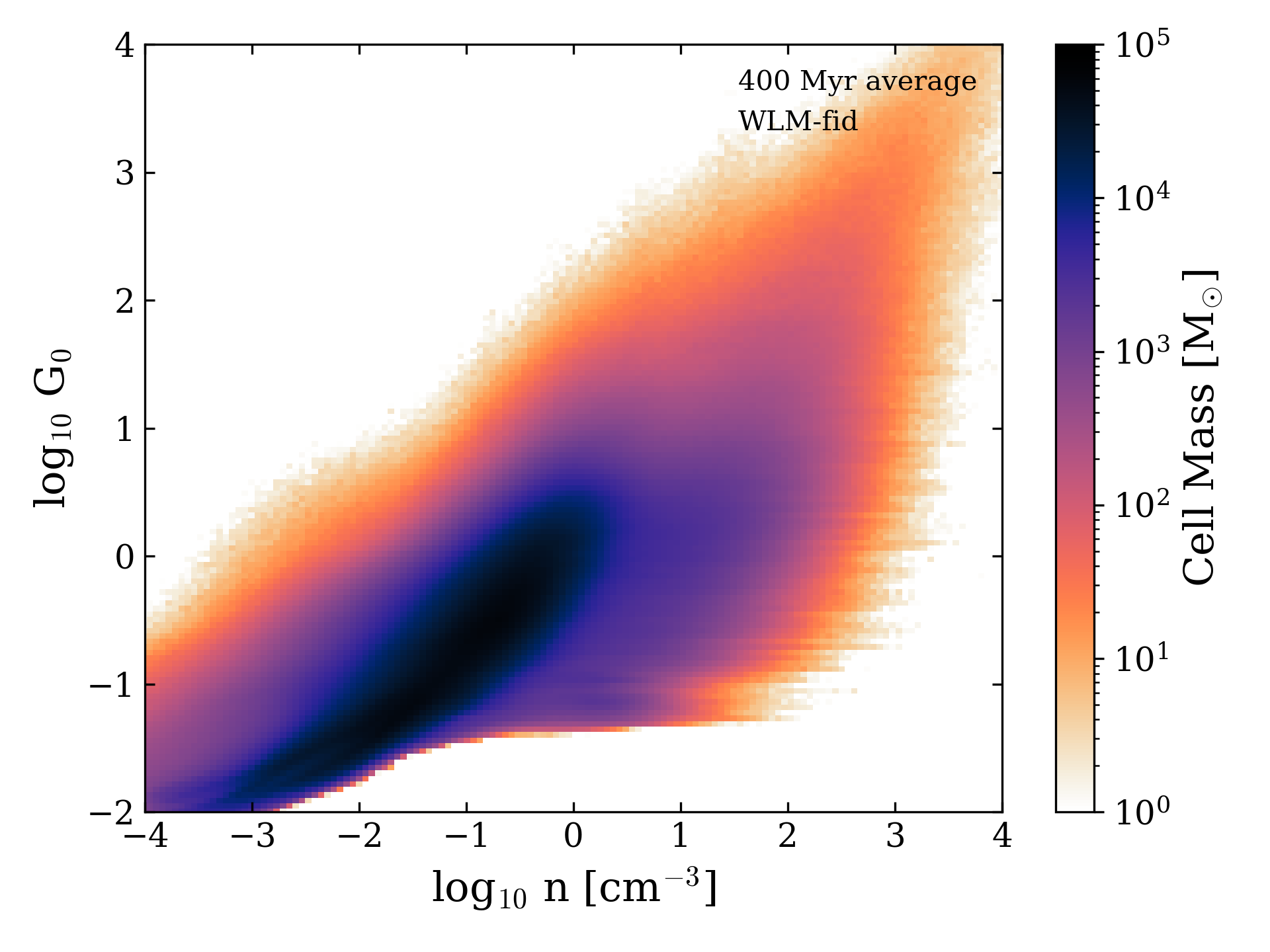}
    \includegraphics[width=0.32\textwidth]{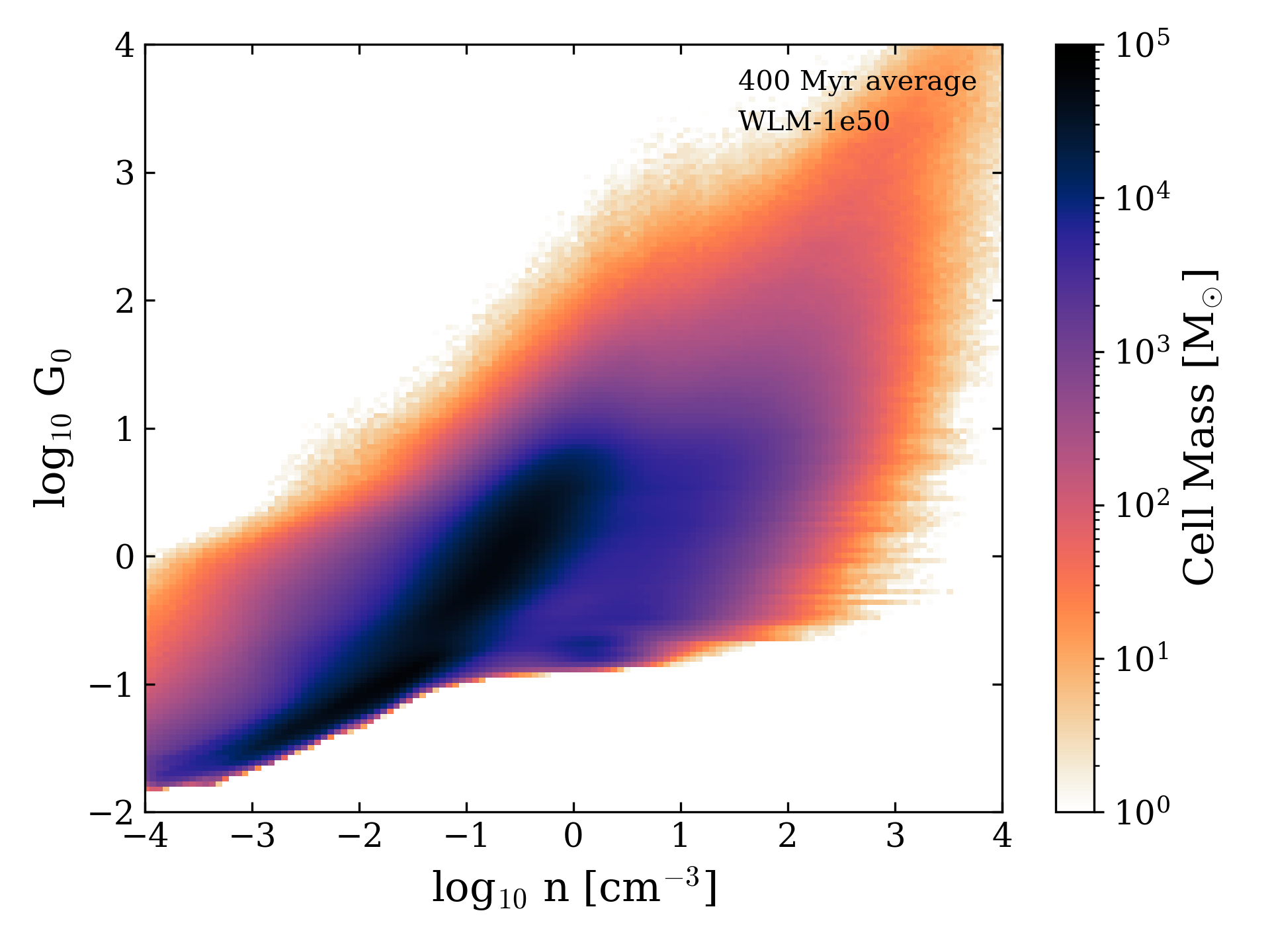}
    \includegraphics[width=0.32\textwidth]{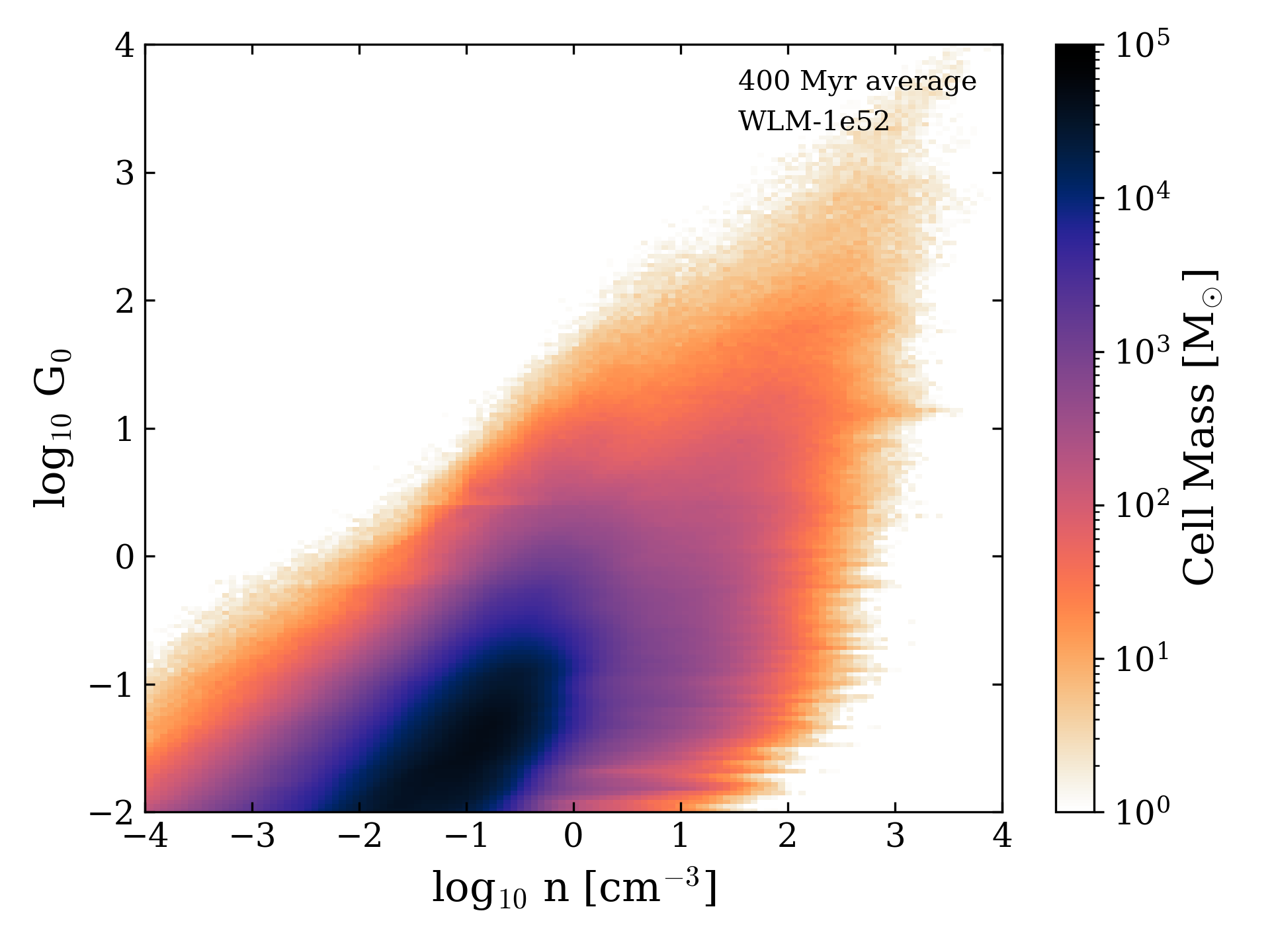}
    \includegraphics[width=0.32\textwidth]{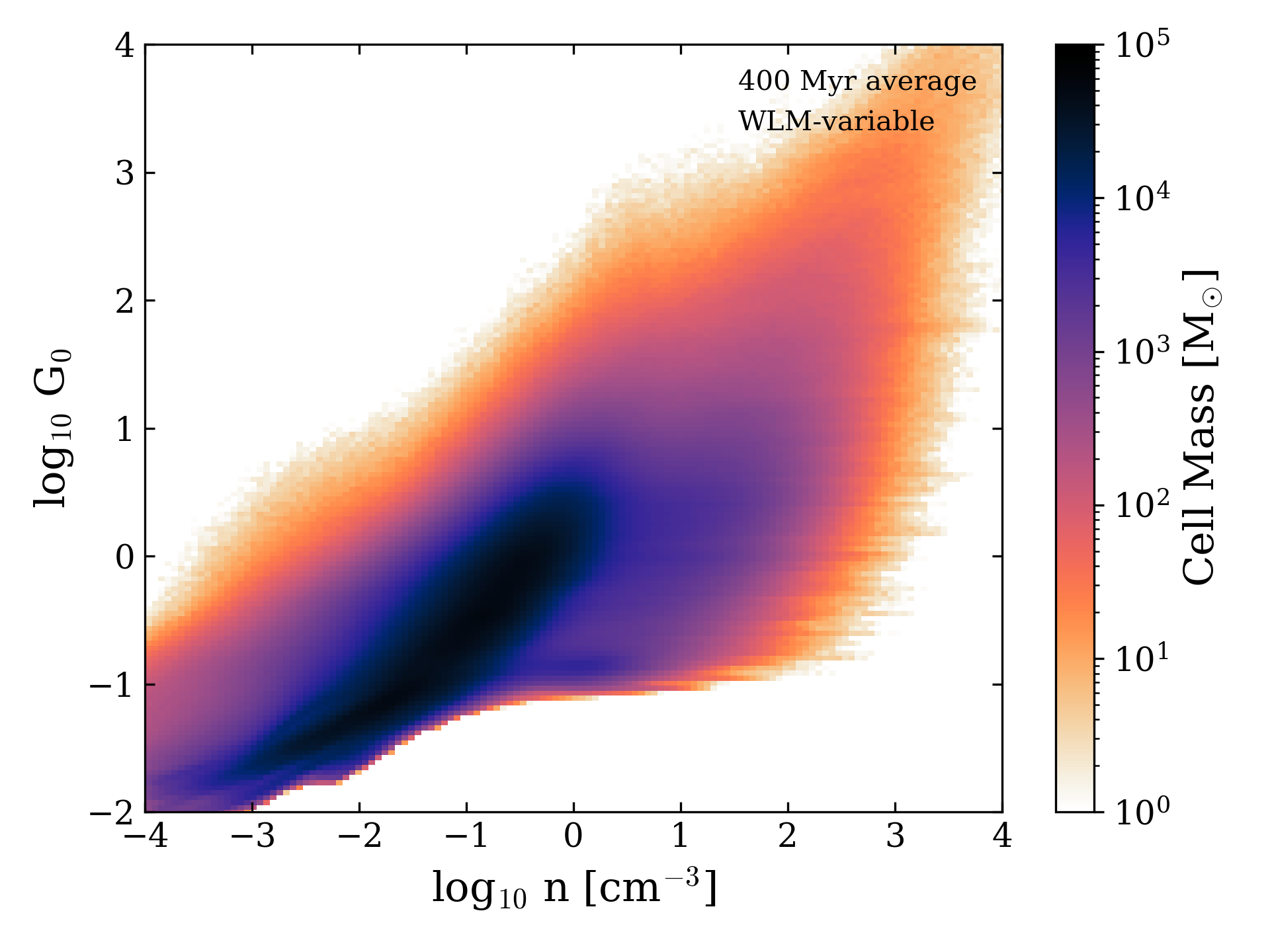}
    \includegraphics[width=0.32\textwidth]{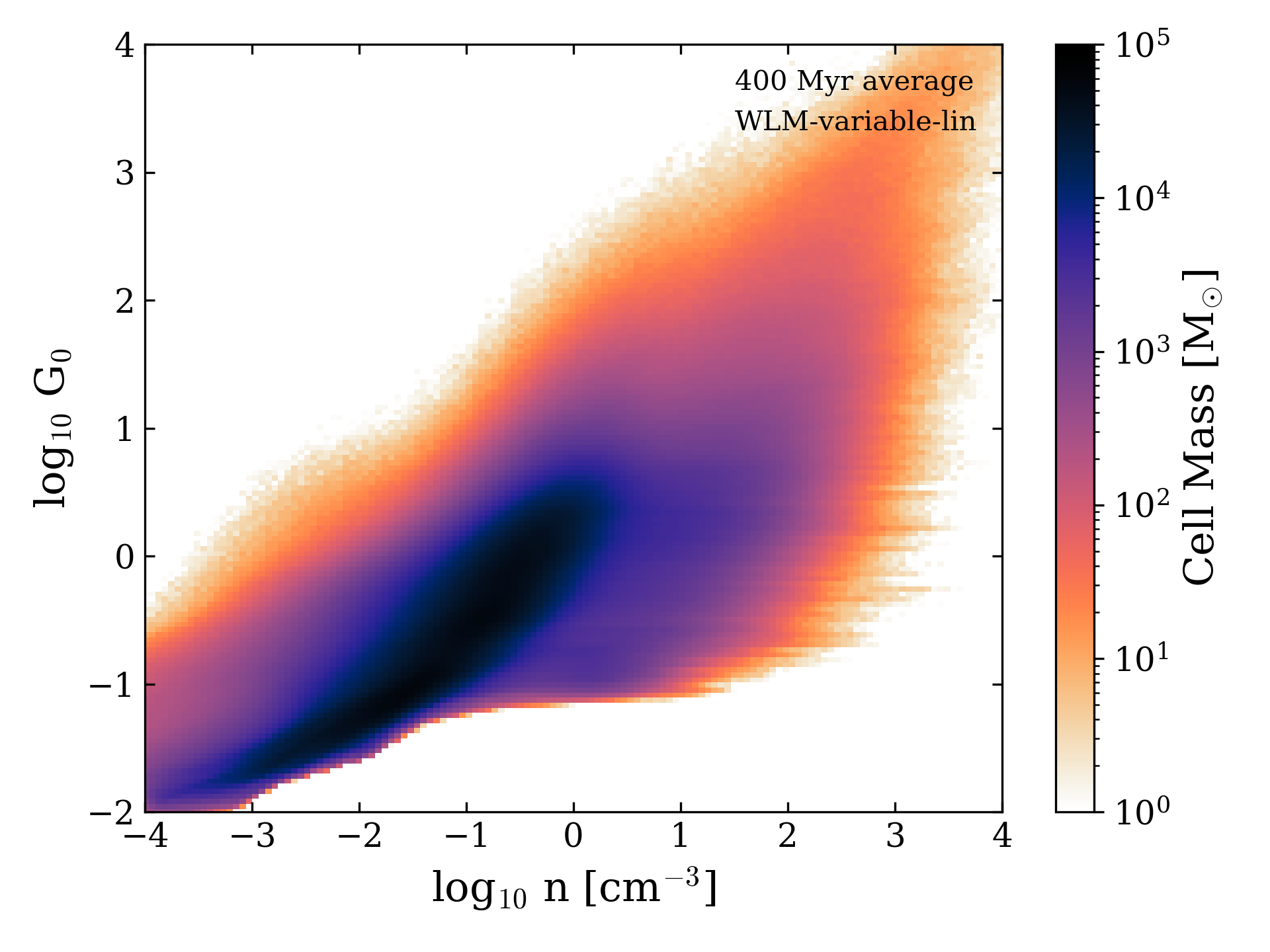}
    \includegraphics[width=0.32\textwidth]{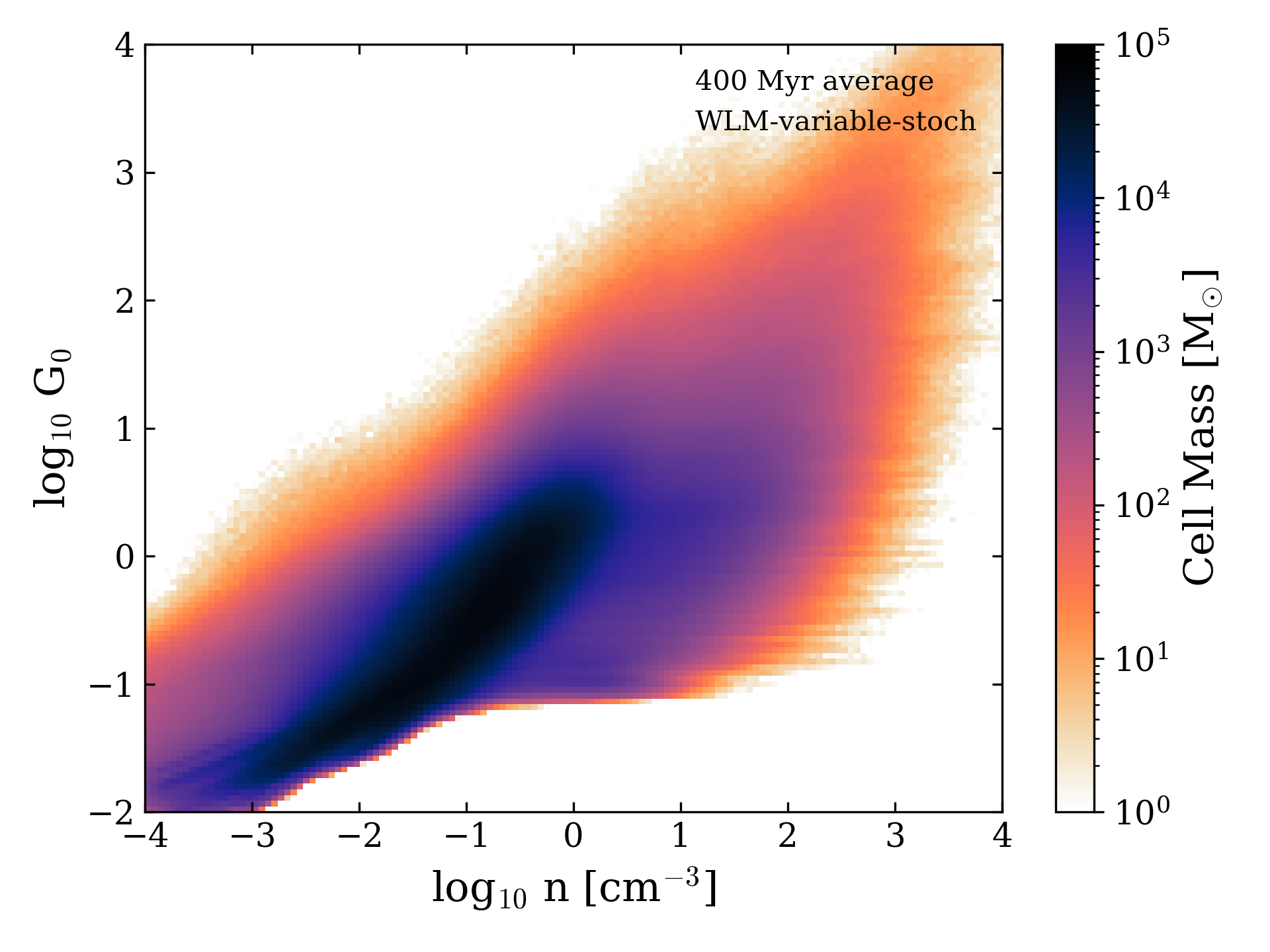}
    \includegraphics[width=0.32\textwidth]{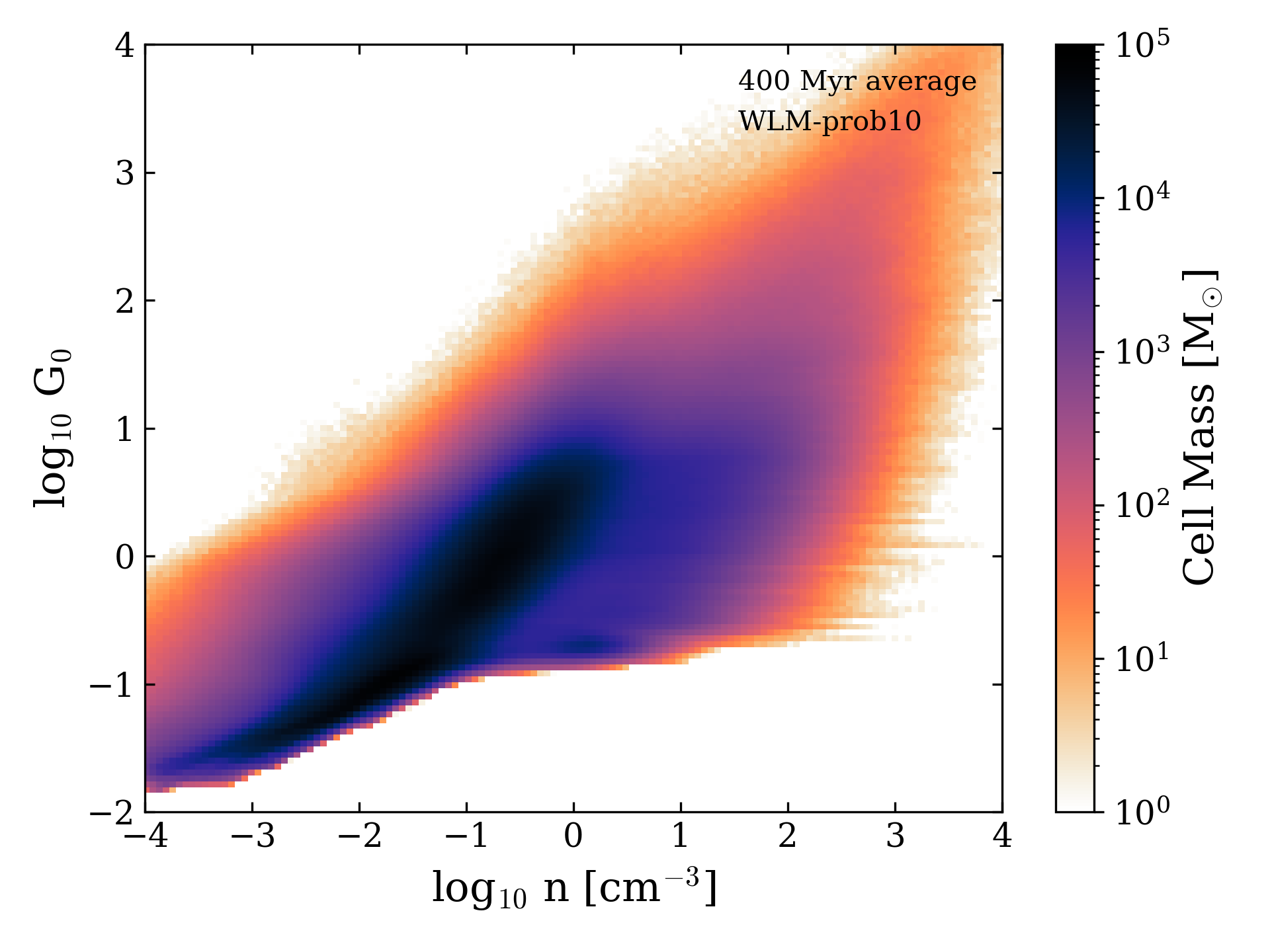}
    \includegraphics[width=0.32\textwidth]{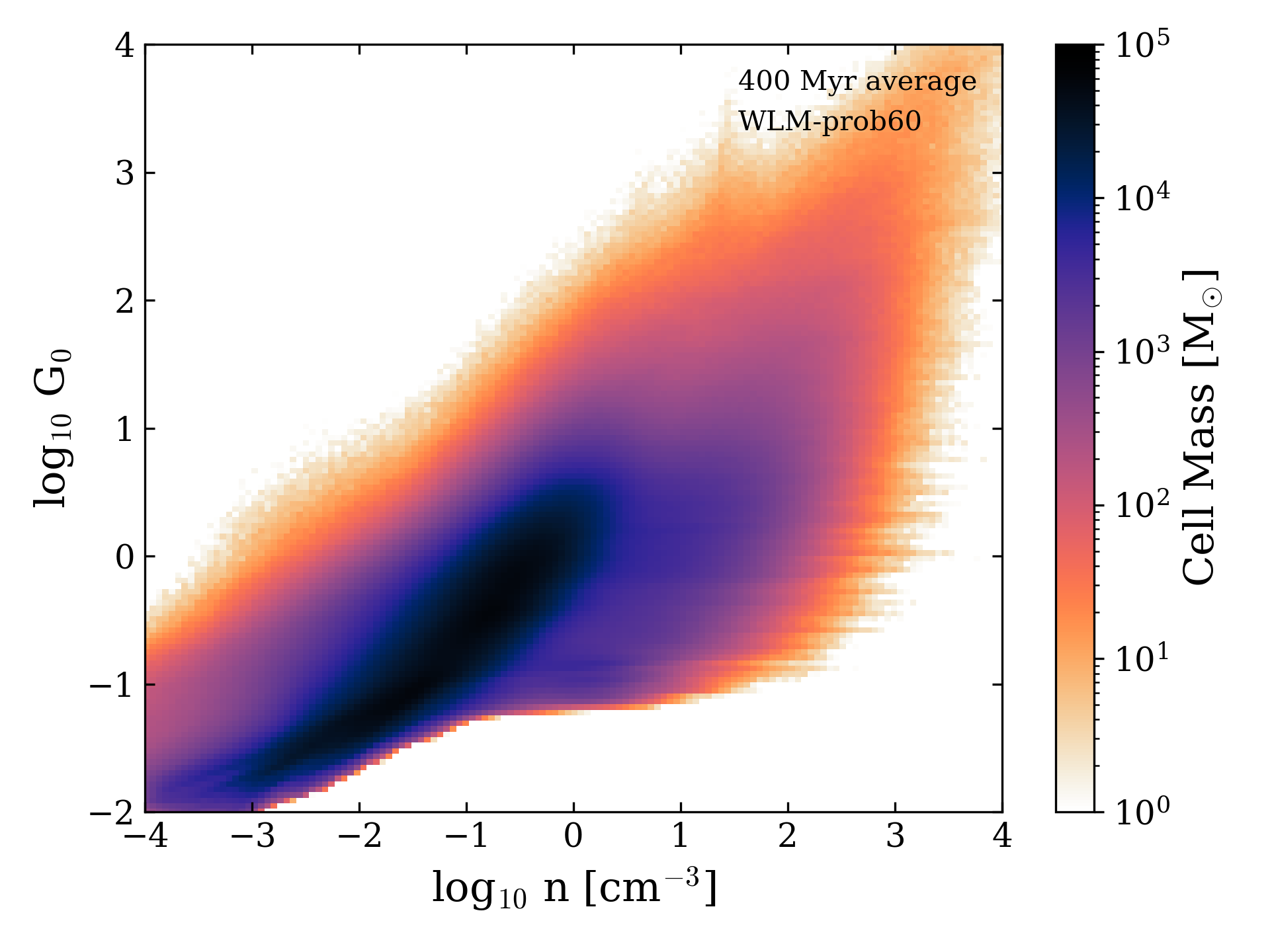}
    \caption{ISM phase space diagrams of density and G0, color coded by mass, for the models WLM-fid (top row, left), WLM-1e50 (top row, center), WLM-1e52 (top row,  right), WLM-variable (middle row, left), WLM-variable-lin (middle row, center), WLM-variable-stoch (middle row, right), WLM-10prob (bottom row, left) and WLM-60prob (bottom row, right).}
    \label{fig:phase_g0}
\end{figure*}

In Fig.~\ref{fig:phase_press} and Fig.~\ref{fig:phase_g0}, we show additional phase-space quantities of interest for the ISM in the form of pressure and the FUV field G0. Both the pressure structure and the structure of G0 are interesting results derived from our simulations. Like the density-temperature structure, the pressure-density structure is not strongly sensitive to the choices made for the feedback parameters in our model (e.g., the SN-feedback energy), and the major distinction among the different models concerns the hot phase (higher explosion energies entail more mass in the hot phase; see, e.g., Fig.~\ref{fig:phase}). The phase-space structure of G0 vs. density shows a similar picture. It is interesting to note that in the run WLM-1e52, we find a reduction in G0, which is related to a reduction in the maximum density that we observe in the dense gas. This implies that in WLM-1e52, the higher feedback energy prevents dense gas from forming more efficiently and more gas is ejected via galactic outflows. This is also evident in the mass loading factors of Fig.~\ref{fig:metal_loadings}; the mass loading factor is greater  in WLM-1e52 than in WLM-fid by a factor of $\sim$ 100; this is offset (though not entirely) by the factor-of-10 reduction in star formation rate exhibited by WLM-1e52. Thus there remains at least a factor of 10 more gas ejected in the case of WLM-1e52 compared to the other models.
 
\subsection{Phase structure of the outflows}

\begin{figure*}
    \centering
    \includegraphics[width=0.32\textwidth]{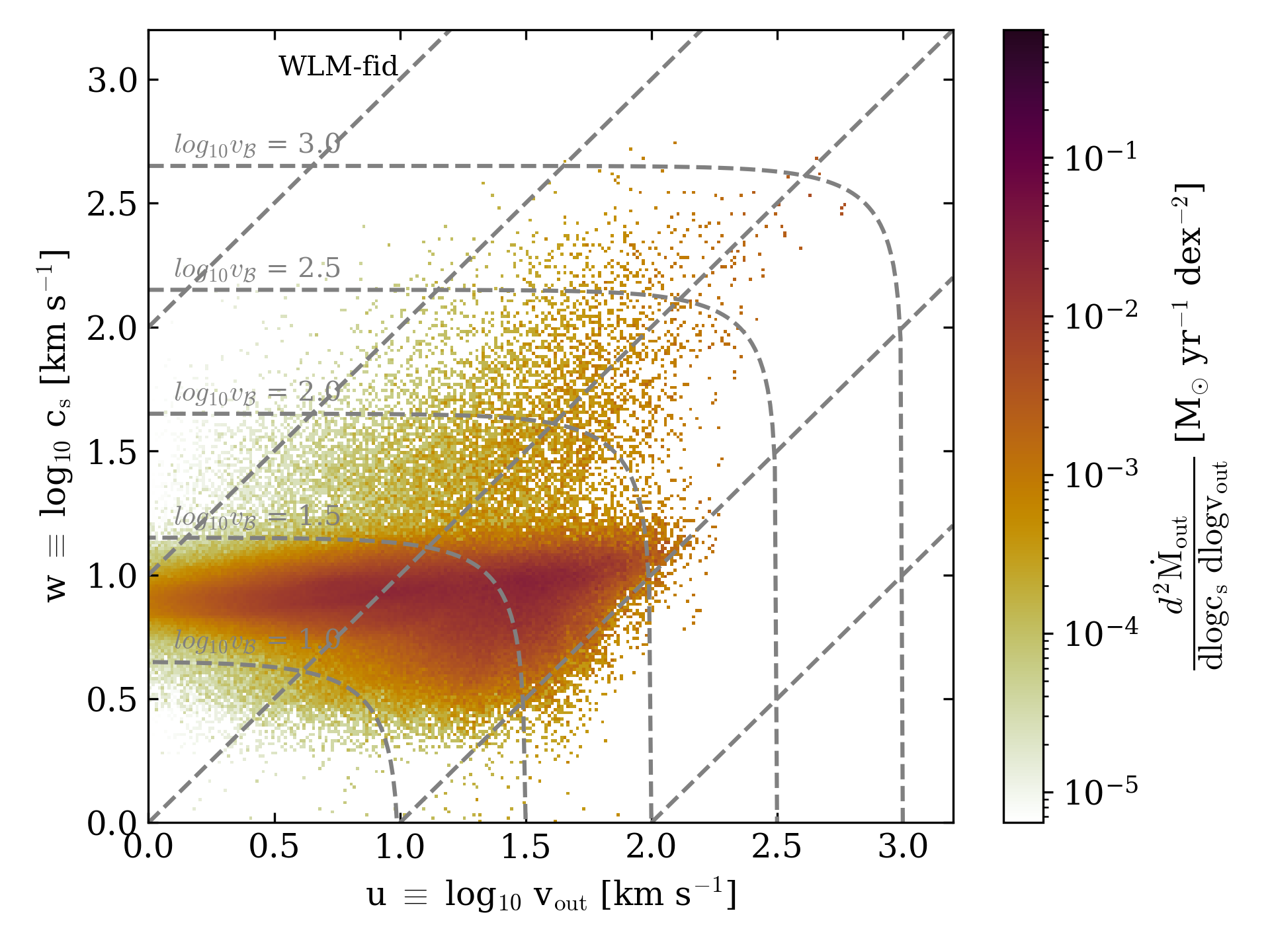}
    \includegraphics[width=0.32\textwidth]{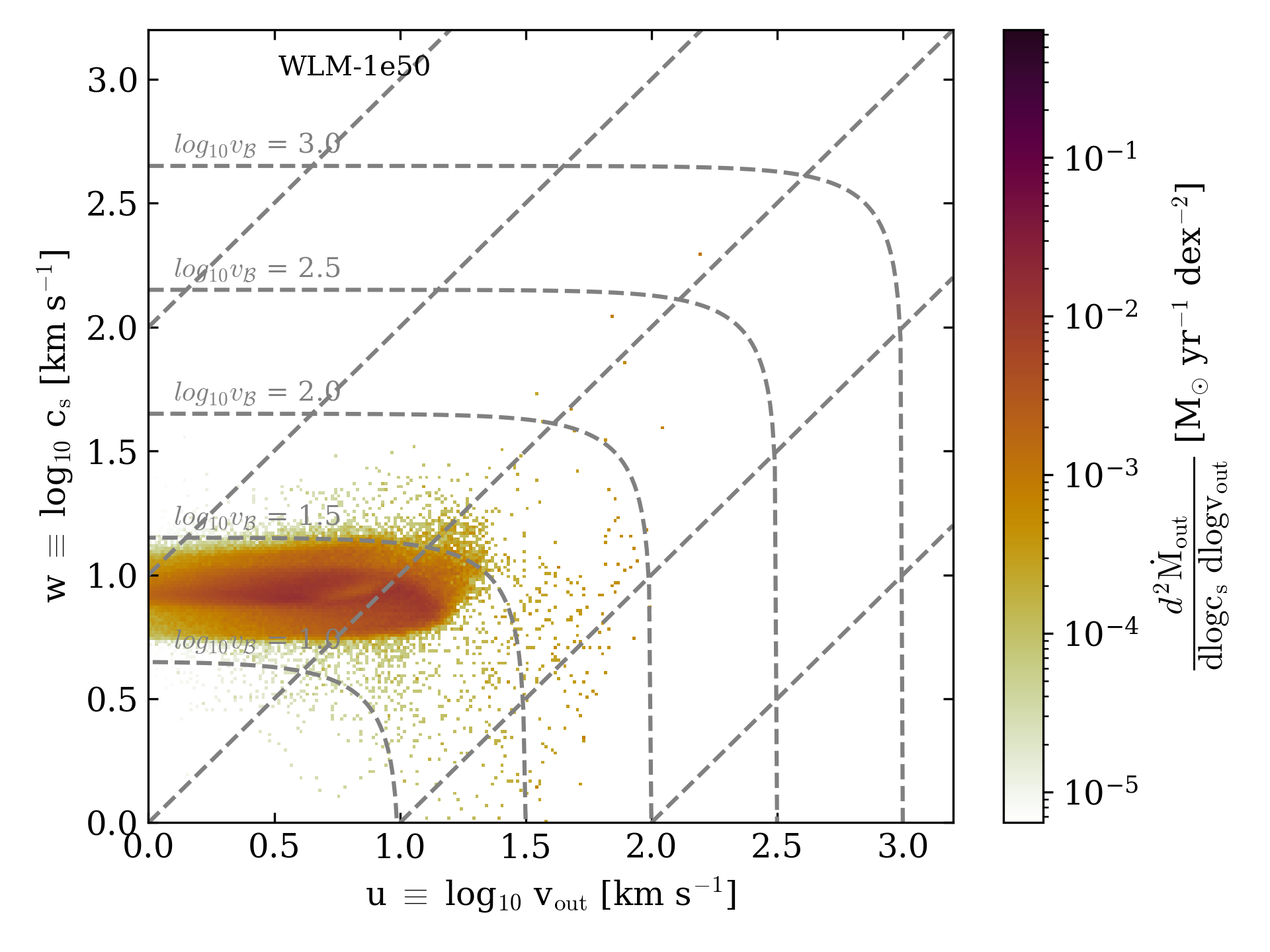}
    \includegraphics[width=0.32\textwidth]{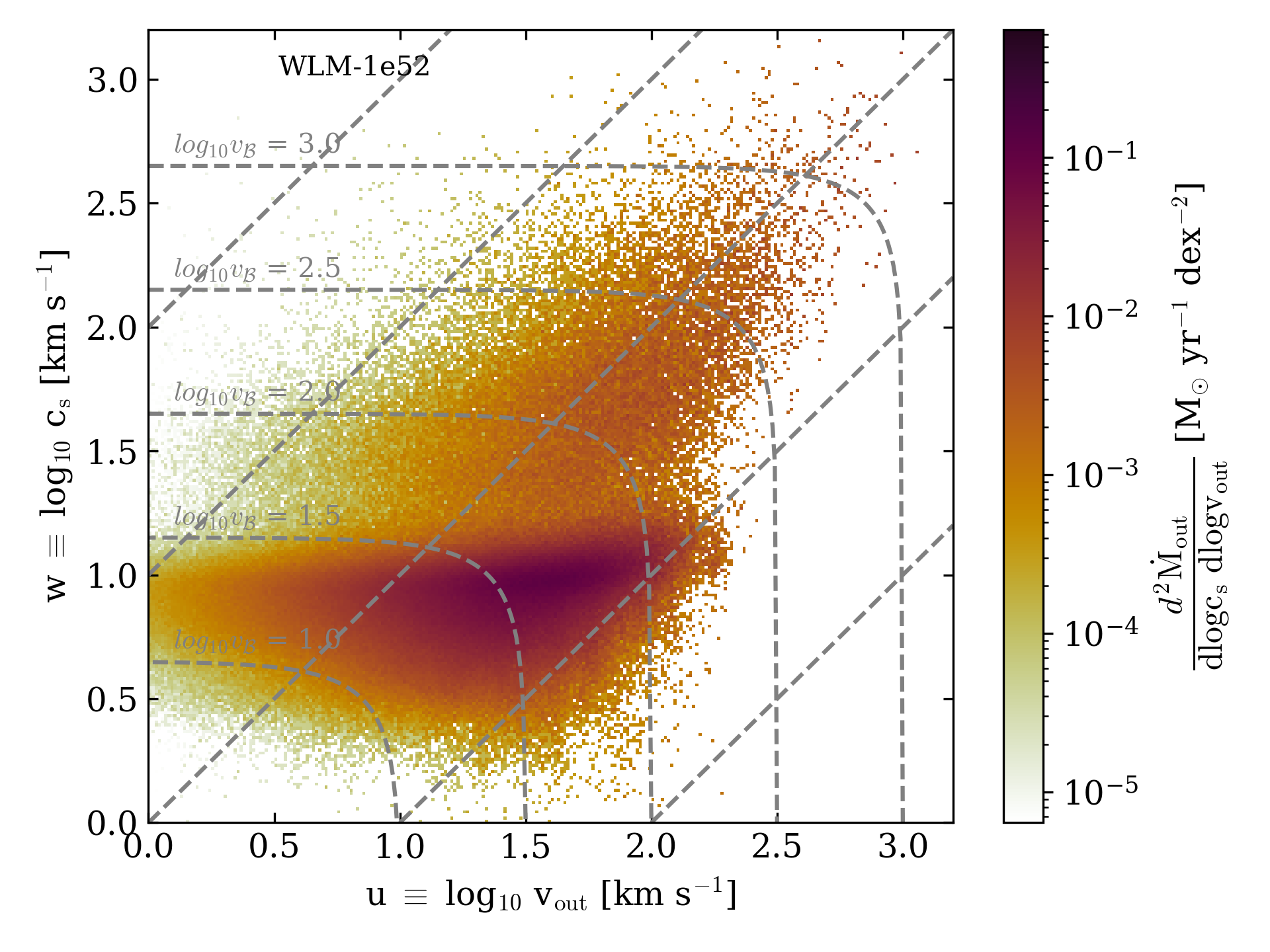}
    \includegraphics[width=0.32\textwidth]{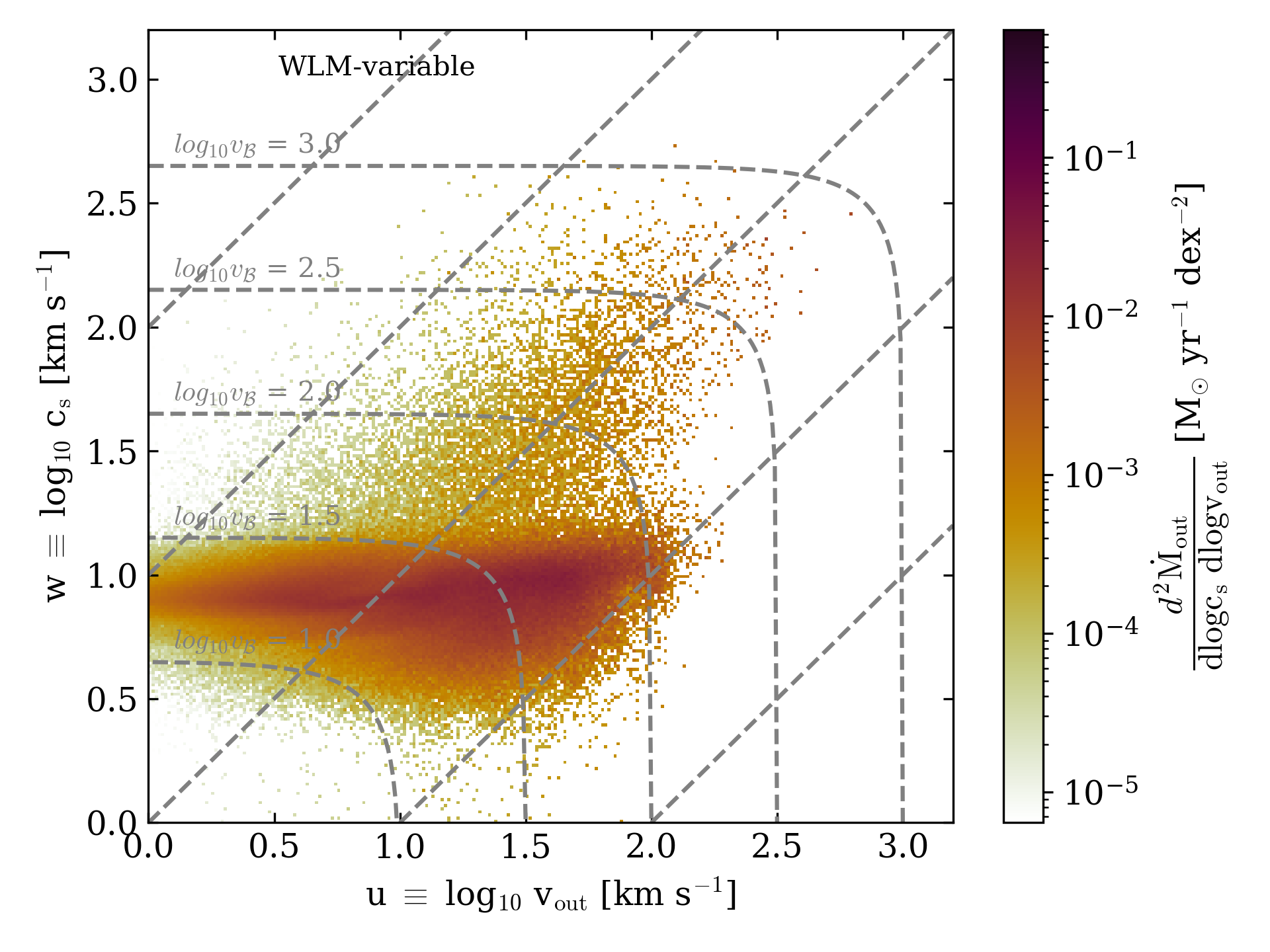}
    \includegraphics[width=0.32\textwidth]{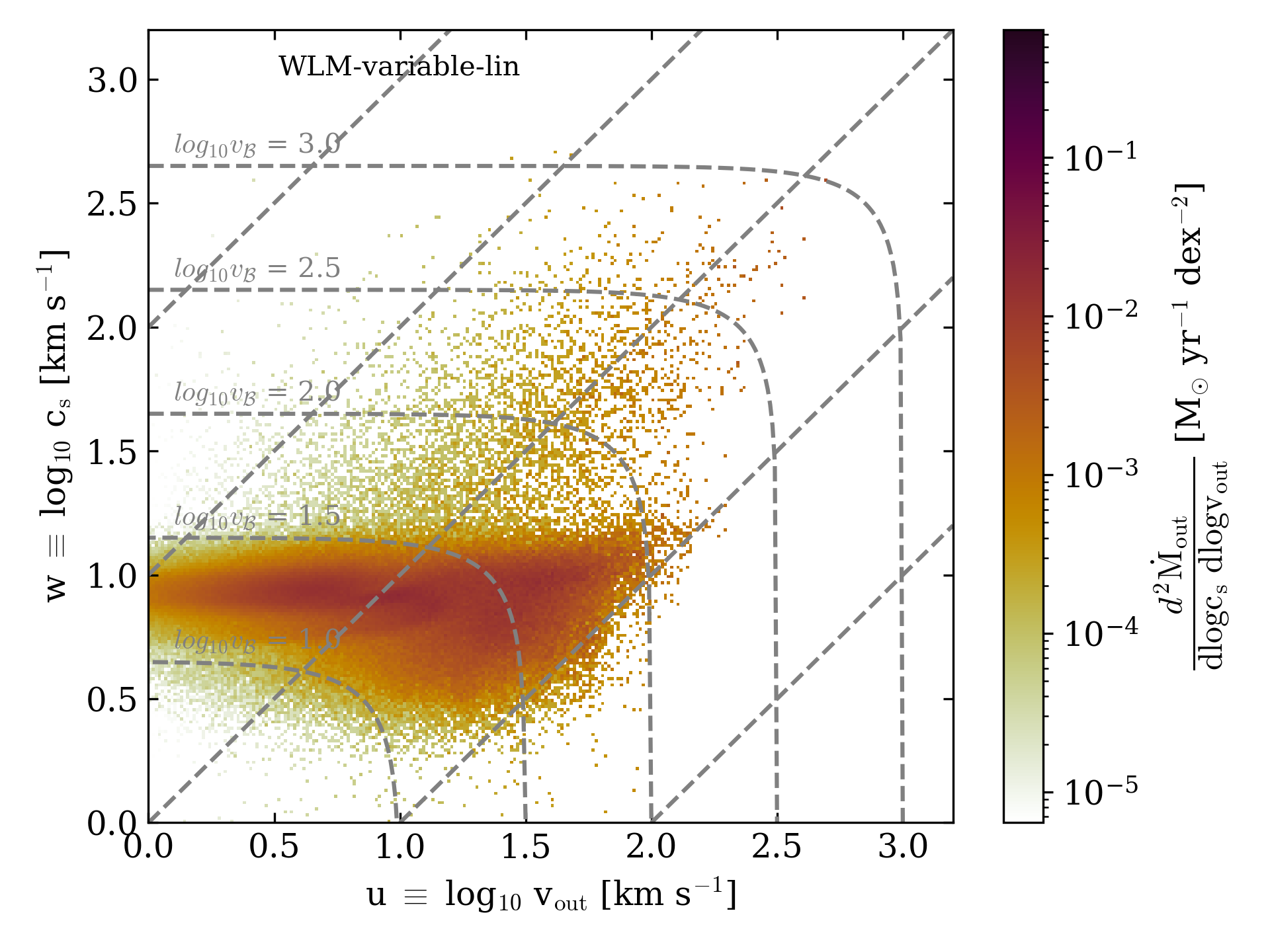}
    \includegraphics[width=0.32\textwidth]{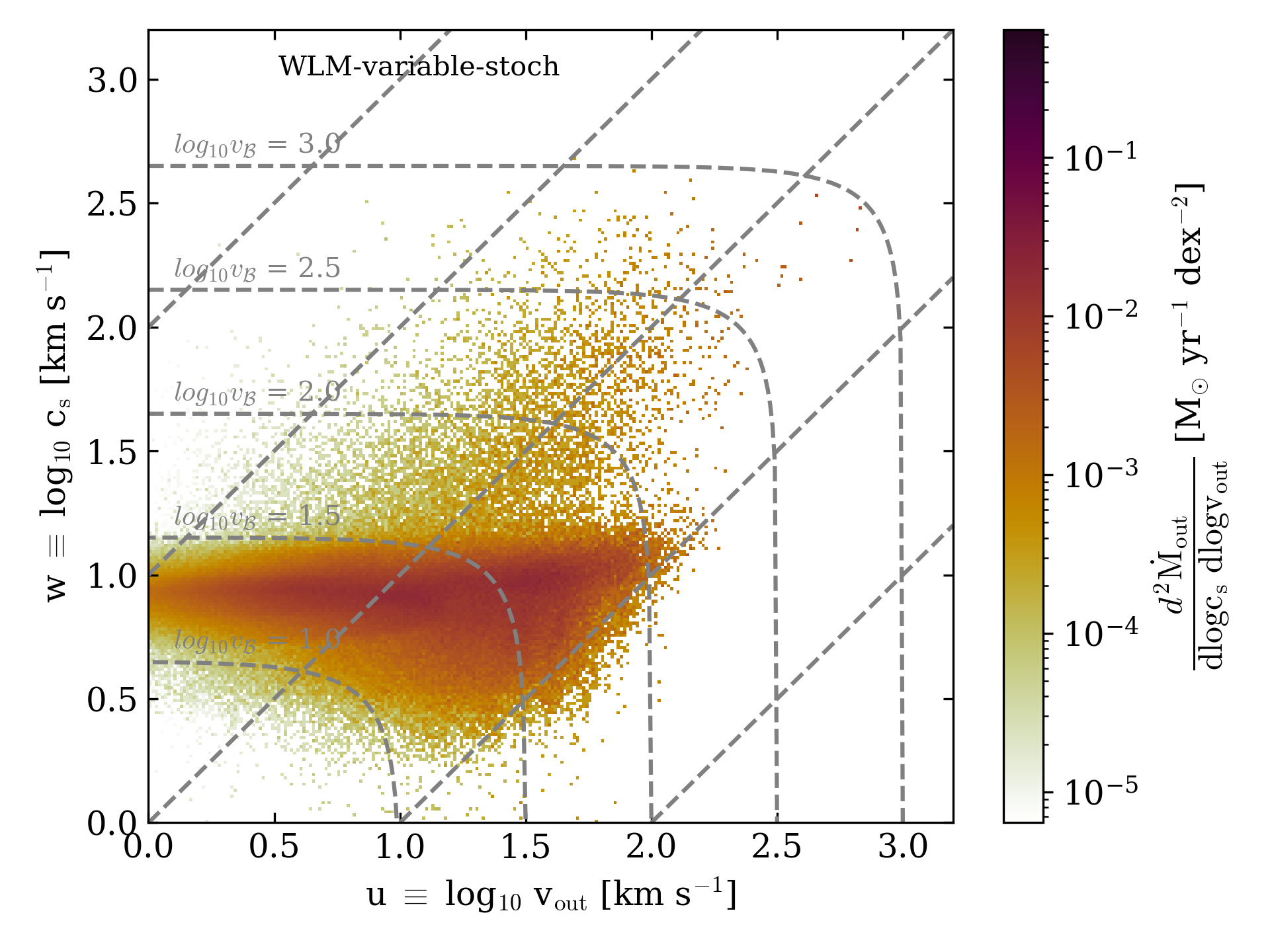}
    \includegraphics[width=0.32\textwidth]{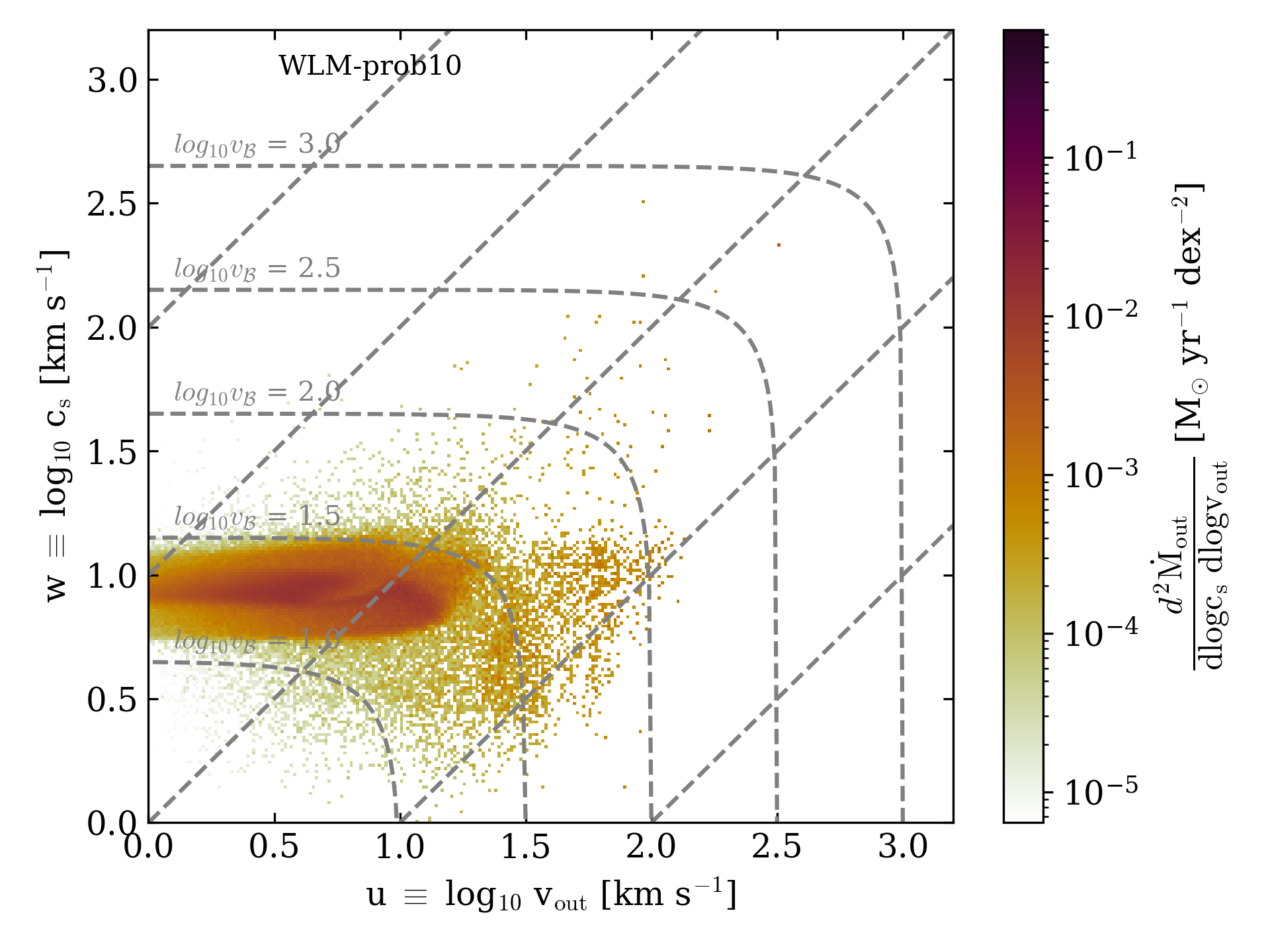}
    \includegraphics[width=0.32\textwidth]{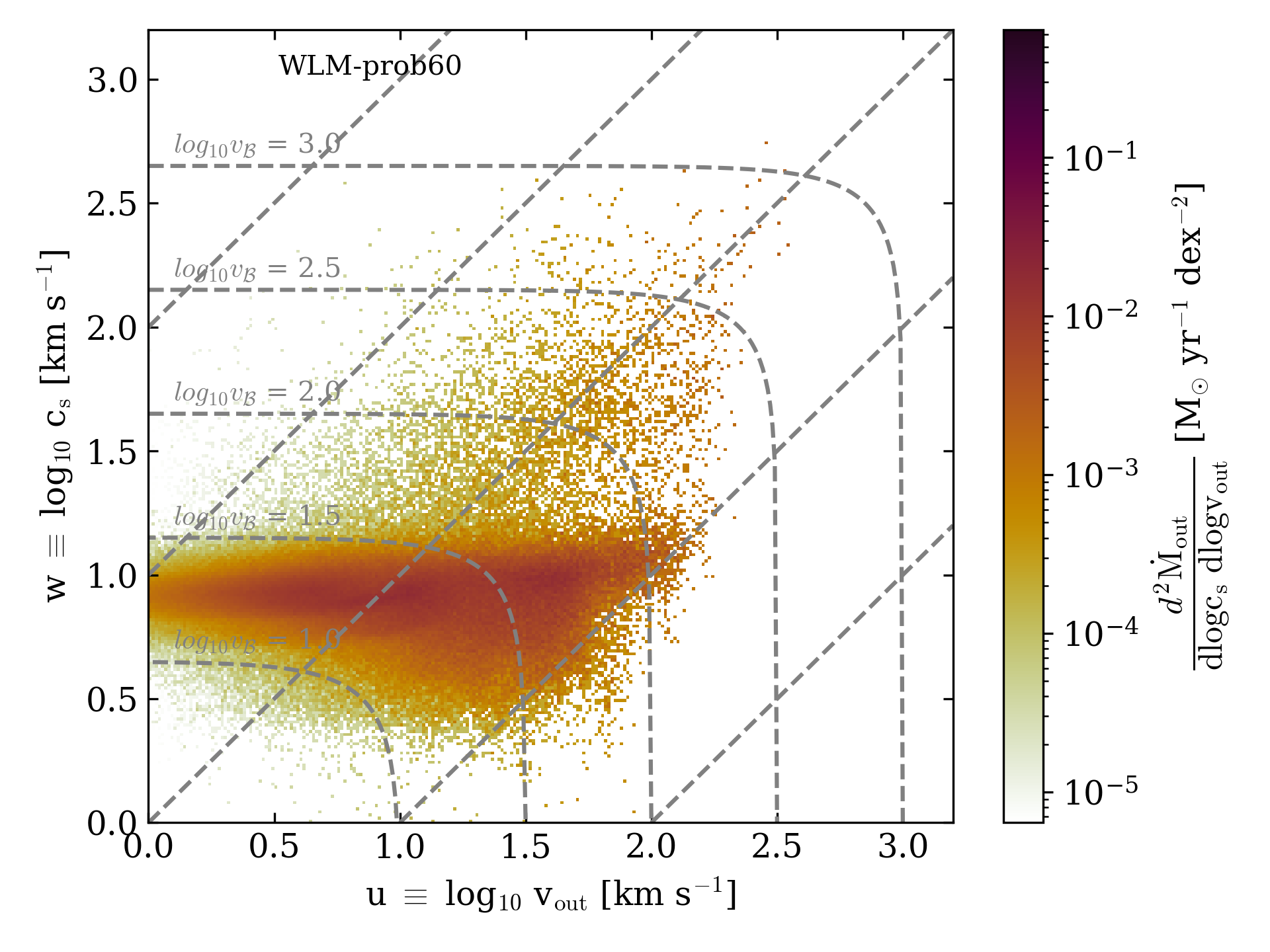}
    \caption{Joint 2D PDFs of outflow velocity and sound speed, color coded by the mass outflow rate for the models WLM-fid (top row, left), WLM-1e50 (top row, center), WLM-1e52 (top row,  right), WLM-variable (middle row, left), WLM-variable-lin (middle row, center), WLM-variable-stoch (middle row, right), WLM-10prob (bottom row, left) and WLM-60prob (bottom row, right).}
    \label{fig:mass_out}
\end{figure*}

\begin{figure*}
    \centering
    \includegraphics[width=0.32\textwidth]{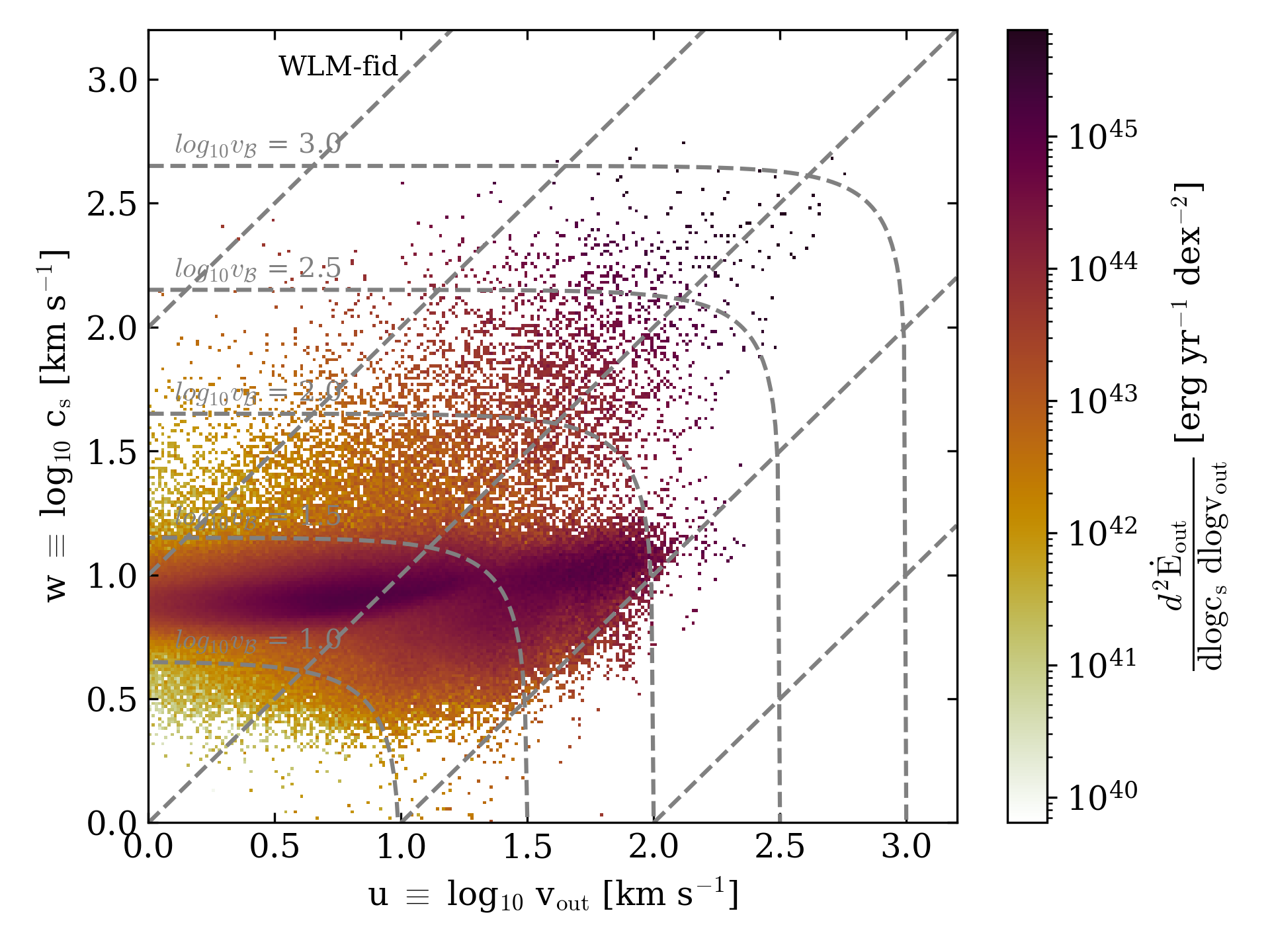}
    \includegraphics[width=0.32\textwidth]{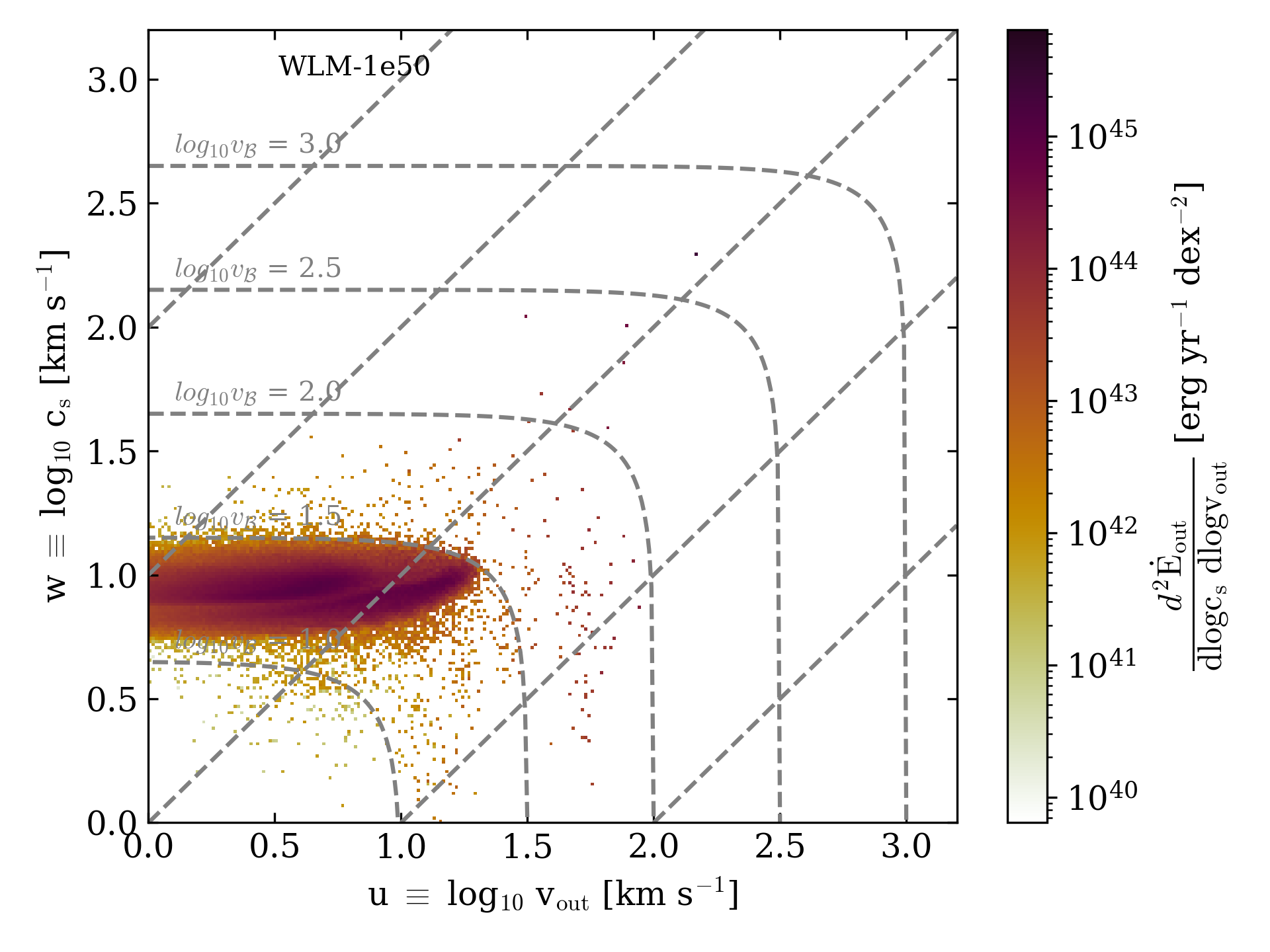}
    \includegraphics[width=0.32\textwidth]{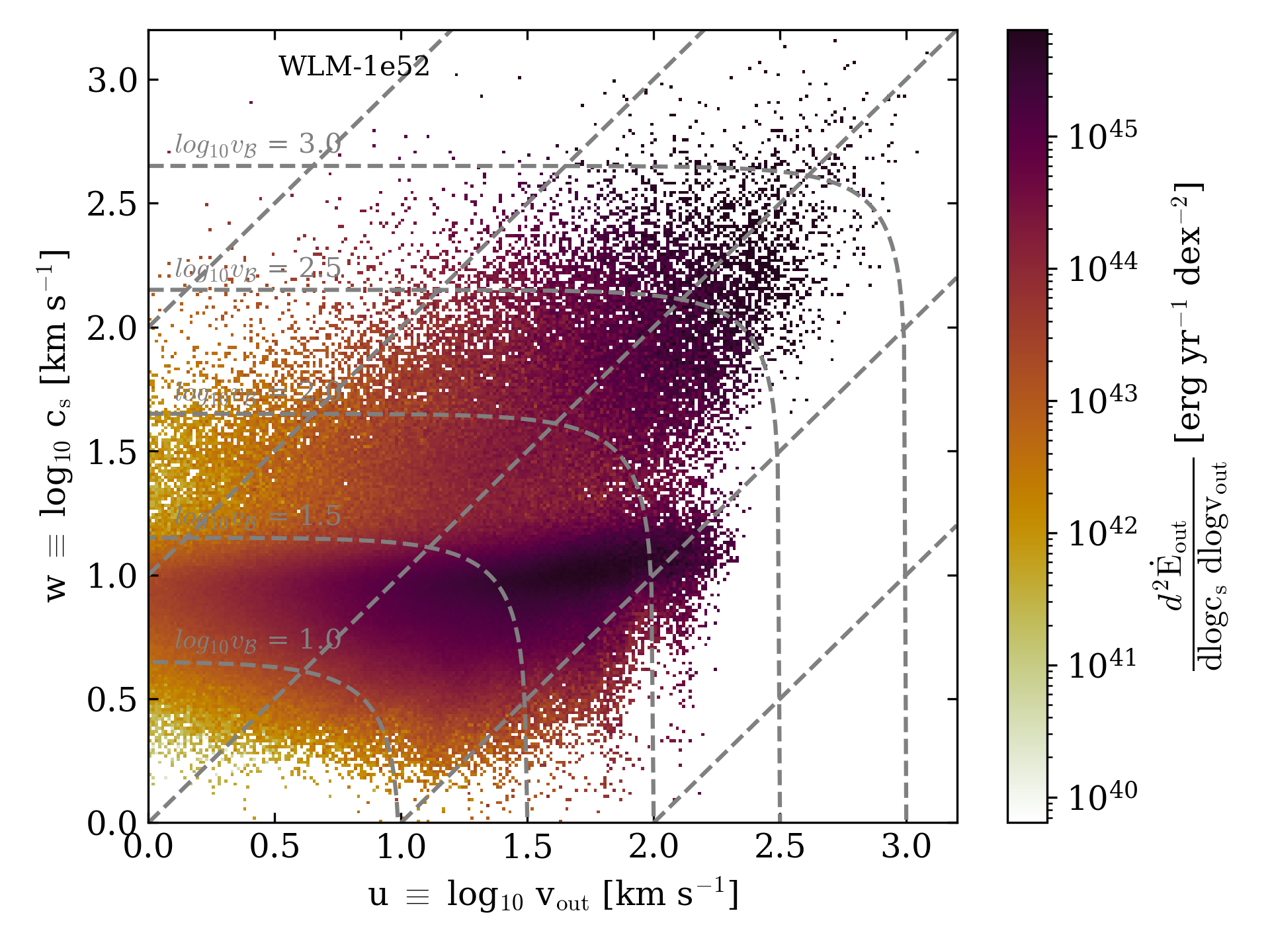}
    \includegraphics[width=0.32\textwidth]{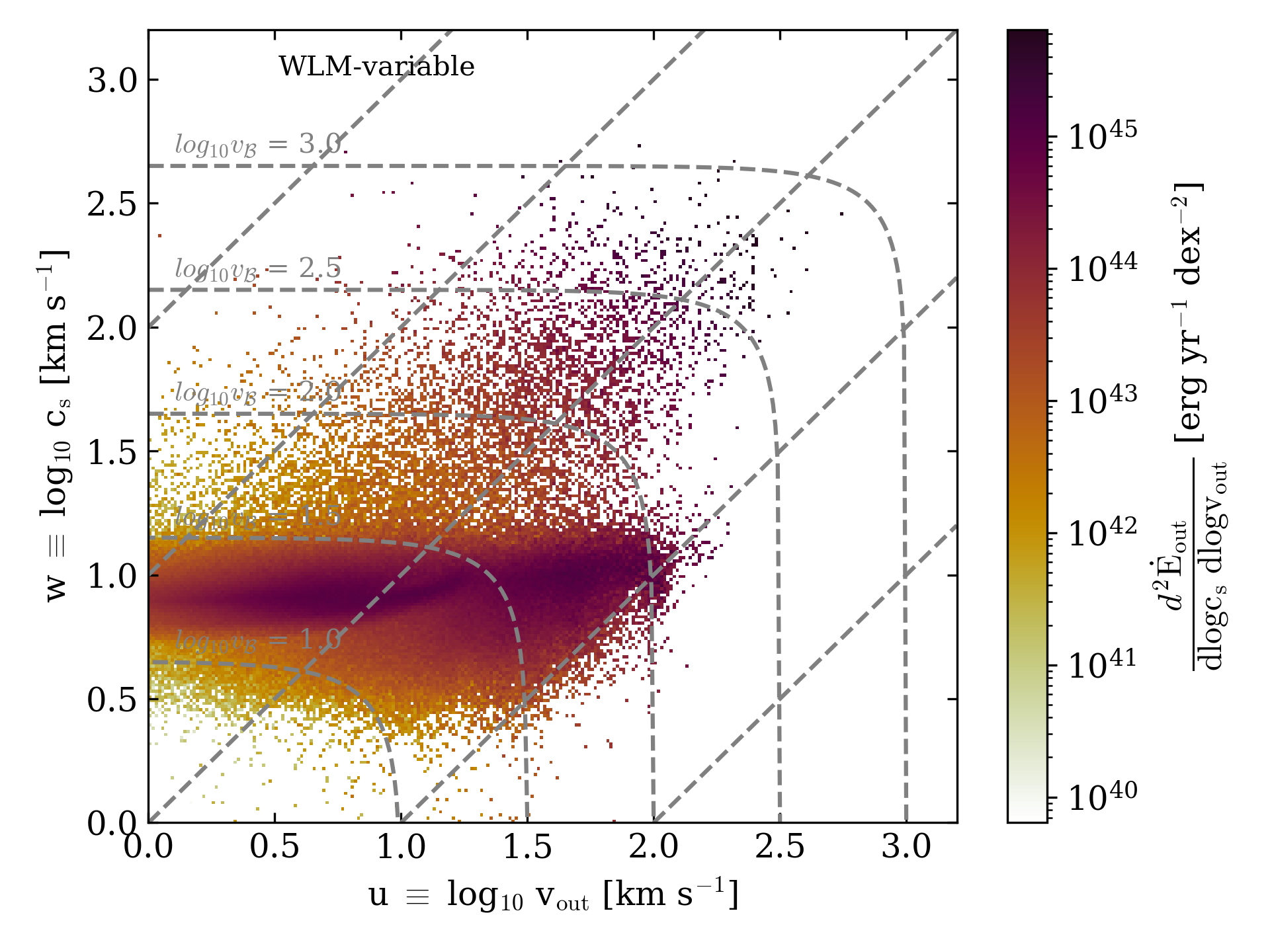}
    \includegraphics[width=0.32\textwidth]{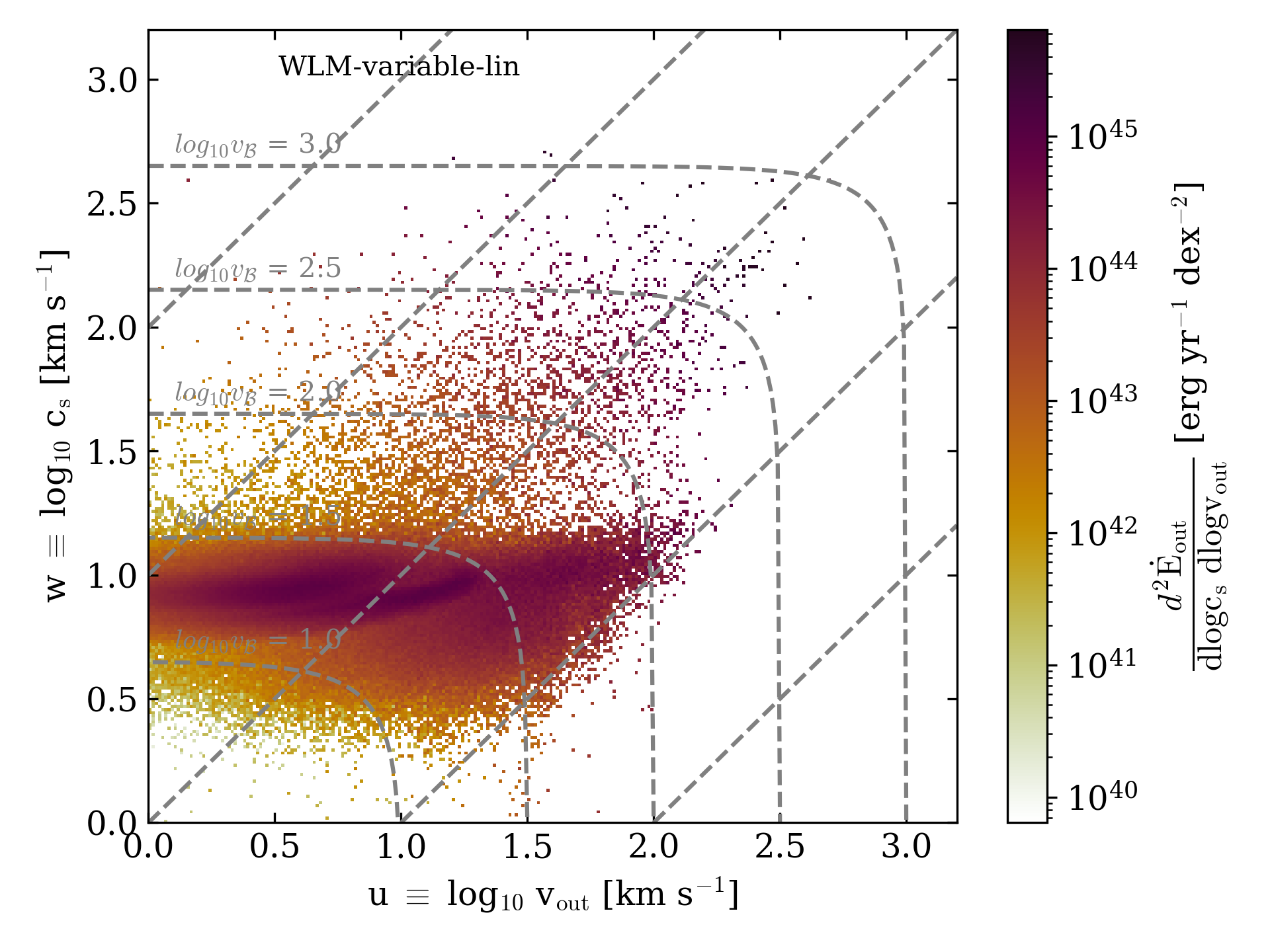}
    \includegraphics[width=0.32\textwidth]{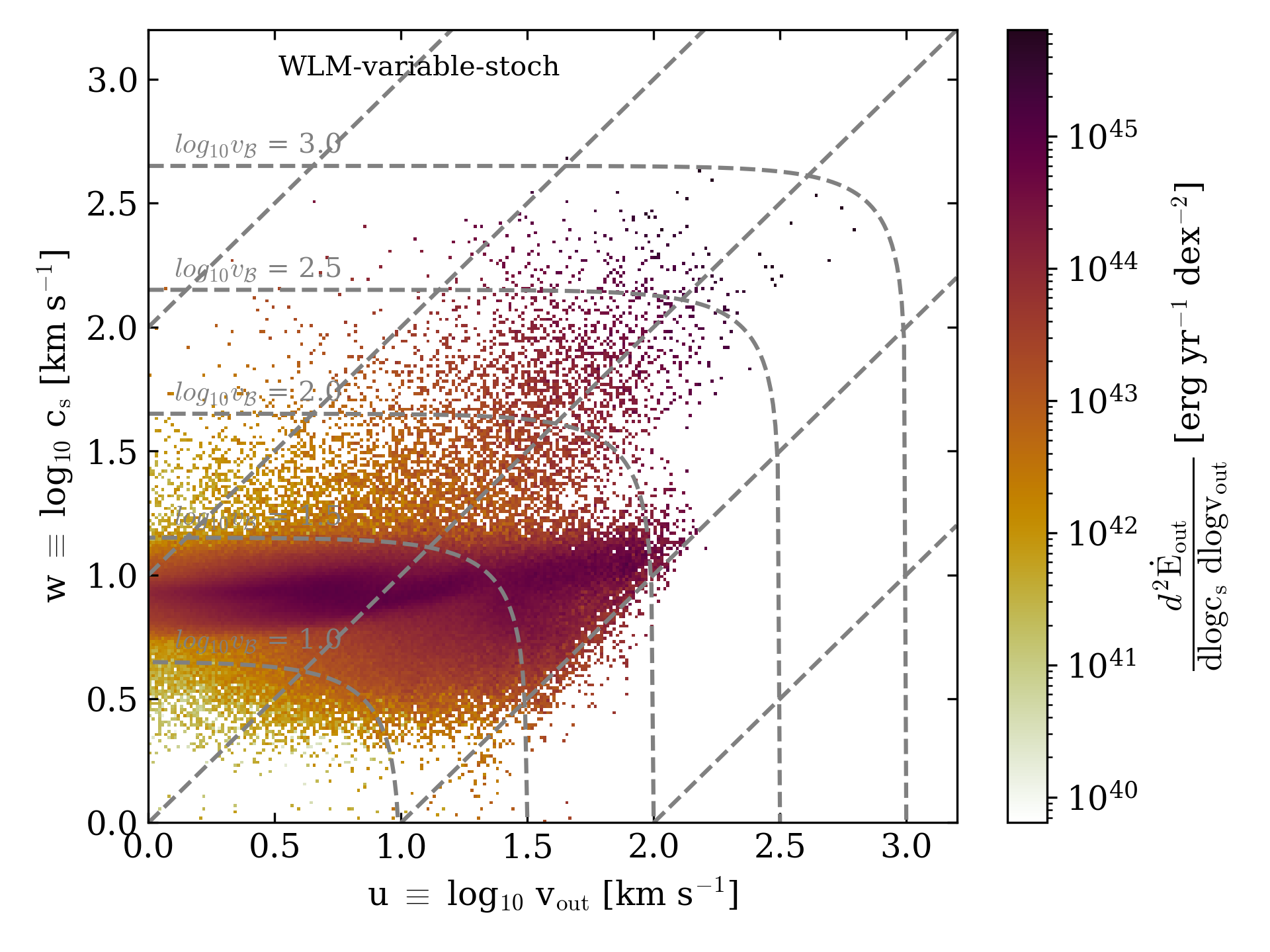}
    \includegraphics[width=0.32\textwidth]{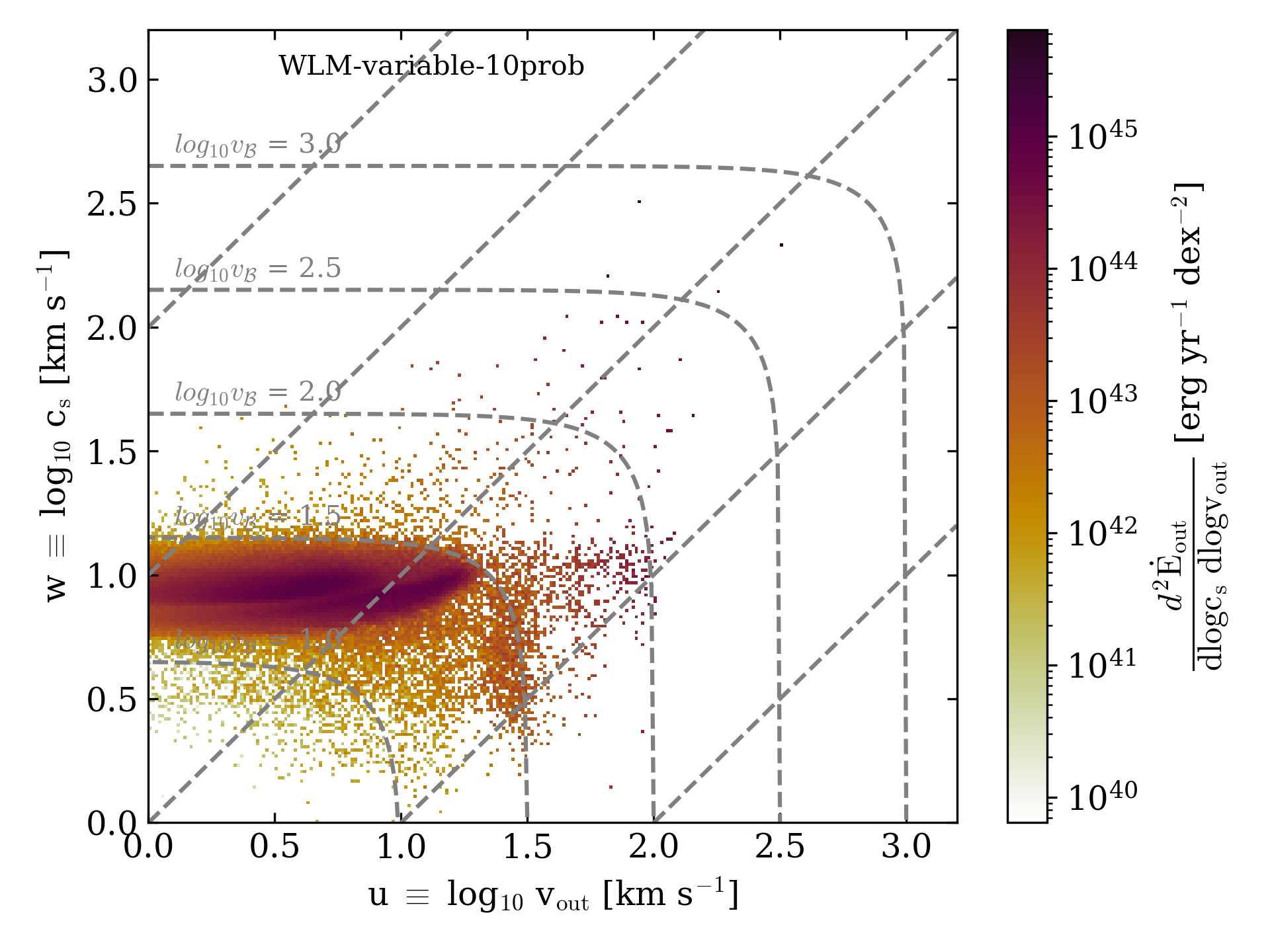}
    \includegraphics[width=0.32\textwidth]{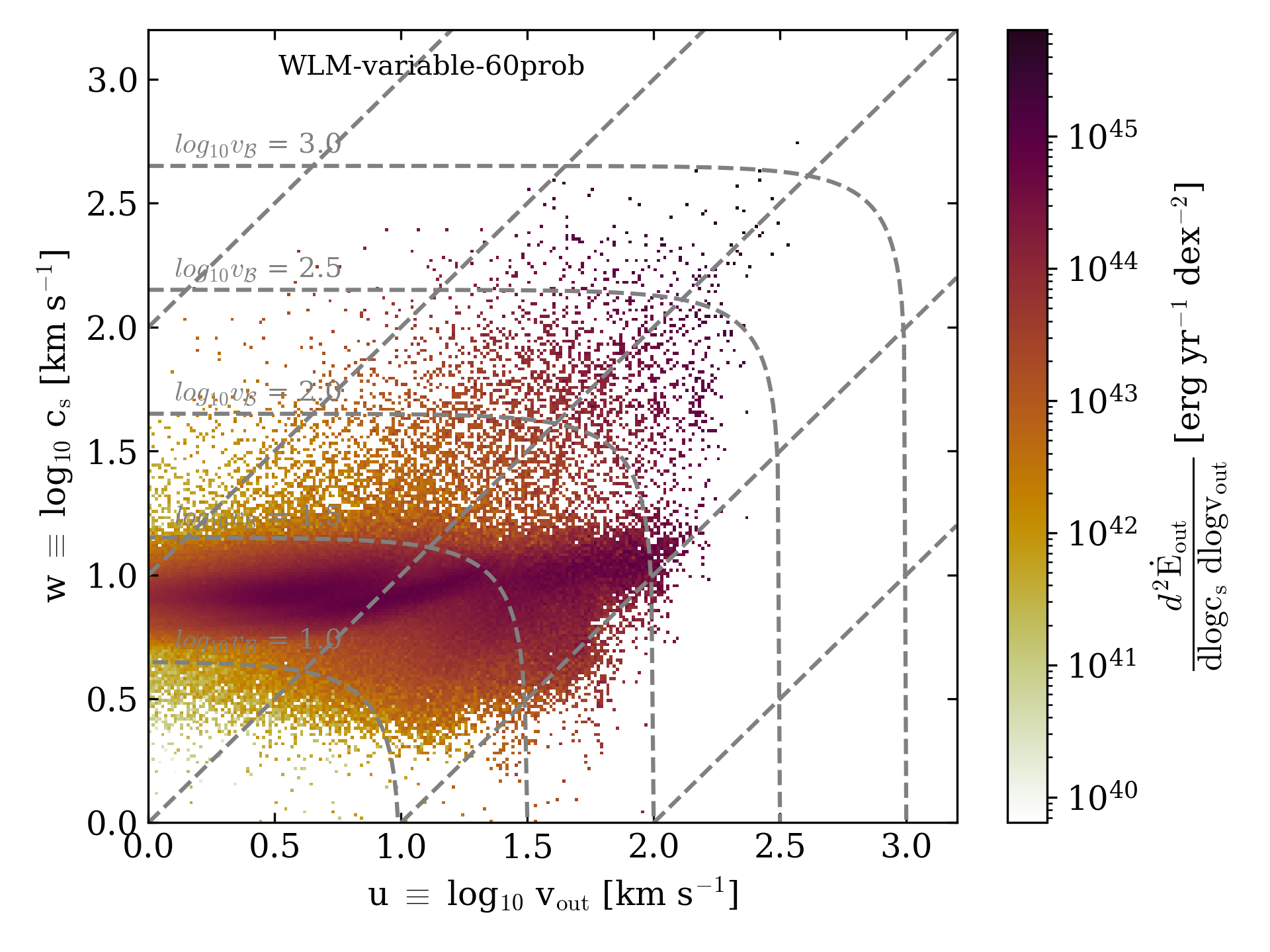}
    \caption{Joint 2D PDFs of outflow velocity and sound speed, color coded by the energy outflow rate for the models WLM-fid (top row, left), WLM-1e50 (top row, center), WLM-1e52 (top row,  right), WLM-variable (middle row, left), WLM-variable-lin (middle row, center), WLM-variable-stoch (middle row, right), WLM-10prob (bottom row, left) and WLM-60prob (bottom row, right).}
    \label{fig:energy_out}
\end{figure*}

In Fig.~\ref{fig:mass_out} and Fig.~\ref{fig:energy_out}, we show the joint 2D PDFs of outflow velocity and sound speed, color coded by mass outflow rate and energy outflow rate, respectively. This analysis is similar to the analysis that has been put forward in recent papers by \citet{Fielding2018}, \citet{Kim2020}, \citet{Steinwandel2022} and \citet{SteinwandelLMC} that study  multi-phase galactic winds. All the panels in these two figures show the mass (Fig.~\ref{fig:mass_out}) and energy (Fig.~\ref{fig:energy_out}) outflow on the colorbar, which we measure at a height of 1 kpc in a plane-parallel slab above and below a disc of  size $dr=100$ pc (in the terminology of equation \ref{eq:massout}). Generally, these 2D PDFs are useful in a time-averaged sense. In our work, we choose to average over a time frame of 400 Myrs -- between 300 Myrs and 700 Myrs of evolution of the system -- including a total of 50 snapshots per simulation.  When comparing the WLM-like systems in this work to more massive systems, such as the simulation of \citet{SteinwandelLMC} or simulations at higher column density as in \citet{Rathjen2022} or \citet{Kim2022}, we come to the striking realization that the hot wind in the WLM-like systems is more transient and thus far more sensitive to changes in the details of the feedback physics. Hence, small changes in SN-energies, as applied in this work, should, really,  ultimately be probed by time-averaging these 2D PDFs, as they show a clear indication of establishing a steady-state hot wind (as shown in our Fig.~\ref{fig:mass_out} and Fig.\ref{fig:energy_out}). However, the PDFs do have relatively similar shapes  in almost all the simulations we carried out, for both the mass and the energy loading, with three exceptions: the runs WLM-1e50, WLM-1e52 and WLM-10prob show very different shapes for the hot phase of the wind. The WLM-1e50 run has no hot wind at all -- in fact, the wind shows almost no sign of multiphase nature -- and there is no gas with a sound speed higher than $\sim 30$ km s$^{-1}$. In contrast, WLM-1e52 shows a hot phase that is very prominent  in both mass and energy. The model WLM-10prob shows little hot wind but still displays a tail towards the higher outflow rate in the warm wind. This tail is absent in the model WLM-1e50. This is important because it probably means that at the current resolution of our simulations, explosions with $10^{50}$ erg, such as those that occur in WLM-1e50, are poorly resolved and  incapable of launching a multiphase galactic wind, unlike explosions with $10^{51}$ erg. This hypothesis is backed up  by the fact that this run (WLM-1e50) has a very low mass (and energy) loading, as demonstrated in Fig.~\ref{fig:mass_out} and Fig.~\ref{fig:energy_out}. However, the ultimate test of this hypothesis can only be realized by another experiment beyond the scope of this work, which we plan to carry out in the future. In such an experiment, one would need to run the WLM-1e50 simulation at 10 times higher resolution and see if a change in  wind-driving capabilities emerges. This would have to be accompanied by an unresolved momentum feedback run at our current resolution based on simulations that do not currently exist in the literature (because one would need to determine the momentum input as a function of number density at the end of shell formation, as is typically done for $10^{51}$-erg explosions, for instance in \citet{Kim2015} and \citet{Steinwandel2020}). 
If that upcoming numerical experiment arrives at the conclusion that there will be a mass- and energy-loaded wind, it would mean that sub-$10^{51}$-erg explosions are poorly resolved in our current framework. If not, it likely means that low-energy explosions around $10^50$ erg are  contributing only to the momentum budget of the ISM, which drives the turbulence and regulates star formation, but not significantly to the driving of galactic outflows. This is important for  realistic simulations of SN feedback in nature, as some lower-energy supernova explosions are  achieved in many self-consistent 1D and 3D SN explosion simulations \citep[see, e.g.,][]{Sukhbold2016,Vartanyan2018,Muller2019,Stockinger2020}, and are also observationally supported by the sub-population of low-luminosity hydrogen-rich SNe \citep[e.g.,][]{Kozyreva2022, Valerin2022}.

In that context, it is interesting to point out that all the WLM-variable+ models based on the tables of \citet{Sukhbold2016} only marginally affect the hot wind structure when compared to the run WLM-fid. This suggests that using $10^{51}$ erg is sufficient for studying multiphase galactic winds in dwarf galaxies, where we find only minor modulations to the star formation rate (by a factor of 2-3) in runs with variable energies. Moreover, we find that the leading cause of this effect is not the modulation of the feedback energy between $0.5 \times 10^{51}$ erg and $2 \times 10^{51}$ erg but rather the absence of explosions for some stars. Now, it is not clear if these unexploded stars really do not contribute to the energy budget or if this is a limitation of the 1D approach taken in \citet{Sukhbold2016}, which assumes that the unexploded stars collapse directly into black holes without any weaker mass and energy ejection. We will discuss this issue in greater detail in Sec.\ref{sec:stellar_models}. 

\section{Discussion} \label{sec:discussion}

\subsection{Comparison to previous work}

The most important points of comparison for our study are the \textsc{arepo} simulations by \citet{Gutcke2021}, who simulated a lower-surface-density dwarf with a  disc-scale length that is larger than ours by a factor of $\sim 1.5$ but with otherwise identical parameters for halo mass and stellar background potential.  Their simulations are also quite comparable to ours in terms of mass and spatial resolution as well as the modeling for the multiphase ISM. The  other main difference between their model and ours is their  use of the tabulated cooling tables of \citet{Ploeckinger2020} while we adopt the non-equilibrium chemistry network based on \citet{Glover2007a}, \citet{Glover2007b}, as implemented in \citet{Hu2017} and \citet{Steinwandel2020}. Additionally, our simulations model the evolution of the FUV field in order to accurately estimate the spatially dependent and time-dependent photoelectric heating rate in the ISM, as well as the photoionizing radiation. Both these processes are omitted in \citet{Gutcke2021}. Nevertheless, it is interesting to report that we find very similar values for mass and metal loading for \textit{all} of our simulations when compared to \citet{Gutcke2021}. Both models find values of around unity for the mass loading and around 0.01 for the metal loading factor. The metal enrichment factors are similar as well, with values of around 2. However, we find some spikes in the metal enrichment factor that appear to be remnants from very clustered single feedback events based on run-to-run variations in the single-star formation model. The biggest differences in terms of loading factors seem to appear in the energy loading, where we find values of around 0.01 to 0.1 while \citet{Gutcke2021} find values between $10^{-4} $ and $10^{-3}$. However, the energy loading factors tend to fluctuate quite a bit with different modeling \citep[see][for a more detailed discussion on this topic]{Hu2023} so this might not be very surprising. More interestingly, we find a much weaker impact on the runs with variable SN-feedback energy (WLM-variable, WLM-variable-lin and WLM-variable-stoch) compared to our fiducial run WLM-fid, which has a fixed SN-energy of $10^{51}$ erg. Whereas \citet{Gutcke2021} find a factor of 2-3 in suppression of the strength of the outflow and almost a dex decrease in star formation, we find an effect of around 30 to 50 percent on the loading factors and  fluctuation in the star formation rate by a factor of up to three for all models. One reason for this might be that, as we have found in some of our past experiments, the lower-surface-density version of WLM that is simulated in \citet{Gutcke2021} is quite sensitive to modest changes in star formation and feedback modeling \citep{Hu2019, Steinwandel2022}, but the higher-column-density version simulated here is not \citep[see also][]{Hislop2022, Lahen2023}. 

There are a few other notable studies of resolved dwarf galaxies, such as \citet{Hu2016}, \citet{Hu2017}, \citet{Hu2019}, \citet{Emerick2019}, \citet{Smith2021}, \citet{Hislop2022} and \citet{Lahen2023}, that generally show very similar results in terms of star formation history, phase structure of the ISM and the evolution of outflow loading factors, where values of around unity for the mass loading and around 0.05 to 0.5 for the metal loading are reported. While the loading factors for mass and metals typically vary by an order of magnitude, the energy loading, as discussed above, typically varies by 2-3 orders of magnitude between different physics implementations. 

The simulations of \citet{Fielding2017} also find values for the mass loading of around unity and find values for the  energy lodging that are lower than ours by roughly a dex. While their dwarf galaxy simulation is similar to ours, their setup is much leaner and does not include first-principle star formation or self-gravity. Hence, they cannot really make a statement about the metal loading or metal enrichment. However, it would be interesting to use such a simplified setup to test the effect of variable SN-energies in a more idealized environment.

Additionally, we can gain insight from so-called ``tallbox'' simulations, where these loading factors tend to be in rather good agreement with our values for mass, metal and energy loading. Specifically, \citet{Li2017}, who simulate a solar-neighborhood environment with the grid code \textsc{enzo} and find mass, metal and energy loading values of around $\sim 6$, $\sim 0.2$ and $\sim 0.1$, respectively, are in quite good agreement with basically \textit{all} of our simulations apart from  WLM-1e50 and WLM-1e52, which yield significantly lower/higher loading factors  than the other models (by roughly a dex). Values similar  to those in\citet{Li2017} are reported in other ``tallbox'' simulations as well, such as \citet{Kim2015_outflows} and \citet{Kim2020}. 

Finally, we would like to note that another interesting approach would be to ``zoom in'' to regions with stellar feedback activity to study the detailed evolution of individual superbubbles in greater detail. To that end, an approach that has  recently been used in simulations of black-hole accretion \citep[][]{Hopkins2023a, Hopkins2023b, Hopkins2023c}, which operates by splitting particles and simultaneously switching on relevant physical processes, could be applied. In that context, our model would work down to  a resolution scale of around 0.1 M$_{\odot}$, which is still 40 times higher in mass resolution than the uniform resolution of 4 M$_{\odot}$ adopted in this work. For higher mass resolution, a model such as STARFORGE \citep{Grudic2022, Guszejnov2022} would be a more appropriate choice. 

\subsection{Supernova-Energy-variability vs. SN-stochasticity}

The central question that can be tackled by our set of simulations is whether the changes in star formation rate in our simulations (by factors of 2-3) and the trend in changes in loading factors (by 30 to 50 percent) are driven by the change in the overall feedback energy or by the stochasticity of the SN-explosions, in the sense that the models with  energy that varies as a function of each individual star's mass also have a good number of massive stars that do not explode to begin with. First of all, if we only consider the models WLM-fid, WLM-1e50 and WLM-1e52, we clearly find that the loading factors we measure are significantly different -- those in the WLM-1e52 case are larger by a factor of 100 than those for WLM-fid, which are again larger by a factor of 100 than those for WLM-1e50. The same is true for metal and energy loading as well. Hence, it is quite clear from those experiments that the exact number for the SN explosion energy does matter for both the star formation, as demonstrated in Fig.~\ref{fig:sfr}, and the evolution of the loading factors, as shown in Fig.~\ref{fig:loadings} and Fig.~\ref{fig:metal_loadings}. 

When we take the models WLM-variable, WLM-variable-lin and WLM-variable-stoch into consideration, the relationship becomes a bit more complicated. First, these models have a star formation rate that is, generally speaking, higher by a factor of 2-3 compared to the model WLM-fid (which appears to have the lowest star formation rate after the model WLM-1e52). This implies that the ISM is self-regulating and is reacting to the exact energy input of stellar feedback by adjusting its star formation rate. This also explains why we observe the highest star formation rate (on average) in the model WLM-1e50. Hence, it seems that the systems compensate for the lower or higher energetics per event by adjusting their star formation rate. This makes sense, as star formation is a self-regulating process. 

In that context, it is interesting to take the runs WLM-10prob and WLM-60prob into consideration. WLM-60prob is the run that would give the same explosion probability that we would get by integrating over the tables of \citet{Sukhbold2016}; however, instead of applying the SN-explosion energy and outcome prescribed by the tables, we simply apply $10^{51}$ erg for every event, but with only a 60 percent probability of having an event to begin with. Interestingly, the star formation rate and the loading factors in the WLM-60prob model  are very similar to those in the models WLM-variable, WLM-variable-lin and WLM-variable-stoch, which would indicate that it does not matter so much if the SN-feedback energy is varied between $0.2 \time 10^{51}$  and $2 \time 10^{51}$ erg based on the tables of \citet{Sukhbold2016}. 
It seems to matter more how sensitive that process is to what \citet{Sukhbold2016} and others often refer to as the ``stochasticity" of the explosion outcome -- more specifically for our purposes, the fraction of massive stars that explode as SNe to begin with.
Moreover, the run WLM-10prob is most interesting when compared to the run WLM-1e50, which should have a similar overall energy budget. While the WLM-10prob run is still driving a mass- and energy-loaded outflow, the run WLM-1e50 does not drive any outflow out to a height of 3 kpc. This means one of two things: one possibility is that the SNe in the WLM-1e50 case are mostly ``stuck'' in the momentum phase, where they generate enough momentum to regulate star formation in the ISM but do not generate enough hot material to drive superbubbles. On the other hand, this might be an indication that the SNe in the WLM-1e50 case are poorly resolved in the current framework at a mass resolution of $4$ M$_{\odot}$. Ultimately, this can only be tested by performing higher-resolution simulations of the system, which is beyond the scope of this work.

\subsection{Stellar physics as sub-grid model for galaxy formation theory \label{sec:stellar_models}}

There is much active research in the stellar astrophysics community that would be valuable to consider in the context of determining galactic evolution through its impact on the outcomes of stellar evolution. 
Central to this effort, in the context of this present work, are two questions: Which stars explode, and with how much energy? If some do not explode, what is the nature of the non-explosions? 

Although there is ongoing debate over which exact ``explosion criteria" determine whether a star will explode as a supernova, there is a general consensus that the radial density profile, the electron fraction, and other physical conditions near the stellar core play  important roles in shaping the explosion outcome \citep[see, e.g.,][and many others]{Janka2012,Fryer2012,Fryer2022,Ertl2016,Muller2016,OConnor2018,Vartanyan2018,Burrows2021,Tsang2022}. 
In fact, a growing number of tools whose development was motivated by the results of detailed 3D simulations can directly carry out calibrated stellar explosions when given the stellar structure at core collapse, such as the semi-analytic framework of \citet{Muller2016}, \citet{Zha2023}, and \citet{Takahashi2023} (among others), the PUSH framework \citep{Perego2015,Ebinger2019,Ebinger2020,Curtis2019,Curtis2021,Ghosh2022} and the STIR framework \citep{Couch2020,Barker2022}. 
In tandem with progress in stellar progenitor modeling, these calibrated explosions will continue to be deployed to refine predictions concerning which stellar masses will or will not explode. 

To that end, 1D (spherically symmetric) stellar evolution models have not only generated grids of models with the goal of mapping from progenitor mass to explosion outcome and energy \citep[e.g.,][]{Woosley2002,Woosley2007,Sukhbold2014,Sukhbold2016,Sukhbold2018,Takahashi2023} but have also shed insight into the strong sensitivity of those grids to uncertainties in stellar input physics. 
In particular, the stellar structure at the time of explosion is sensitive to the treatment and extent of the convective boundary mixing in the stellar core \citep[e.g.,][]{Farmer2016,ADavis2019,Wagle2019,Wagle2020}, the treatment of nuclear reactions and nuclear reaction rates \citep[e.g.,][]{Farmer2016,Sukhbold2020,Farag2022},  stellar rotation \citep[e.g.,][]{Akiyama2005,Li2023}, and stellar mass loss \citep[e.g.,][]{Renzo2017,Ertl2020}. Moreover, the stellar core structure and, therefore, the explosion outcome are sensitive to 3D effects not readily captured in 1D models, such as the angular momentum transport in convective and rotating shell-burning regions \citep{Fields2020,Fields2021}. 

As most massive stars are born in multiple-star systems and may experience interaction within their lifetime \citep[e.g.,][]{deMink2014}, binarity is also becoming an increasingly important ingredient in any recipe for stellar outcomes. Binary stellar evolution has been shown to greatly impact stellar lifetimes and, therefore, explosion timing (e.g., \citealt{Zapartas2017,Stanway2018}), as well as core structure \citep[e.g.,][]{Laplace2021, Schneider2021,Zapartas2021,Zapartas2021b} and binding energy \citep[e.g.,][]{Renzo2023}, which directly affect the explosion outcome. These effects, in the context of successful and failed explosions, can potentially be probed by binary black hole populations seen by LIGO \citep{Schneider2023}. Ongoing efforts are also directed towards implementing these findings in the context of rapid binary population synthesis \citep[see, e.g.,][]{Zapartas2017,Eldridge2022} and direct binary stellar evolution population synthesis modeling, e.g., with the recently-developed POSYDON code \citep[][]{Fragos2023} built upon models constructed with the MESA stellar evolution software \citep{Paxton2011,Paxton2013,Paxton2015,Paxton2018,Paxton2019, Jermyn2023}. This can be leveraged to directly predict stellar explosion outcomes and energies in binary stellar populations \citep{Patton2020,Patton2022}. Such efforts can suggest potential sub-grid models not just for single-star-resolution simulations in low-mass galaxies such as those presented here, but also for population-averaged outcomes for use in larger-scale galactic simulations. 

Moreover, there has been considerable effort put forth in mapping out where the ``edge cases" in explosion outcomes lie. On the lower-mass ($\sim$ 8-10$M_\odot$) end, super-asymptotic giant branch stars will explode as electron-capture-induced supernovae \citep[e.g.,][]{Nomoto1987,Poelarends2008,Poelarends2017,Jones2016,Cinquegrana2023}, which differ in their chemical yields and energetics depending on the explosion mechanism \citep[][]{Jones2019}. On the high-mass end, very massive stars ($\gtrsim80M_\odot$) will eject some of their envelopes during their dying days and explode as pulsational- and pair-instability supernovae \citep[e.g.,][]{Rakavy1967,Woosley2002,Chatzopoulos2012,Farmer2019,Leung2019,Renzo2020tdc,Renzo2020hfree,Farag2022}. 

Finally, there is a growing body of work dedicated to what happens to the stars that do not explode. 
One scenario of interest is that weak shocks arising in failed explosions may cause energetic outbursts, induced either by the neutrinos from the core collapse \citep[e.g.,][]{Nadezhin1980, Lovegrove2013, Coughlin2018, Quataert2019, Schneider2022} or by the circularization of the inner layers of the infalling envelope, which has substantial random angular momentum \citep{Quataert2019, Antoni2022, Antoni2023}. These span mass and energy scales ranging from $\sim10^{47}$-$10^{49}$ erg and from fractions of a solar mass to tens of solar masses for partial or complete envelope ejection. Such transients are of particular interest in the context of this work, as the results here are shown to be sensitive to the zero-energy feedback of the failed supernovae (as $0\ll10^{51}$ erg, it is still the case that $10^{49}\gg0$ erg).
If these feedback episodes do occur at the site of a failed core-collapse explosion, perhaps galaxy simulations that resolve them can be compared to observations in order to constrain the energetics of outbursts. 

While the discussion here has focused on the final collapse and SN explosion feedback, it should also be mentioned that massive stars are observationally known to have violent outbursts that resemble SN imposters spectroscopically and energetically \citep[e.g.,][]{Smith2011,Smith2020,Andrews2021,Aghakhanloo2023},
so it might also be the case that some of these dying massive stars have multiple lower-energy feedback input episodes. This too is worth exploring in resolved galactic simulations. For example, it has been shown by \citet{Keller2022_stellar_haloes} that the exact mode of injection of the feedback energy matters. The same statement would hold for multiple feedback episodes of massive stars. 

Advanced grids of stellar models and their explosions continue to emerge, built with the goal of advancing knowledge in the field of stellar physics. We want to highlight the power of resolved galactic-scale simulations, such as the ones presented in this work, to test the predictions emerging from these stellar models by allowing  detailed comparisons with observed data from nearby galaxies and building upon theoretical breakthroughs to better constrain feedback processes in the ISM. 

\section{Conclusions} \label{sec:conclusions}

\begin{table*}
    \caption{Summary of the mean loading factors at hot wind formation (3 kpc height)}
    \centering
    \label{tab:loadings}
    \begin{tabular}{lccccccc}
      \hline
	  Name & SFR [M$_{\odot}$ yr$^{-1}$] & $\eta_\mathrm{M}^\mathrm{out}$ & $\eta_\mathrm{E}^\mathrm{out}$ & $\eta_\mathrm{Z}^\mathrm{out}$ & $y_\mathrm{Z}^\mathrm{out}$\\
	  \hline
      WLM-fid & 0.0012 & 0.90 & 0.019 & 0.081 & 2.14\\
      WLM-1e50 & 0.0025 & 0.011 & 0.00092 & 0.00068 & 1.65 \\
      WLM-1e52 & 0.00021 & 94.5 & 0.95 & 5.02 & 1.34 \\
      WLM-variable & 0.0016 & 0.42 & 0.020 & 0.043 & 2.37 \\
      WLM-variable-lin & 0.0016 & 0.30 & 0.013 & 0.031 & 2.35 \\
      WLM-variable-stoch & 0.0017 & 0.33 & 0.014 & 0.035 & 2.38 \\
      WLM-10prob & 0.0026 & 0.25 & 0.005 & 0.027 & 1.78 \\
      WLM-60prob & 0.0016 & 0.40 & 0.008 & 0.045 & 2.55 \\
      \hline
      \end{tabular}
\end{table*}

We have presented a set of isolated disk galaxy simulations with variations in the SN-feedback energy and the explosion outcome at fixed metal yield to isolate the role of the SN-feedback energy in galaxy evolution. The key findings of our study can be summarized as follows:
\begin{enumerate}
    \item The exact value of the SN-energy  has a minor effect on the morphological structure of the ISM. However, at lower feedback energy, the ISM appears more filamentary than at high energy, where the structure is more porous.
    \item Varying the SN-energy based on the tables of \citet{Sukhbold2016} has a minor effect on the phase structure of the ISM: we find that the build-up of the hot phase of the ISM in density-temperature phase space is unaffected. However, the star formation rate is affected and increases by around 30 percent in the mean and by around a factor of two in the peaks in the simulations that follow the data of \citet{Sukhbold2016}. However, this seems to be an effect that is  driven not by the energy variation but rather by the fact that there are a number of non-exploding SNe in the tables of \citet{Sukhbold2016} that we explicitly tested in our study.
    \item The extreme runs WLM-1e50 and WLM-1e52 show that explosions with $10^{52}$ erg produce a highly mass- and energy-loaded wind, whereas explosions with $10^{50}$ erg produce a wind that is weaker by at least 2 orders of magnitude in mass and energy loading when compared to the mean values recovered for \textit{all} our other simulations. 
    \item In contrast, the run WLM-10prob, which is energetically similar to WLM-1e50, does produce a wind that is only slightly less mass- and energy-loaded than the reference run WLM-fid. Both runs react to the missing energy in the ISM by increasing their respective star formation rates. However, in the case of the run WLM-1e50, the SNe remain ineffective at producing a wind. This implies that the $10^{50}$-erg explosions may be unresolved in the framework, and further testing with more simulations  will be necessary to ultimately determine if this is true.
\end{enumerate}

Simulations of galaxy evolution in the dwarf galaxy regime have progressed substantially over the course of the last five years, to the level where single-star resolution is achievable. Thus, the ability to use the input of stellar evolution theory to build sub-grid models in this new kind of ``resolved feedback'' simulation of galaxy evolution is essential to push the envelope. In turn, these simulations are a great testbed for stellar evolution theory in the context of galaxy and ISM observations, as they can help establish which physical processes and outcomes of stars affect their large-scale environment and which do not. 

\section*{Acknowledgements}
We acknowledge useful discussions with Mathieu Renzo. UPS and JAG are supported by the Flatiron Institute. The Flatiron Institute is supported by the Simons Foundation. Fig.~\ref{fig:splash} was generated using \textsc{splash} \citep[][]{Price2007}. 

\vspace{5mm}
\facilities{SuperMUC-NG (LRZ Garching, pn72bu), Rusty (Flatiron Institute)}

\software{P-Gadget3 \citep[][]{springel05, Hu2014, Steinwandel2020}, astropy \citep{2013A&A...558A..33A,2018AJ....156..123A}, CMasher \citep{cmasher}, matplotlib \citep{Matplotlib}, scipy \citep[][]{scipy}, numpy \citep[][]{numpy}}

\bibliography{sample631}{}
\bibliographystyle{aasjournal}

\end{document}